\documentclass{report}
\usepackage[a4paper, left=3cm, right=3cm, top=4.0cm, bottom=2.5cm, headsep=1cm]{geometry}
\usepackage{epsfig}
\usepackage{subfigure}
\usepackage[T1]{fontenc}
\usepackage{graphicx}
\usepackage{amsfonts} 
\usepackage{latexsym} 
\usepackage{amsmath,amssymb}
\usepackage{bm}
\def\openone{\leavevmode\hbox{\small1 \normalsize \kern-.64em1}}
\begin{document}
\sloppy
\vspace*{3cm}
\begin{center}
{\Huge{\bf{Quantum Entanglement in Some Physical Systems}}}\\
\vspace*{1cm}
\large
Dissertation\\
of Marcin Wie\'sniak, M. Sc.,\\
Institute of Theoretical Physics and Astrophysics,\\
University of Gda\'nsk, Gda\'nsk, Poland,\\
written under the supervision of\\ 
prof. Marek \.Zukowski\\
\vspace*{12cm}
Gda\'nsk, 20.09.2007.
\end{center}
\newpage
\vspace*{4cm}
\begin{center}
{\Huge \textbf{Abstract}}
\end{center}

Quantum entanglement has been recognized as a precious resource in various information processing tasks. Many of its applications rely on the discrepancy between Quantum Mechanics and Local Realism. This contrast was first shown by Bell \cite{bell}. 

For this reason a large part of the following dissertation has been devoted to the history of the discussion on the falsification of Local Realism. We begin with recalling the consideration of Einstein, Podolsky, and Rosen \cite{epr}. Next, we resume the answer of Bohr \cite{bohr} and the papers of Bohm \cite{Bohm1,Bohm2}, in which he shows a possibility of existence of hidden variables in Quantum Mechanics. Finally, we reach the argument of Bell \cite{bell}, who proved that if such parameters indeed exist, they cannot have a local character.

The Bell argument, originally drawn for two qubits and two measurements per qubit, has been generalized to more complicated experimental cases. We present inequalities by Mermin \cite{mermin}, Ardehali \cite{ardehali}, Belinskii and Klyshko \cite{belinskiiklyshko}, Werner and Wolf \cite{wernerwolf}, Weinfurter and \.Zukowski \cite{weinfurterzukowski}, and \.Zukowski and Brukner \cite{zukowskibrukner}. All these derivations were done for an arbitrary number of qubits, but still with only two alternative measurements on each subsystem. The Bell expressions, which utilize more local measurement were presented inter alia by Laskowski, Paterek, \.Zukowski and Brukner \cite{patlaszukbruk}. 

Another approach to the problem of finding new Bell inequalities is through an analysis of a convex hull, for example, in the space of inter-qubit correlations. This problem has been addressed by  e.g. \.Zukowski \cite{qph0611086}. His method was further developed for three qubits by Wie\'sniak, Badzi\k{a}g, and \.Zukowski \cite{w7}. Therein, we derive versions of the Bell theorem being special forms of the criterion presented in \cite{wuzong}, but also an inequality, in which all three observers perform one three measurements on their particles. Such an inequality cannot be equivalent to any previously known ones.

We also consider consequences of taking the rotational invariance of the correlation function as an additional constrain on Local Realism \cite{w3}. Upon these assumptions, we introduce a Bell inequality, in which every observer performs a continuum of measurements. It is violated by GHZ states stronger than any other known inequalities.

Another problem addressed in this dissertation is the channel transparency necessary to falsify Local Realism in the scheme Bj\"ork, Jonsson, and S\'anchez-Soto \cite{bjorkjonssonsanchezsoto}. It is a refinement of the proposal of Tan, Walls, and Collett \cite{tanwallscollett}, in which a carrier of non-classicality is a single photon. We show in \cite{w7} that the channel transparency (which in a two-qubit scheme could be associated with the quantum efficiency of dtectors), above which the Clauser-Horne inequality \cite{clauserhorne} can be violated is about $17.2\%$. This is significantly less than than the threshold on the detection efficiency in two-photon experiments, found by Garg and Mermin \cite{gargmermin} to be about $82.8\%$

We then abandon the problem of the Bell theorem, and pass to the question of entanglement in bulk bodies. Because of the high complexity of the physical system and our ignorance about the state, quantum correlations must be confirmed with entanglement witnesses \cite{witness}. After clarifying the notion of a witness we recall some of arguments that low-temperature values of the internal energy reveal entanglement \cite{pla301,bruknervedralzeil} and genuine multi-partite quantum correlations \cite{guehnetothbriegel}. We argue that also non-linear functions of the state can serve as entanglement witnesses. This is, for example, the case of the magnetic susceptibility for systems with rotationally invariant Hamiltonians \cite{w4}. The generality of the magnetic susceptibility as an entanglement witness is demonstrated with various examples. We show that also the heat capacity can reveal entanglement \cite{w5}. The last result is related to the Third Law of Thermodynamics.

\newpage
This research was supported by following institutions:
\begin{itemize}
\item
University of Gda\'nsk\\
Stipend for Ph. D. students\\
Grant No. BW/5400-5-0256-3
\item
Austrian-Polish projects \emph{Quantum Communication and Quantum Information}
\item German-Polish project 
\emph{Novel Entangled States for Quantum Information Processing: Generation and Analysis}
\item
Foundation for Polish Science\\
Stipends under the Professorial Subsidy of Marek \.Zukowski
\item
State Committee for Scientific Research\\
Grant No. PBZ-MIN-008/P03/03\\
Grant No. 1 P03B 04927
\item
Austrian Science Foundation\\
Project SFB 1506
\item
The Erwin Schr\"odinger International Institute for Mathematical Physics\\
Junior Research Fellowship
\item
European Union\\
QAP programme Contract No. 015848
\item
National University of Singapore\\
A*STAR Grant. No. 012-104-0040
\item
Foundation for Polish Science\\
START Scholarship for Young Researchers
\end{itemize}

\tableofcontents
\chapter{Introduction}
\section{Historical Prelude and Motivation}

\hspace*{5mm}
The formulation of Quantum Mechanics started with introducing the concept of a portion of energy in the black body radiation by Planck \cite{planck}. Subsequently, Einstein deduced the existence of the quantum of light \cite{einstein1905} from the behavior of entropy of the electromagnetic field in a cavity when one varies its volume. The emerging new theory of Nature, based on wave-functions as a description of the system, allowed to explain many phenomena. 
 A lot of attention in the development of the theory has been attracted by a phenomenon first discussed in more details by Einstein, Podolsky, and Rosen \cite{epr}. 

Schr\"odinger \cite{schroedinger} named this unique feature of quantum-mechanical systems {\em entanglement}\footnote{Original german: Verschr\"ankung.} and captured its nature (at least for pure states) in an observation that while we may posses the maximal knowledge about a composite quantum system treated as a whole, in the extreme case measurable quantities related to individual parts of the system may be completely undetermined. The subsystems are describes only in a reference to each other. 

Quantum Mechanics allows to arbitrarily superpose wave-functions, even if there exists a Nature-preferred basis of distinguishable situations. For instance, an atom can be driven to an arbitrary superposition of the ground and excited states. The feature of entanglement directly follows from the principle of the superposition. If a joint pure state of two or more quantum systems cannot be expressed as a tensor product of the states of each subsystem, but it is necessary to use a superposition of such products, such a state is called entangled\footnote{Throughout the thesis, the corresponding subsystem shall be denoted by an upper square-bracketed index behind mathematical objects, e.g. operators or states.}
;
\begin{eqnarray}
|\psi\rangle^{[12]}\quad\textbf{entangled}\Leftrightarrow|\psi\rangle^{[12]}\neq|\psi\rangle^{[1]}|\psi\rangle^{[2]}.
\end{eqnarray}
A mixed state is entangled if it cannot be decomposed into a convex combination of mixed product states,
\begin{eqnarray}
\label{mixedentangled}
\rho^{[12]}\quad\textbf{entangled}\Leftrightarrow\rho\neq\sum_iP_i\rho^{[1]}_i\otimes\rho^{[2]}_i.\\
\forall_i P_i\geq 0,\sum_iP_i=1.\nonumber
\end{eqnarray}
The right-hand side of the inequality in (\ref{mixedentangled}) is called a separable state.

Einstein, Podolsky, and Rosen \cite{epr} began a discussion on the the possibility of excluding any local and realistic description of the Universe and posed a question about the completeness of Quantum Mechanics. The violation of Local Realism by entangled states was some time later elegantly demonstrated by Bell \cite{bell} in a form of an inequality. This inequality was thereafter experimentally confirmed, first by Aspect {\em et al.} \cite{Aspect1982}, and generalized for systems of higher complexity, for instance, in \cite{mermin,ardehali,belinskiiklyshko,wernerwolf,weinfurterzukowski,zukowskibrukner,wuzong,patlaszukbruk}, or more recently in \cite{qph0611086}. 
This intensive search for new entanglement tests has been motivated by identifying correlations of purely quantum nature as a resource for a teleportation of an unknown quantum state \cite{teleportation}, quantum computation \cite{grover,shor,deutschjozsa}, as well as decreasing the communication complexity in computational tasks \cite{complexity}. It has been also assigned a role in certain protocols for distributing a secret cryptographic key \cite{bennettbrassard88,ekert91,cryptography} (The relation between entanglement and communication complexity problems or qunatum key distribution is through the violation of Bell inequalities). However, protocols for the last two tasks were proposed and experimentally realized also with a use of a single photon \cite{singlephotoncrypto,singlephotoncomplex}.

However, even if an entangled state does not violate Local Realism, it can be useful from the point of of view  of quantum information processing. It was shown by Bennett {\em et al.} \cite{distillation} that weakly entangled states can be locally post-processed with local actions classical communication communication between partners possessing the state (in many copies) in order to distill maximally entangled pairs of qubits. Even if it is not possible for a given entangled state, that is the state is bound entangled \cite{horodeckippt}, its correlations can be used to activate non-classicality of other physical systems \cite{activation}. In any case, non-separability of a state can be confirmed with positive, but not completely positive maps \cite{peresppt,horodeckippt} or entanglement witnesses \cite{witness,jamiolkowski}-(originally) hermitian operators, extreme eigenvalues of which are associated to entangled states. 

\section{Summary of Results}
\hspace*{5mm}

For the reasons explained in the first section, a large part of the research, in which the Author of the following dissertation has participated over last years, has been devoted to the problem of new versions of the Bell Theorem. Results with potentially practical applications are presented in Chapter 3. Therein we exemplify a method of generating new Bell expression from the analysis of a convex hull in the correlation space for the simplest non-trivial case of three qubits\footnote{A qubit (quantum bit) is a physical system described by a two-dimensional Hilbert space, for example a spin-$\frac{1}{2}$ or a polarization of a photon, as well as a unit amount of quantum information, that can be stored in such a system.} and a choice of up to three observables for each observer. For most of the newly-found inequalities, we present a sufficient condition on a state for satisfying the inequality. 
The Chapter ends with considering consequences of taking the rotational invariance as an additional constraint on Local Realism. Assuming this we are able to present a Bell condition, that can be violated exponentially stronger by GHZ states than any thus far known standard inequalities. 

In Chapter 4 we argue that the so-called ``single-photon'' Bell-type experiment, initially proposed by Tan, Walls, and Collett \cite{tanwallscollett} and theoretically refined by Bj\"ork, Jonsonn, and S\'anchez-Soto \cite{bjorkjonssonsanchezsoto}, is very robust against photon losses. 

Entanglement witnesses allow to investigate non-separability in bulk solid systems, arbitrarily large lattices of spins in a thermal equilibrium. We then aim to detect quantum correlations with certain values of thermodynamical quantities, like the internal energy, the heat capacity, and the magnetic susceptibility. The pioneering work due to Wang and Zanardi \cite{pla301} links the internal energy with a measure of two-qubit entanglement, the Wootters concurrence \cite{concurrence}. In spite of possible experimental difficulties in measuring the internal energy, many authors \cite{qph0406040,tothinternalenergy,dowling,wu,guehnetothbriegel} followed the original idea. On the other hand, Brukner, Vedral, and Zeilinger \cite{bruknervedralzeil} have shown that entanglement is implied by low-temperature values of the magnetic susceptibility for a specific substance, cupric nitrate (Cu(NO$_3$)$_2$). In section 5.6 we will repeat the argument from \cite{w4} that the magnetic susceptibility is a valid entanglement criterion for thus far the widest class of materials, all those in which spin interaction is isotropic. Also in Section 5.8 we aim to prove that independently from symmetries possessed by the Hamiltonian, entanglement can be revealed by low-temperature behavior of the heat capacity \cite{w5}. We stress that this result is strongly related to the unattainability of the the absolute zero temperature, which is possibly equivalent to the Third Law of Thermodynamics \cite{nernst}.

\chapter{Bell Theorem-Introduction}
\section{Einstein-Podolsky-Rosen Paradox}

\hspace*{5mm}
In 1935 Einstein, Podolsky, and Rosen \cite{epr} noticed the discrepancy between the point of view that Quantum Mechanics is a compete theory and the probable from their point of view description of the Universe based on two main assumptions:
\begin{itemize}
\item{{\em Locality} is defined from Special Relativity \cite{einstein1905}. It states that no two events can influence each other if they are spacelike separated. 
The mutual independence not necessarily means that the events might not have had the common cause.}
\item{The other important notion, a new meaning of which was introduced in \cite{epr}, is {\em Realism}. The Authors themselves, avoiding complicated philosophical considerations, conclude that:
\begin{quote}
If, without in any way disturbing a system, we can predict with certainty (i.e., with probability equal to unity) the value of a physical quantity, then there exists an element of physical reality corresponding to this physical quantity.
\end{quote}
Thus Realism means that the results do not appear at the moment	 of the measurement, but rather wait to be revealed by an observer. Such a requirement is natural in the macroscopic world. For example, objects have their dimensions and masses, although they could have never been measured or weighted. The act of the measurement only reveals facts unknown to the observer. }
\end{itemize}

The theories based on these two assumptions are called local realistic theories  or theories with local hidden variables (LHV). Local Realism can be interpreted as a situation in which every part of a quantum system carries a set of information about which result would be yielded under any measurement. 

Non-avoidable randomness in Quantum Mechanic forced the Authors to face the following dilemma: either (i) the quantum-mechanical description of the Universe based on a wave-function cannot be considered complete, that is not all the elements of the reality are described, or (ii) quantities related to non-commuting observables cannot be simultaneous elements of the same reality.

The potential paradox was illustrated with the following wave-function of two particles with coordinates $x^{[1]}$ and $x^{[2]}$\footnote{It is enough to consider particles living in one-dimensional geometric space.}:
\begin{equation}
\label{eprstate1}
\Psi(x^{[1]},x^{[2]})=\int_{-\infty}^{+\infty}e^{i(x^{[1]}-x^{[2]}+x_0)p}dp,
\end{equation}
where $x_0$ is a constant and $p$ here denotes a momentum of either of particles. The integration gives
\begin{equation}
\label{eprstate2}
\Psi(x^{[1]},x^{[2]})\propto\delta(x^{[1]}-x^{[2]}+x_0),
\end{equation}
and after the double Fourier Transform we get
\begin{equation}
\label{eprstate3}
\Psi(p^{[1]},p^{[2]})\propto\delta(p^{[1]}+p^{[2]}-p_0)
\end{equation}
with $p_0$ being another constant. The particles are thus separated by a constant distance, and in the same time the sum of their momenta is equal to $p_0$. As noticed by Schr\"odinger \cite{schroedinger}, when we trace out one of the particles it turns out that the other particle is completely delocalized, in both the position and the momentum representations. Particles can be equally likely found in every point of the Universe and their mean kinetic energy is infinite. 

According to (\ref{eprstate2}) if one of the observers (hereafter called Alice) detects a particle at some given point $x$, she and only she knows instantaneously that the second particle can be found at $x_0+x$. At the same time the other observer (called Bob) would find it exactly there in spite of the fact that due to the assumption of locality, no information about the outcome of the Alice's measurement could have propagated with superluminal speed. Thus, following the definition, the position of Bob's particle is an element of the reality. But Alice could have decided to measure the momentum of her particle rather than position and get some value $p$ and learn that Bob's particle is has the momentum $-p+p_0$, which also seems to be an element of the reality.

One must remember, however, that operator $\hat{x}\:^{[1(2)]}$ does not commute with $\hat{p}\:^{[1(2)]}$, and thus Quantum Mechanics forbids to have well--defined values of the position and the momentum of the same object at the same time. If we agree that both the position and the momentum of the second particle are elements of the same reality, the Quantum Theory cannot be complete, as according to this theory only one of the values can be fully defined at a time. If we insist that a wave-function is the most complete description of a system, the position and the momentum cannot be simultaneous elements of the reality. On the other hand, different representations of the same wave-function are related to each other by transforms. Einstein, Podolsky, and Rosen had come to the conclusion that Quantum Mechanics could not have been considered complete.

\section{Anwser of Bohr}

\hspace*{5mm}

A direct response to the EPR letter \cite{epr} was an article of Bohr \cite{bohr}. Using the notion of complementarity, which was introduced by him, he explained that considered criterion of completeness of Quantum Mechanics is not adequate due to fundamental differences between classical and quantum-mechanical descriptions of the reality.

The paper starts with introducing two theoretical situations. In the first one, a particle falls on a diaphragm with a slit and, provided that it was aimed at the slit, its trajectory is distracted by a diffraction. The diaphragm is a part of a larger measuring device and is fixed with respect to the coordinate system of the laboratory. The only uncertainty of the position of the particle just behind the slit is related to its width, here denoted as $\Delta q(=\sqrt{\langle q^2\rangle-\langle q\rangle^2}$, $\langle\cdot\rangle$ denotes the mean value). The diffraction causes a momentum interchange between the particle and the diaphragm of the magnitude $\Delta p(=\sqrt{\langle p^2\rangle-\langle p\rangle^2})$. This momentum is, however, transported to the optical table, and further to the laboratory. It is hence impossible to precisely determine $\Delta p$. Therefore we cannot predict the exact point of a far screen at which the photon would be detected.

In contrast, we can consider the case in which the diaphragm is a freely movable part. Before and after the photon passes the slit, we can measure the momentum of the diaphragm with help of a stream of probe particles. Again, any such measurement would mean a displacement of the diaphragm due to collisions with probe beams. Although we have observed the momentum interchange, we deal with the uncertainty of the position. Still, we are unable to predict the exact point of the detection.

How does this correspond to the EPR pair? Instead of one, let us consider two slits in a diaphragm which are both narrow in comparison to the distance between them. In case of movable slits, when we are able to measure momentum of the diaphragm, the  state of the two particles is close to the EPR state. We can obtain knowledge of the difference of positions $|q^{[1]}-q^{[2]}|$ as precisely as narrow are the slits, and, under certain assumptions, we can learn about the amount of the momentum interchanged with an arbitrary precision.  However, since the initial position of the plate was unknown, we cannot determine $q^{[1]}+q^{[2]}$, and as it was a three-body collision, it is also impossible to have a precise value of $p^{[1]}-p^{[2]}$. Now it is up to the observer which quantity he measures (on one particle) and which corresponding property of the second particle will be defined.

The difference, from the point of view of Bohr, lies in the context of the measurement. In the first case, the diaphragm was a fixed element of a measuring device. We can arbitrarily precisely determine the position of the particle in the initial state. In case of a movable diaphragm, it was rather a part of a quantum system. In a sense, a single slit experiment is equivalent to the EPR gedanken-experiment, with the diaphragm playing a role of the other particle. In both cases Quantum Mechanics allows to predict certain properties of the particle without any disturbance, thus it cannot be considered incomplete. The division of the system to a microscopical quantum system and a macroscopic measuring device is the basis of Bohr's Copenhagen interpretation. Significantly, we are always forced to deal with an unavoidable and uncontrollable interchange of certain quantities, like the momentum, between the two systems.

\section{Bohmian Interpretation of Quantum Mechanics}

\hspace*{5mm}
The discussion on completeness of Quantum Mechanics returned on the occasion of papers by Bohm \cite{Bohm1,Bohm2} from 1952, in which he proposed his own interpretation of the theory. He argued that the formulation of the Quantum Theory is fully consistent and it is the most possible complete theory of nature, however it is still possible to introduce additional parameters. They are, by definition, not accessible for the observer. These parameters shall allow to predict results of all possible measurements, as Einstein, Podolsky, and Rosen were considering. He based his statement on the analogy that Thermodynamics correctly describes the behavior of gases, although it neglects positions and momenta of individual atoms and just gives their statistical distributions. Similarly, Quantum Mechanics may contain some elements, which carry information about outcomes of all possible measurements, but since they are never known, the theory gains a statistical nature.

The Bohmian interpretation is based on the new understanding of the Schr\"odinger equation. If we write the wave-function $\psi(x)$ of a particle of mass $m$ in a potential $V(x)$ using two real-valued functions $\psi(x)=R\exp(iS/\hbar)$, the Schr\"odinger equation,
\begin{equation}
i\hbar\frac{\partial\psi(x)}{\partial t}=-\frac{\hbar^2}{2m}\nabla^2\psi(x)+V(x)\psi(x),
\end{equation}
can be expressed as
\begin{equation}
\frac{\partial R}{\partial t}=-\frac{1}{2m}(R\nabla^2S+2\nabla R\nabla S),
\end{equation}
\begin{equation}
\frac{\partial S}{\partial t}=-\left(\frac{(\nabla S)^2}{2m}+V(x)-\frac{\hbar^2}{2m}\frac{\nabla^2R}{R}\right).
\end{equation}
It is convenient to introduce the probability density $P(x)=R^2(x)$:
\begin{equation}
\label{bohm1}
\frac{\partial P}{\partial t}+\nabla\cdot\left(P\frac{\nabla S}{m}\right)=0,
\end{equation}
\begin{equation}
\label{bohm2}
\frac{\partial S}{\partial t}+\frac{(\nabla S)^2}{2m}+V(x)-\frac{\hbar^2}{4m}\left(\frac{\nabla^2 P}{P}-\frac{(\nabla P)^2}{2P}\right)=0.
\end{equation}
Then, in the classical limit of $\hbar\rightarrow 0$, the phase function $S$ becomes a solution of the Hamilton-Jacobi equation, so that we can interpret its gradient as the momentum. (\ref{bohm1}) expresses the probability conservation. We thus see that that the Schr\"odinger evolution differs from the classical one in such a way that besides the ``classical" potential we have the ''quantum-mechanical" one, 

\begin{equation}
U(x)=-\frac{\hbar^2}{4m}\left(\frac{\nabla^2P}{P}-\frac{(\nabla P)^2}{2P^2}\right)=-\frac{\hbar^2}{2m}\frac{\nabla^2R}{R}.
\end{equation}

This greatly resembles the situation in Electrodynamics. In both cases the field, the electromagnetic field or the wave-function satisfy certain equations (Maxwell's or Schr\"odinger's), from which we compute forces. Thus knowing the state at a certain instant of time we are able to trace the whole evolution, back forth and back in time. 

The essential difference between the electromagnetic field and the wave-function is the form of the equation. The Schr\"odinger equation is homogeneous with respect to $\psi$, whereas the Maxwell equations are not with respect to the field. Bohm sees this difference as a possibility to modify Quantum Mechanics, such that significant effects would appear only at very short distances.

The inhomogeneous modification of the Schr\"odinger equation suggested Bohm emerged from the fact that at the time some phenomena in the subatomic scale did not have proper descriptions. Quantum Mechanics might turn out to not be valid at the Planck scale, that is distances of the order of $10^{-15}$ meters and would then need to be modified. Obviously, a more general theory, correct also at these distances, must reproduce results of Quantum Mechanics. Bohm now suggests to abandon his idea of seeing $\nabla S/m$ as a velocity $v$ of the particle, as well as to assume that the homogeneous and linear with respect to $\psi$ Schr\"odinger equation is not an evolution equation. Giving up these two hypotheses implies that we cannot treat the wave-function as a statistical description of an ensemble of particles.

An example of such a correction in the papers of Bohm is done by introducing a force, which would equalize the physical velocity $v$ of a particle and $\nabla S/m$. This should happen in a very short time $\tau$ of the order or $10^{-15}m/c$\footnote{$m$ here denotes meters, not the mass.} or $10^{-33}s$. A modified classical equation of motion has a form of
\begin{equation}
m\frac{d^2x}{dt^2}=\nabla\left(V(x)-\frac{\hbar}{2m}\frac{\nabla^2R}{R}\right)+f(v-\nabla S/m),
\end{equation}
and the non-homogeneous Schr\"odinger equation reads
\begin{equation}
i\hbar\frac{\partial\psi}{\partial t}=H\psi+\xi(p-\nabla S),
\end{equation}
where $f(0)=\xi(0)=0$. The quantum evolution returns to its homogeneous form after time $\tau$. Until then, all basic principles of Quantum Mechanics, such as the superposition principle, might not be valid. However, this or other corrections of such a character would allow to extend the theory with hidden variables predicting simultaneously the position and the momentum with an arbitrarily high precision.

In \cite{Bohm2} the Author explains his interpretation of the Einstein-Podolsky-Rosen gedankenexperiment. According to Bohm, since the wave-function (\ref{eprstate2}) is real, both particles remain at rest. Their positions are described by a probability distribution, such that always $x^{[2]}-x^{[1]}=x_0$. The observer measures the position of one of the particles and disturbs the momentum of the other in an uncontrollable way, and, similarly, the act of measuring the momentum of one particle unpredictably disturbs the position of the other. In this way Bohm is able to save the statement that one cannot measure two arbitrary quantities, even though in case of the EPR experiment one acts on two spatially separated subsystems. The disturbance of field $S$ may happen with a superluminal speed. This hypothesis must be of course questioned with respect to its agreement with the Special Relativity, as predictions of the latter are not in a conflict with Quantum Mechanics. An instant change of the quantum-mechanical potential cannot result in transfer of any useful information between two distant observers. Without a measurement neither of them has any information about the position or the momentum of his particle until he gets a classical information about the other half of the EPR pair. The transfer of the classical information only via the EPR channel would be possible if there was a way to predict the position or the momentum without performing any measurement. This is clearly impossible in Schr\"odinger's formulation of Quantum Mechanics, which, as Bohm argues, does not hold at Planck scales. One can construct theories, which, as it was shown, allow to overcome the Heisenberg bound for uncertainties, but also these, which permit superluminal propagation of informations. Then, we have two more options to avoid a discrepancy between Quantum Mechanics and Special Relativity. Firstly, there might exist a law, which simply prohibits superluminal changes of the controllable parts of the quantum-mechanical potential also below the Planck length. The other option is that the Lorenz invariance can be just irrelevant at nanoscales. It would be an important extension of the ideas of General Relativity to consider the metric tensor $\hat{g}$, and thus the curvature of the spacetime, dependent on the wave-function $\psi(x)$, which describes the matter and the energy in this space. The postulate of the speed of light as the universal constant, the largest possible speed in the Universe, could then be relaxed.
\section{Bell Theorem}

\hspace*{5mm}

Bohm had argued \cite{Bohm1,Bohm2}, that Quantum Mechanics under some assumptions can be interpreted as a deterministic theory based on hidden variables. However the paper of Bell \cite{bell} from 1964 showed that these additional parameters, if they exist, cannot have a local character.

Bell proposes that Alice and Bob can locally perform one of two possible dichotomic measurements, $A_1^{[1]}, A_2^{[1]}$ for Alice, $A_1^{[2]}, A_2^{[2]}$ for Bob, which yield ``$+1$" or ``$-1$" as results. Let us denote the results as $a_1^{[1]},a_2^{[1]}$ for Alice and $a_1^{[2]},a_2^{[2]}$ for Bob. The assumption of Realism allows us to use all four quantities in the same time and Locality excludes mutual influences between events at Alice's and Bob's side. Out of two expressions, $a_1^{[1]}+a_2^{[1]}$ and $a_1^{[1]}-a_2^{[1]}$, exactly one is equal to $\pm 2$, the other vanishes and similarly for $a_1^{[2]}+a_2^{[2]}$ and $a_1^{[2]}-a_2^{[2]}$. As a consequence, for any given set of lcal hidden variables determining results $\{a_1^{[1]},a_2^{[1]},a_1^{[2]},a_2^{[2]}\}$, only one of four expressions $(a_1^{[1]}\pm a_2^{[1]})(a_1^{[2]}\pm a_2^{[2]})$ is equal to $\pm 4$, while three others are 0. After averaging over many runs of the experiment we can write
\begin{equation}
\label{belloriginal}
\sum_{s^{[1]},s^{[2]}=\pm 1} \left|\left\langle(A_1^{[1]}+s^{[1]}A_2^{[1]})(A_1^{[2]}+s^{[2]}A_2^{[2]})\right\rangle_{\lambda}\right|\leq 4,
\end{equation}
where the subscript $\lambda$ stresses that (\ref{belloriginal}) is valid for a given set $\lambda$ of local hidden variables. 

One may also consider theories, in which various possible sets of LHV are distributed with a probability density $\rho(\lambda)$. This is reflected in our uncertainty about measurement outcomes. To compute averages in such stochastic theories one must integrate over all possible sets $\lambda$ with $\rho(\lambda)$ as an integration kernel: 
\begin{equation}
\label{belloriginal2}
\sum_{s^{[1]},s^{[2]}=\pm 1}\left|\int\left\langle(A_1^{[1]}+s^{[1]}A_2^{[1]})(A_1^{[2]}+s^{[2]}A_2^{[2]})\right\rangle_{\lambda} \rho(\lambda)d\lambda\right |\leq 4,
\end{equation}
where $d\lambda$ is a metric in the set of $\lambda$.

Since the Bell inequality is linear with respect to the mean values, its local realistic bound can be concluded from the first kind of theories. We hereafter assume that the theory under our consideration is deterministic and denote local realistic predictions by a subscript $LHV$.

Since taking the modulus changes at most the global sign of the expression, (\ref{belloriginal}) can be also written with use of a sign function $S(s^{[1]}=\pm 1,s^{[2]}=\pm 1)=\pm 1$ \cite{zukowskibrukner}:
\begin{equation}
\label{bellsignfunction}
-4\leq\sum_{s^{[1]},s^{[2]}=\pm 1}S(s^{[1]},s^{[2]})\left\langle(A_1^{[1]}+s^{[1]}A_2^{[1]})(A_1^{[2]}+s^{[2]}A_2^{[2]})\right\rangle_{LHV}\leq 4.
\end{equation}

From the point of view of the discussion on the possibility of local hidden variables only these eight inequalities (\ref{bellsignfunction}) are important, in which the sign function is not factorisable, $S(s^{[1]},s^{[2]})\neq S_1(s^{[1]})S_2(s^{[2]})$. The remaining eight produce trivial bounds $\left|\langle A_i^{[1]}A_j^{[2]}\rangle\right|\leq 1$. In the first case, the inequalities are equivalent to the one introduced by Clauser, Horne, Shimony, and Holt \cite{chsh}:
\begin{equation}
\label{chshinequality}
-2\leq\langle A_1^{[1]}A_1^{[2]}+A_1^{[1]}A_2^{[2]}+A_2^{[1]}A_1^{[2]}-A_2^{[1]}A_2^{[2]}\rangle_{LHV}\leq 2.
\end{equation}

\begin{figure}
\centering
\includegraphics[width=7cm]{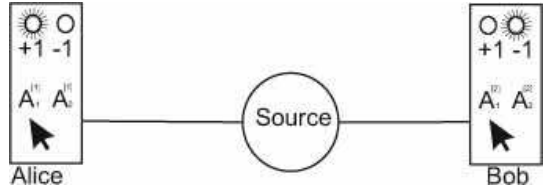}
\label{bellexperiment}
\caption{A scheme of a Bell experiment. Two partners, Alice and Bob, receive particles from a common source and perform one of two measurements of their choice. The mean values are plugged into (\ref{belloriginal}.)}
\end{figure}

These inequalities may not be valid anymore when one uses the formalism of Quantum Mechanics to compute mean values of measurements. Local measurements with results ``$\pm 1$" on two-dimensional quantum systems are associated with observables in a form of a scalar product of a normalized three-dimensional vector built of real components $\vec{a}\:^{[i]}_j$ with a vector of Pauli matrices $\vec{\sigma}^{[i]}$ \footnote{These observables, together with $\sigma_0=\left(\begin{array}{cc}1&0\\0&1\end{array}\right)$ will appear often in our further considerations. They constitute an orthogonal basis of $2\times 2$ matrices, as $\textbf{Tr}\sigma_i\sigma_j=2\delta_{ij}$. It turns out that it is extremely convenient to parametrize an $N$-qubit state by mean values of products of Pauli matrices,
\begin{eqnarray}
T_{a...n}=\textbf{Tr}\rho\sigma^{[1]}_a...\sigma^{[N]}_n;&\rho=\frac{1}{2^N}\left(\sum_{a,...,n=0}^3T_{a...n}\sigma_{a}^{[1]}...\sigma_{n}^{[N]}\right).
\end{eqnarray}
The set of numbers $\{T_{a...n}\}_{a,...,n=0}^3$ shall be called a correlation tensor. Strictly speaking, only $\{T_{a...n}\}_{a,...,n=1}^3$ transforms properly under $O(3)$ rotations, and originally the name has been given to this subset. The rest of coefficients describes reduced states. This two meanings of the notion of the correlation tensor can be met in this dissertation.}
,
\begin{eqnarray}
\sigma_1=\left(\begin{array}{cc}0&1\\1&0\end{array}\right),&\sigma_2=\left(\begin{array}{cc}0&-i\\i&0\end{array}\right),&\sigma_3=\left(\begin{array}{cc}1&0\\0&-1\end{array}\right),
\end{eqnarray}
so that Alice's observables are given by $A_i^{[1]}=\vec{a}_i\:^{[1]}\cdot\vec{\sigma}\:^{[1]}$ and Bob's-$A_i^{[2]}=\vec{a}_i^{[2]}\cdot\vec{\sigma}^{[2]}$. The the eigenstates of the $\sigma_z$ operator are represented by $|0\rangle\leftrightarrow\left(\begin{array}{c}1\\0\end{array}\right)$ and $|1\rangle\leftrightarrow\left(\begin{array}{c}0\\1\end{array}\right)$, respectively.

Now, let us assume that the pair of spins$-\frac{1}{2}$ is in the singlet state $|\Psi^-\rangle=(|10\rangle-|01\rangle)/\sqrt{2}$, where in the state $|0\rangle$ the spin is parallel to the $z$-axis and anti-parallel in $|1\rangle$. Then the correlation function is given by $\langle A^{[1]}_iA^{[2]}_j\rangle=-\vec{a}\:^{[1]}_i\cdot\vec{a}\:^{[2]}_j$. For convenience, we will restrict ourselves to observables in the $xy$-plane, which are given by
\begin{equation}
A^{[1]}(\varphi_A)=\cos\varphi_A\sigma_1^{[1]}+\sin\varphi_A\sigma_2^{[1]}=\left(\begin{array}{cc}0&e^{-i\varphi_A}\\e^{i\varphi_A}&0\end{array}\right)\otimes\left(\begin{array}{cc}1&0\\0&1\end{array}\right),
\end{equation}
\begin{equation}
A^{[2]}(\varphi_B)=\cos\varphi_B\sigma_1^{[2]}+\sin\varphi_B\sigma_2^{[2]}=\left(\begin{array}{cc}1&0\\0&1\end{array}\right)\otimes\left(\begin{array}{cc}0&e^{-i\varphi_B}\\e^{i\varphi_B}&0\end{array}\right).
\end{equation}
The correlation function is then equal to
\begin{equation}
E(\varphi_A,\varphi_B)=\langle \Psi^-|A^{[1]}(\varphi_A)A^{[2]}(\varphi_B)|\Psi^-\rangle=-\cos(\varphi_A-\varphi_B).
\end{equation}
Let us now choose two observables at each side. It turns out that the optimal choices are $A_1^{[1]}=A^{[1]}(0),A_2^{[1]}=A^{[1]}(\pi/2),A_1^{[2]}=A^{[2]}(\pi/4),A_2^{[2]}=A^{[2]}(-\pi/4)$. The middle of (\ref{chshinequality}) takes then the value of $-3\cos\pi/4+\cos 3\pi/4=-2\sqrt{2}$. This is less than any theory with local hidden variables can predict. The Bell theorem thus states that there exist measurements, for which quantum mechanical predictions are in disagreement with all possible predictions based on Local Realism. The natural question arises whether this can be confirmed experimentally.
\section{Additional Assumptions for Bell Theorem}
\hspace*{5mm}

Apart from the {\em Locality} and {\em Realism} we need to take one more very important, but often forgotten assumption. We need to assume that the observers choose local observables not long before the measurements and independently from each other. This could be guaranteed if both of them use fast {\em random} (or even pseudo-random) number generators to determine which local measurement shall be performed. If the choice of measurements for each run of the experiment was predefined much before the emission of the entangled pair, a local realist could claim that the hidden parameters of the particles were influenced by this predetermination. 

The conditions of Randomness and Locality were strictly satisfied in the experiment performed by Weihs {\em et al.} \cite{weihs}. In this realization of the experiment, each measuring station was connected with a source by 500 meters of an optical fiber, but one half of each cable was rolled up next to the source, while the rest was strechted to Alice and Bob, separated by about 400 meters. Entangled pairs remained for about 1 $\mu$s close to the source, and propagated to the observers for roughly the same time. During the propagation time local measurements were randomly chosen using ultra-fast opto-electronic switches. Random number generators consisted of a light emitting diode, a filter, a beam splitter, and two detectors behind it. The filters attenuated the emitted light to about one photon per pulse, which was detected at one of the sides behind the beam splitter. The results of these measurements determined the position of an opto-electronic transducer, through which an entangled photon passed, thus the choice of a local observable. Such a decision was made $10^7$ times per second, finally when photons were not further than 25 m from the stations. Information which observables were chosen could have reached in time neither the source, nor the second particle.

\begin{figure}
\centering
\includegraphics[width=4cm]{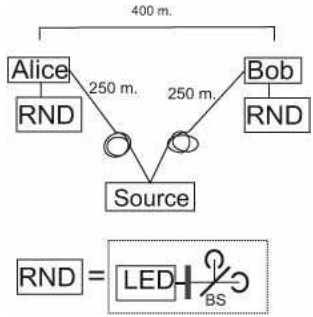}
\label{weihsfigure}
\caption{The experiment of Weihs {\em et al.} \cite{weihs}. The intervals between generations of random numbers by RND is about 10 times shorter than the time of the propagation of photons in fibers.}
\end{figure}

In spite of rigorously satisfying the Locality and Randomness conditions, the experiment of Weihs {\em et al.} \cite{weihs} was not ultimately convincing, since the quantum detection efficiency, that is the probability that the particle was detected given that it actually reached the detector, was insufficiently high.

Let us denote the detection efficiency of every detector used in the experiment as $\eta$. Following Garg and Mermin \cite{gargmermin}, for the sake of view of advocating Local Realism it is convenient to assume that each non-registered photon would produce always the same result, (say, ``$+1$"). Let us, moreover, assume that the state is a Werner state, that is a mixture of a maximally entangled state, $|\Psi^-\rangle\langle\Psi^-|=\frac{1}{2}(|01\rangle-|10\rangle)(\langle 01|-\langle 10|)$, with a maximally mixed state, $\frac{\openone_{4\otimes 4}}{4}$\footnote{By $\openone_{x\times x}$ we denote a unit matrix of dimension $x$.}: 
\begin{eqnarray}
&\rho=V|\Psi^-\rangle\langle\Psi^-|+(1-V)\frac{\openone_{4\times 4}}{4},&\\
&0\leq l\leq 1.&\nonumber
\end{eqnarray}
Such states are created when one of the photons from an entangled pair is being a subject to a depolarizing channel described by a transformation $\rho\rightarrow V\rho+(1-V)\sum_{i=0}^3\sigma^{[1]}_i\rho\sigma^{[1]}_i$. Thus after taking into account the detection efficiency, we have a effective correlation function $E_{eff}(\varphi_A,\varphi_B)=-V\eta^2\cos(\varphi_A-\varphi_B)+(1+\eta)^2$. After plugging into (\ref{chshinequality}), one gets
\begin{eqnarray} 
-2\leq&-V\eta^2(\cos(\varphi_A-\varphi_B)+\cos(\varphi'_A-\varphi_B)+&\nonumber\\ &\cos(\varphi_A-\varphi'_B)-\cos(\varphi'_A-\varphi'_B))+2(1-\eta)^2&\leq 2,
\end{eqnarray}
 and optimizing over angles we obtain a necessary condition for a violation of the inequality:
\begin{equation}
V\eta^2\sqrt{2}+(1-\eta)^2> 1.
\end{equation}
This is a condition for critical values of $V$ and $\eta$ only above which the Bell inequality can be violated:
\begin{equation}
\label{critpar}
\eta_{CRIT}=\frac{2}{\sqrt{2}V_{CRIT}+1}.
\end{equation}
The inequality (\ref{chshinequality}) cannot be violated if $V\leq\frac{1}{\sqrt{2}}$. If the photons reach detectors without any loss of coherence, the required detection efficiency to violate a CHSH inequality is $\eta_{CRIT}=2(\sqrt{2}-1)\approx 82.8\%$. 

The relation (\ref{critpar}) can be also obtained from the Clauser-Horne (CH) inequality \cite{clauserhorne}  for event probabilities:
\begin{eqnarray}
\label{chinequality}
&P(\varphi_A,\varphi_B)+P(\varphi'_A,\varphi_B)+P(\varphi_A,\varphi_B')-P(\varphi'_A,\varphi'_B)-P(\varphi_A)-P(\varphi_B)&\leq 0.\nonumber\\
\end{eqnarray}
The joint probabilities take a form $P(\varphi_A,\varphi_B)=\frac{\eta^2}{4}(1-V\cos(\varphi_A-\varphi_B))$, whereas local probabilities are given by $P(\varphi_A)=P(\varphi_B)=\frac{\eta}{2}$. Applying this forms to (\ref{chinequality}) and after the optimization we get
\begin{equation}
\frac{\eta^2}{2}(1-\sqrt{2}V)-\eta\leq 0,
\end{equation}
and finally obtain (\ref{critpar}).

It was shown by Eberhard \cite{eberhard} that the critical efficiency can be lowered to $66.7\%$. This is done, however, for states arbitrarily close to product ones.

One should also consider false detections, called dark counts. As we assume, they occur with the probability $P_D$ within the time gate. Now, the effective probabilities are $P_{eff}(\varphi_A,\varphi_B)=\frac{\eta^2}{4}(1-V\cos(\varphi_A-\varphi_B))+\eta(1-\frac{\eta}{2}) P_D+(1-\frac{\eta}{2})^2P_D^2$ and $P_{eff}(\varphi_A)=P(\varphi_B)=\frac{\eta}{2}+(1-\frac{\eta}{2})P_D$, which put into (\ref{chinequality}) gives 
\begin{equation}
V_{CRIT}=(1-P_D)(2-\eta_{CRIT})(P_D(2-\eta_{CRIT})+\eta_{CRIT})/(\sqrt{2}\eta_{CRIT}^2).
\end{equation}
 The dark counts have their origin in thermal excitations of electrons in a photosensitive element, which are not distinguishable from photoexcitations. To avoid this problematic effect, detectors are often cooled down to the temperature of liquid nitrogen, i.e., 72K. As a controllable parameter, the dark count probability is not as relevant as the detection efficiency.

Out of many realizations of the Bell experiment, only in the one of Rowe {\em et al.} \cite{rowe} the detection loophole was overcome. They have realized an entangled state with trapped ions $^9$Be$^+$, what has allowed to achieve almost perfect measurement efficiency (about 99$\%$). However, the two entangled ions were placed in the same trap, and separated from each other only by micrometers. Thus the locality loophole was left wide open.

\chapter{Generalizations of Bell Theorem}
\section{GHZ paradox}
\hspace*{5mm}

In 1989 the two-qubit Bell theorem \cite{bell} had already been broadly discussed in literature (e. g. \cite{clauserhorne,chsh,bennettbrassard88}), and there have been realizations a Bell-type experiment \cite{Aspect1982} by then. Although up to date no such experiment had all the possible loopholes closed, already at the time the majority of the physical community accepted the possibility of having two quantum objects correlated in a strictly non-classical way. One natural question was if entanglement can be demonstrated with more than two mutually correlated objects. Another interesting problem was to concentrate the research on systems described by Hilbert spaces of dimensions higher than two. At the first sight, both limits, the limit of large numbers of subsystems and the one of many degrees of freedom, may suggest reaching macroscopic, thus fully classical regime. However, it is just enough to consider two spin-$l$ particles, which isotropically interact with each other,
\begin{equation}
H=\hat{\vec{S}}^{[1]}\cdot\hat{\vec{S}}^{[2]}.
\end{equation}
The ground state \footnote{The absolute zero temperature is not attainable in a finite process \cite{nernst}, but temperature can be made arbitrarily close and hence the actual thermal state can arbitrarily well approximate the ground state} is a singlet state, in which no information about individual spins is available. The von Neumann entropy of the reduced states, which since 1996 is known to be a unique measure of entanglement for pure, bipartite systems \cite{bennettbernsteinpopescuschumacher}, reads
\begin{eqnarray}
S(\textbf{Tr}_2|g\rangle\langle g|)=-\textbf{Tr}(\textbf{Tr}_2|g\rangle\langle g|)\log_2(\textbf{Tr}_2|g\rangle\langle g|)\nonumber\\
=-\textbf{Tr}\frac{\openone_{(2l+1)\times(2l+1)}}{2l+1}\log_2\frac{\openone_{(2l+1)\times(2l+1)}}{2l+1}=\log_2 (2l+1)
\end{eqnarray}
and grows logarithmically with the magnitude of the spin. Here $\textbf{Tr}_2$ denotes a partial trace over degrees of freedom of the second particle. One can also show, for example using an argument from \cite{tothinternalenergy} or \cite{w4}, that the higher the spins are, the higher is the critical temperature, above which known criteria do not reveal entanglement. Can more qubits also lead to more non-classical predictions than two qubits?

The first positive answer to this question was given by Greenberger, Horne, and Zeilinger \cite{greenbergerhornezeilinger}, who have exemplified an even stronger discrepancy between predictions of Local Realism and Quantum Mechanics than found in \cite{bell}. Surprisingly, this did not require a new inequality. Their original derivation was done for four qubits, nevertheless, let us discuss the version for three particles in more details.		

Let each of three observers, Alice, Bob, and Charlie, receive a spin-$\frac{1}{2}$ coming from a common source. Each of them chooses one of two dichotomic observables ($A^{[1]}_1,A_2^{[1]},A^{[2]}_1,A^{[2]}_2,A^{[3]}_1,A^{[3]}_2$) with spectra $\{+1,-1\}$. Let these observables be projections of the spins onto axes in the $xy$-plane, as in the case of two-qubit Bell theorem. Assume that the results of these measurements were deterministically predicted by local hidden variables. For example local hidden variables may set three different products, $A^{[1]}_1A_2^{[2]}A_2^{[3]},A_2^{[1]}A_1^{[2]}A_2^{[3]},A_2^{[1]}A^{[2]}_2A_1^{[3]}$ to be equal 1. Then from these three results we can compute the result for $A_1^{[1]}A_1^{[2]}A_1^{[3]}$:
\begin{equation}
\times\frac{\begin{array}{ccccc}A_1^{[1]}&A_2^{[2]}&A_2^{[3]}&=&1\\
A_2^{[1]}&A^{[2]}_1&A^{[3]}_2&=&1\\
A_2^{[1]}&A_2^{[2]}&A_1^{[3]}&=&1\end{array}}{\begin{array}{ccccc}A_1^{[1]}&A_1^{[2]}&A_1^{[3]}&=&1\end{array}}.
\end{equation}
This is due to the fact, that since the spectra of all observables are $\{+1,-1\}$, their squares are just unity operators. On the other hand, if Alice, Bob, and Charlie measure $x$ or $y$ components of their spins, for a certain quantum mechanical state they can have

\begin{eqnarray}
\label{ghzarg}
&\langle\sigma_1^{[1]}\sigma_2^{[2]}\sigma_2^{[3]}\rangle&=1,\nonumber\\ 
&\langle\sigma_2^{[1]}\sigma_1^{[2]}\sigma_2^{[3]}\rangle&=1,\nonumber\\ 
&\langle\sigma_2^{[1]}\sigma_2^{[2]}\sigma_1^{[3]}\rangle&=1,\nonumber\\ 
&\langle\sigma_1^{[1]}\sigma_1^{[2]}\sigma_1^{[3]}\rangle&=-1,\nonumber\\ 
\end{eqnarray}
what follows from

\begin{eqnarray}
\times\frac{\begin{array}{ccc}\sigma_1^{[1]}&\sigma_2^{[2]}&\sigma_2^{[3]}\\ \sigma_2^{[1]}&\sigma_1^{[2]}&\sigma_2^{[3]}\\ \sigma_2^{[1]}&\sigma_2^{[2]}&\sigma_1^{[3]}\end{array}}{\begin{array}{ccccc}\sigma^{[1]}_1&-\sigma^{[2]}_1\sigma^{[3]}_1\end{array}}.
\end{eqnarray}

A quantum state, for which mean values are equal to those assumed in (\ref{ghzarg}) takes a form of $\frac{1}{\sqrt{2}}(|000\rangle-|111\rangle)$ and is referred to as the 3-qubit GHZ state. Similar calculations hold for other possible results of first three products. The calculations based on Local Realism give results opposite to  quantum-mechanical ones.

This way of showing the contrast between Local Realism and Quantum Mechanics is called ``All-versus-Nothing" Bell theorem, or the Bell theorem without inequalities. It can be easily generalized to more qubits. A possible set of products, which allow to make predictions for all non-vanishing mean values in $xy$-planes consists of $\prod_{i=1}^{N}\sigma_1^{[i]}$ and all the combinations, in which Alice and one other observer choose $\sigma_2$, while the rest of them measures $\sigma_1$, e.g. $\sigma_2^{[1]}\sigma_2^{[2]}\prod_{i=3}^N\sigma_1^{[i]}, \sigma_2^{[1]}\sigma_2^{[3]}\prod_{i=2,4}^N\sigma_1^{[i]}$, etc.. 

All-versus-Nothing paradoxes can be also derived for hyperentangled \footnote{By hyperentangled we mean entangled in more a pair of degrees of freedom, such photons are entangled both in polarization modes and spatial modes referring to different paths in Mach-Zehnder interferometers, in which they propagate.} pairs of photons \cite{cabello}. 

It is also possible to derive All-versus-Nothing paradoxes for higher dimensional systems. For example, for three qutrits\footnote{Systems of the Hilbert space dimensionality 3.} Alice, Bob, and Charlie choose one of two observables with spectra $\{1,\alpha,\alpha^2\}$ $(\alpha=\exp(2\pi i/3))$, so that $(A_i^{[j]})^3=\openone_{3\times 3} (i=1,2,j=1,2,3)$. In such a case, a local realistic product
\begin{equation}
\label{avsnqutrits}
\times\frac{\begin{array}{ccc}A_1^{[1]}&A_1^{[2]}&A_1^{[3]}\\A_2^{[1]}&A_1^{[2]}&A_1^{[3]}\\A_1^{[1]}&A_2^{[2]}&A_1^{[3]}\\A_1^{[1]}&A_1^{[2]}&A_2^{[3]}\end{array}}{\begin{array}{ccc}A_2^{[1]}&A_2^{[2]}&A_2^{[3]}\end{array}}
\end{equation}
is relevant. If we find a set of observables and a state for which, e. g., $\langle A_1^{[1]}A_1^{[2]}A_1^{[3]}\rangle=\langle A_2^{[1]}A_1^{[2]}A_1^{[3]}\rangle=\langle A_1^{[1]}A_2^{[2]}A_1^{[3]}\rangle=\langle A_1^{[1]}A_1^{[2]}A_2^{[3]}\rangle=1$, but $\langle A_2^{[1]}A_2^{[2]}A_3^{[2]}\rangle\neq 1$, one has an explicit violation of Local Realism. Nevertheless, paradoxes derived for higher-dimensional systems are usually not as strong as those for qubits. 
\section{Mermin Inequalities}
\hspace*{5mm}

Mermin \cite{mermin} observed that the whole non-classicality of the $N$-qubit state 
\begin{equation}
\label{ghzstate}
|GHZ_N\rangle=\frac{1}{\sqrt{2}}\left(|0\rangle^{\otimes N}+|1\rangle^{\otimes N}\right),
\end{equation}
the correlation function in $xy$-planes of which reads
\begin{eqnarray}
\label{ghzcorrelation}
E(\phi_1,...,\phi_N)=\left\langle GHZ_N\left|\prod_{i=1}^N(\cos\phi_i\sigma_1^{[i]}+\sin\phi_i\sigma_2^{[i]})\right|GHZ_N\right\rangle
=\cos\left(\sum_{i=1}^N\phi_i\right),
\end{eqnarray}
is related to the fact that it is an eigenstate of the operator 
\begin{equation}
\label{mermin1}
M=\frac{1}{2}\left(\prod_{j=1}^{N}(\sigma_1^{[j]}+i\sigma_2^{[j]})+\prod_{j=1}^N(\sigma_1^{[j]}-i\sigma_2^{[j]})\right).
\end{equation}
In (\ref{mermin1}) only these terms survive in which $\sigma_2$ appears $k=0,2,4,..,2Int(N/2)$\footnote{By $Int(x)$ let us here denote the integer part of x, e.g. $Int(2.5)=2$.} times. If $k/2$ is even, there is a $``+"$ sign in front of a product of Pauli matrices, and $``-"$ otherwise. The corresponding mean values for the GHZ state (\ref{ghzstate}) have the same sings and each of them has a modulo 1. Thus the eigenvalue of (\ref{mermin1}) corresponding to (\ref{ghzstate}) is equal to the number of non-vanishing terms in $M$, $2^{N-1}$. 

On the other hand, if we assume Local Realism, the mean value of $A$ can be written as
\begin{equation}
\label{merminf}
F=\langle M\rangle_{LHV}=\textbf{Re}\left(\prod_{j=1}^N(a^{[j]}_1+ia^{[j]}_2)\right)_{LHV}.
\end{equation}
By $a_l^{[j]}$ we here denote the outcome of the $\sigma_l^j$ measurement determined by LHV. $F$ is a real part of a product of $N$ complex numbers, each of the modulo $\sqrt{2}$ and the argument $\pm\pi/4\pm\pi$. The vector representing in a complex plane this product can be parallel to the real axis for $N$ even, or create angle $\pi/4$ with the axis for $N$ odd. The classical limit of this expression is hence $2^{N/2}$ and $2^{(N-1)/2}$, respectively. Thus Mermin \cite{mermin} gave the first quantitative argument that the GHZ state for $N$ qubits leads not only to exponentially many (with $N$) All-versus-Nothing paradoxes, but also to an exponentially strong violation of Bell inequalities.

\section{Ardehali Inequalities}
\hspace*{5mm}

The first generalization of the Mermin inequalities was introduced by Ardehali \cite{ardehali}. He suggested to consider two operators for the $N$-qubit case:
\begin{eqnarray}
M_1=&(\sigma_1^{[1]}\sigma_1^{[2]}...\sigma_1^{[N-1]}\nonumber\\
&-(\sigma_2^{[1]}\sigma_2^{[2]}\sigma_1^{[3]}...\sigma_1^{[N-1]}+\sigma_2^{[1]}\sigma_1^{[2]}\sigma_2^{[3]}...\sigma_1^{[N-1]}+...)\nonumber\\
&+(\sigma_2^{[1]}\sigma_2^{[2]}\sigma_2^{[3]}\sigma_2^{[4]}\sigma_1^{[5]}...\sigma_1^{[N-1]}+...)\nonumber\\
&-...+...)(\vec{a}\:^{[N]}_1+\vec{a}\:^{[N]}_2)\cdot\vec{\sigma}^{[N]},\\
\nonumber\\
M_2=&((\sigma_2^{[1]]}\sigma_1^{[2]}...\sigma_1^{[N-1]}+\sigma_1^{[1]}\sigma_2^{[2]}...\sigma_1^{[N-1]}+...)\nonumber\\
&-(\sigma_2^{[1]}\sigma_2^{[2]}\sigma_2^{[3]}\sigma_1^{[4]}...\sigma_1^{[N-1]}+\sigma_2^{[1]}\sigma_2^{[2]}\sigma_1^{[3]}\sigma_2^{[4]}...\sigma_1^{[N-1]})\nonumber\\
&+...-...)(\vec{a}\:^{[N]}_1-\vec{a}\:^{[N]}_2)\cdot\vec{\sigma}^{[N]}.
\end{eqnarray}
As before, the scalar product of real unit vectors $\vec{a}\:^{[N]}_1$ and $\vec{a}\:^{[N]}_2$ with the Pauli matrices vector defines the first and the second observable of the $N$th observer.
The first operator is built only of terms in which $\sigma_2$ appears $k=0,2,4,...,2Int((N-1)/2)$ times for qubits 1 to $N-1$. Is $k/2$ even, the sign in front of the term is $``+"$ rather than $``-"$, which appears for $k/2$ odd. Similarly, in the second operator $\sigma_2$ appears $k=1,3,...,2Int((N-1)/2)+1$ times within this set of qubits. There is a $``+"$ sign for $(k-1)/2$ even and we have $``-"$ for $(k-1)/2$ odd. Thus the operators can be written as
\begin{eqnarray}
\label{ardehali1}
M_1=&\textbf{Re}\left(\prod_{j=1}^{N-1}(\sigma_1^{[j]}+i\sigma_2^{[j]})\right)(\vec{a}_1\:^{[N]}+\vec{a}_2\:^{[N]})\cdot\vec{\sigma}^{[N]},\\
\label{ardehali2}
M_2=&\textbf{Im}\left(\prod_{j=1}^{N-1}(\sigma_1^{[j]}+i\sigma_2^{[j]})\right)(\vec{a}_1\:^{[N]}-\vec{a}_2\:^{[N]})\cdot\vec{\sigma}^{[N]}.
\end{eqnarray}
Quantum mechanically, if vectors $\vec{a}\:^{[N]}_1$ and $\vec{a}\:^{[N]}_2$ are chosen to be $(1/\sqrt{2},1/\sqrt{2},0)$ and $(1/\sqrt{2},-1/\sqrt{2},0)$, respectively, we can reconstruct the original result of Mermin, that is that (\ref{ghzstate}) can be made an eigenstate of $M_1+M_2$ with the respective eigenvalue $2^{N-1/2}$. 

Local and realistic calculations are also similar to the ones made by Mermin. It is enough to notice that, as in case of the original derivation, for local realistic theories only one of two expressions, $\langle A^{[N]}_1\rangle_{LHV}+\langle A^{[N]}_2\rangle_{LHV}$ and $\langle A^{[N]}_1\rangle_{LHV}-\langle A^{[N]}_2\rangle_{LHV}$, is equal to $\pm 2$, the other vanishes. It then suffices to repeat the bounding argument of Mermin for (\ref{ardehali1}) and (\ref{ardehali2}) to obtain that the local realistic bound is $2^{(N-1)/2}$ for $N$ odd, or $2^{N/2}$ for $N$ even. 

Thus in the case of the GHZ states Ardehali's inequalities can be violated with a strength up to $\sqrt{2}$ times higher than Mermin's inequalities. This advantage was achieved in 
\cite{ardehali} by giving the last observer a possibility to measure projections of the spin onto two {\em arbitrary} axes, rather than onto two {\em fixed} ones. This allows Adrehali inequalities to be violated by both the $N$-partite GHZ state and a product of the $(N-1)$-qubit GHZ state and a single qubit pure state. The derivation of the inequality also refers to Bell's original idea of pairs of in some sense complementary quantities. The complementarity is expressed in the fact that if one of the expressions takes its maximal value, the other vanishes. 

\section{Mermin-Ardehali-Belinskii-Klyshko Inequalities}
\hspace*{5mm}

The idea of such a complementarity was exploited in the inequalities derived by Belinskii and Klyshko \cite{belinskiiklyshko}. Let us consider two series of operators, $S^{N}_1,S^{N}_2$, defined in an iterative way:
\vspace{5pt}
\begin{equation}
\label{klyshko}
S_1^{N+1}=\frac{1}{2}\left(S_1^{N}(A^{[N+1]}_1+A^{[N+1]}_2)+S^{N}_2(A^{[N+1]}_1-A^{[N+1]}_2)\right).
\end{equation} 
\vspace{5pt}
To initialize this iteration, we need to assume that the seeds are $S^0_1=S^0_2=1$, or $S^1_1=A^{[1]}_1,S^1_2=A_2^{[1]}$. The structure of $S^N_2$ is the same as of $S^N_1$, but with $A^{[k]}_1$ and $A^{[k]}_2$ interchanged for all $k\leq N$.  Let us write down few first expressions:
\vspace{5pt}
\begin{eqnarray}
\label{klyshko2}
S^2_1=&\frac{1}{2}(A_1^{[1]}(A^{[2]}_1+A_2^{[2]})+A^1_2(A^{[2]}_1-A_2^{[2]})),\\
\nonumber\\
\label{klyshko3}
S^3_1=&\frac{1}{2}(A_1^{[1]}A^{[2]}_1A^{[3]}_2+A^{[1]}_1A^{[2]}_2A^{[3]}_1+A^{[1]}_2A^{[2]}_1A^{[3]}_1-A^{[1]}_2A^{[2]}_2A^{[3]}_2),\\
&...\quad .\nonumber
\end{eqnarray}
\vspace{5pt}
Trivially, the extreme values for $S_1^1$ and $S_2^1$ are $\pm 1$. Then it is easy to argue from the construction of the sequence that also all mean values of $S^N_{1(2)}$ above 1 or below $-1$ cannot be explained with Local Realism. In deterministic theories based on LHV one of terms in (\ref{klyshko}) vanishes, the magnitude of the other is cancelled to 1 with the fractional factor $\frac{1}{2}$ appearing in each iteration.

The quantum-mechanical bound was shown in \cite{belinskiiklyshko} to be $2^{(N-1)/2}$. Thus so-called Mermin-Ardehali-Belinskii-Klyshko (MABK) inequalities turn out to be violated in some cases with a higher ratio than ever possible for Mermin or Ardehali inequalities. 

\section{WWW\.ZB Inequalities}
\hspace*{5mm}

The derivations described above, especially due to Belinskii and Klyshko \cite{belinskiiklyshko}, utilize the  complementarity mentioned above, however, only a certain configuration of such pairs of such complementary\footnote{In Local Realism.} expressions is used and each consideration leads to one certain inequality. 

The complete (in the sense explained further) set of Bell inequalities for $N$ qubits given that each observer chooses between two dichotomic observables  was derived by Werner and Wolf \cite{wernerwolf}, and independently by Weinfurter and \.Zukowski \cite{weinfurterzukowski} and by \.Zukowski and Brukner \cite{zukowskibrukner}. They are known as WWW\.ZB inequalities. 

The construction given in \cite{zukowskibrukner} is similar to the original argument in \cite{bell}. Let us consider all possible products $\left\langle\prod_{i=1}^N(A^{[k]}_1+s^{[k]}A^{[k]}_2)\right\rangle$ with $s^{[k]}=\pm 1$. Local-realistically, only one such expression at the time has non-zero value at the time. For a particular choice of signs, $s^{[k]}=\langle A^{[k]}_1\rangle_{LHV}\langle A^{[k]}_2\rangle_{LHV}$ (we remind that in these theories these mean values are also $\pm 1$), the product is equal to $\pm 2^N$. Thus the general Bell inequality for $N$ qubits and 2 observables per site reads
\begin{equation}
\label{zbinequalitygeneral}
\sum_{s^{[1]},...,s^{[N]}=\pm 1}\left|\left\langle\prod_{k=1}^{N}(A^{[k]}_1+s^{[k]}A^{[k]}_2)\right\rangle\right|\leq 2^N,
\end{equation}
and is equivalent to $2^{2^N}$ inequalities
\begin{equation}
\label{zbinequalityspecific}
\sum_{s^{[1]},..,s^{[N]}=\pm 1}S(s^{[1]},...,s^{[N]})\left\langle\prod_{k=1}^{N}(A^{[k]}_1+s^{[k]}A^{[k]}_2)\right\rangle\leq 2^N,
\end{equation}
where the sign function $S(.)$ depends on signs $s^{[k]}$ and takes values $\pm 1$. Within the set, we also find trivial inequalities, like $\left|\left\langle\prod_{k=1}^N A_1^{[k]}\right\rangle\right|\leq 1$. But also true non-trivial Bell expressions. The highest violation ratio is observed when (\ref{zbinequalityspecific}) reconstructs the MABK inequalities. 

It is possible to give a sufficient condition for an arbitrary state of $N$ qubits to satisfy (\ref{zbinequalityspecific}). Let us first rederive the necessary and sufficient condition for two qubits originally presented by the Horodecki Family \cite{3horodeckiprl85}, but rather in the formalism of \cite{zukowskibrukner}. The general two-qubit inequality can be written as 
\begin{equation}
\label{2qubitgeneral}
\sum_{l^{[1]},l^{[2]}=1}^2\left|\left\langle(A_1^{[1]}+(-1)^{l^{[1]}-1}A^{[1]}_2)(A^{[2]}_1+(-1)^{l^{[2]}-1}A_{[2]}^2)\right\rangle\right|\leq 4.
\end{equation}
Recall that in Quantum Mechanics the mean values of our interest are computed from the correlation tensor, $E_{j_1,..,j_N}=\langle A^{[1]}_{j_1}...A^{[N]}_{j_N}\rangle=\hat{T}\cdot(\vec{a}^{[1]}_{j_1}\otimes...\otimes\vec{a}^{[N]}_{j_N})\!(j_1,...,j_N=1,2)$ \footnote{The scalar product of two tensors is here understood as $\hat{A}\cdot\hat{B}=\sum_{i_1,...,i_N=1}^3A_{i_1...i_N}B_{i_1...i_N}$.}. Both observers can choose such local coordinate systems that the unit vectors defining their observables will satisfy $\vec{a}\:^{[k]}_{1}+\vec{a}\:^{[k]}_2=2\vec{e}\:_1^{[k]}\cos\alpha_k/2$ and $\vec{a}\:^{[k]}_{1}-\vec{a}\:^{[k]}_2=2\vec{e}\:_2^{[k]}\sin\alpha_k/2$, given that 
$\vec{e}_1,\vec{e}_2,\vec{e}_3$ are unit vectors parallel to axes $x,y$ and $z$, respectively, and $\alpha_k$ is the angle between $\vec{a}\:^{[k]}_1$ and $\vec{a}\:^{[k]}_2$. Putting this into (\ref{2qubitgeneral}) we get
\begin{eqnarray}
\label{2qubitc1}
(|T_{11}|,|T_{12}|,|T_{21}|,|T_{22}|)\cdot\nonumber\\
(|\cos\alpha_1\cos\alpha_2|,|\cos\alpha_1\sin\alpha_2|,
|\sin\alpha_1\cos\alpha_2|,|\sin\alpha_1\sin\alpha_2|)\leq 1, 
\end{eqnarray}
where $T_{ij}$s are elements of the two-qubit correlation tensor (see Section 2.4).
This means the following. One can take a correlation tensor of a state, initially given in any local bases. Then one constructs a modified tensor, $\hat{T}^{mod}$ in which all elements have been replaced by their moduli. If there are local coordinate systems in which any element of $\hat{T}^{mod}$ exceeds 1, that is if $\hat{T}^{mod}$ is not a correlation tensor of any physical state,  the state violates (\ref{2qubitgeneral}). This greatly resembles Peres-Horodecki criterion \cite{peresppt,horodeckippt} (to be explained in Section 5.2), in which physicality of a state is also questioned after a certain operation, i. e., a partial transposition. 

The two vectors appearing in (\ref{2qubitc1}) can be made parallel by a proper choice of $\alpha_1$ i $\alpha_2$ and coordinate systems. Moreover, the second vector built of trigonometric functions is always normalized. Hence, on the basis of the Cauchy inequality, $|\vec{a}\cdot\vec{b}|\leq|\vec{a}||\vec{b}|$, the necessary and sufficient condition for a state to satisfy (\ref{2qubitc1}) reads
\begin{equation}
\label{zb2q}
\sum_{a,b=1}^2T_{ab}^2\leq 1.
\end{equation}

In the same fashion one can consider (\ref{zbinequalitygeneral}) for more qubits, up to the point in which we parallelize the two vectors. This is in general not possible for more qubits due to the linear growth of free parameters and the exponential growth of of vector components. The analogous condition,
\begin{equation}
\label{zbc2}
\sum_{a,...n=1}^2T_{a...n}^2\leq 1,
\end{equation}
is only sufficient. If (\ref{zbc2}) holds in all coordinate systems, the state can never violate any of inequalities (\ref{zbinequalityspecific}). However, the condition with the modified condition, that is that
\begin{equation}
\label{zbc3}
\max\left|T^{mod}_{11..1}\right|\leq 1
\end{equation}
holds  in all local bases is still necessary and sufficient for (\ref{zbinequalitygeneral}) to be satisfied. The maximum is taken over all three-dimensional rotations, both before and after the modification of the tensor, for all local coordinate systems (each observer fixes his/her own Cartesian system).

\section{Drawbacks of WWW\.ZB Inequalities \cite{w1,w2}}
\hspace*{5mm}

An example of a state, which neither satisfies (\ref{zbc2}), nor violates (\ref{zbinequalitygeneral}) is a so-called noisy W state, given by
\begin{eqnarray}
\label{noisyw}
\rho_{W,N}(V_N)=V_N|W_N\rangle\langle W_N|+(1-V_N)\frac{\openone_{2^N\times 2^N}}{2^N},\\
|W_N\rangle=\frac{1}{N}(|10...0\rangle+|01...0\rangle+...+|0...1\rangle),\nonumber
\end{eqnarray}
which is certainly entangled for $V_N>\frac{N}{(\sqrt{2}-1)2^{N-1}+N}$. Our paper \cite{w2} shows that for this interval of $V_N$ noisy W states reveal a violation of Local Realism in a certain protocol. Namely, we demand that $N-2$ observers perform $\sigma_3$ measurements on their qubits and obtain ``$+1$" as their results. Such a result came with a high probability from the W admixture and with a relatively low likelihood from the white noise. Thus the overall W-state-to-noise ratio increases with every projection. If this ratio is higher than $\frac{1}{\sqrt{2}}$ after $N-2$ projections, the state of remaining 2 qubits, on which no measurements were made (possessed in many copies), violates a Bell inequality. Using this protocol we have shown that noisy W states lead to a stronger violation of Local Realism than noisy GHZ states. For $N\geq 11$ we need a smaller admixture of the W state to white noise in order to violate a Bell inequality than a required amount of the GHZ state. This advantage grows exponentially with $N$. For example, for 3 qubits, the threshold for $V_N$ is $\frac{3}{4\sqrt{2}-1}\approx 1/1.5523\approx 0.644212$. 

More importantly, in \cite{w2} we show that the maximum of $\sum_{a,b,c=1}^2T_{abc}^2=\frac{7}{3}V_3^2$ for a $\rho_{W,3}(V_3)$ or more generally, $(3-\frac{2}{N})V_N^2$ for an arbitrary value of $N$. However, we have performed a numerical optimization to compare the value of $V_3$, above which (\ref{zbinequalitygeneral}) is not satisfied to the one, above which (\ref{chshinequality}) is violated after the application of the protocol. We have shown that a pure 3-qubit W state violates (\ref{zbinequalitygeneral}) by a factor of $1.523<\sqrt{\frac{7}{3}}\approx 1.527$. Thus there exists an interval $0.6547\approx\sqrt{\frac{3}{7}}\leq P_3<\frac{1}{1.523}\approx 0.657$, in which the condition (\ref{zbc2}) is not satisfied, but (\ref{zbinequalitygeneral}) is.

Gisin \cite{gisinpla} and Gisin and Peres \cite{gisinpla16215} have shown that all pure entangled  states violate the CHSH inequality. The general two-qubit state reads
\begin{equation}
|\psi\rangle=c_{00}|00\rangle+c_{01}|01\rangle+c_{10}|10\rangle+c_{11}|11\rangle.
\end{equation}
The state is normalized to $\sum_{i,j=0}^1|c_{ij}|^2=1$.
Now, we can find a normalized product state  $|0'0'\rangle$, so that the scalar product $\langle 0' 0'|\psi\rangle$ is real and maximal. This implies that $\langle 1'0'|\psi\rangle=\langle 0'1'|\psi\rangle=0$, where $|1'\rangle$ are orthogonal to $|0'\rangle$ in both local Hilbert spaces. Finally, since the global phases determining $|1'\rangle$ are yet unset, we can choose $\langle 1'1'|\psi\rangle$	to be real and positive. Thus we can rewrite $|\psi\rangle$ as
\begin{equation}
\label{2qschmidt}
|\psi\rangle=\cos\alpha|0'0'\rangle+\sin\alpha|1'1'\rangle,
\end{equation}
with $0\leq\alpha\leq\pi/4$. Such a procedure is called the Schmidt decomposition. It aims to minimize the number of non-zero coefficients of a state, or, more generally, a vector living in a tensor product of Hilbert spaces. It can be generalized to any bipartite system, or any number of qubits.

It is simple to argue that the Bloch vectors of (\ref{2qschmidt}), $\langle\vec{\sigma}^{[1]}\rangle$ and $\langle\vec{\sigma}^{[2]}\rangle$, have only $z$-components non-vanishing, $T_{03}=T_{30}=\cos^2\alpha-\sin^2\alpha=\cos 2\alpha$, and that the correlation tensor has only three non-zero elements: $T_{11}=-T_{22}=2\cos\alpha\sin\alpha=\sin 2\alpha, T_{33}=1$ (see Section 2.4 for the definition of $T_{ij}$s). If we apply (\ref{zb2q}), we get $T_{11}^2+T_{33}^2=1+\sin^22\alpha\geq 1$ and $=1$ only for the case of a product state, for which $\alpha=0$. By applying transformation reverse to the ones used in the Schmidt decomposition this proof can be generalized for all pure entangled states.

Despite the result of Gisin and Peres, one can ask whether the WWW\.ZB inequalities, despite of forming a complete set of inequalities for $N$-partite correlation functions with two alternative observables per site, are violated by all pure states.
\.Zukowski, Brukner, Laskowski, and Wie\'sniak cite{w1} have shown, this is not the case. 
We have studied a family of states, which, despite of their obvious non-classicality, do not violate any of (\ref{zbinequalityspecific}) inequalities. The family are generalized GHZ states, 
\begin{equation}
\label{ghzgeneral}
|GHZ(\alpha,N)\rangle=\cos\alpha|0\rangle^{\otimes N}+\sin\alpha|1\rangle^{\otimes N}
\end{equation}
with $\alpha$ as before. Scarani and Gisin \cite{jpa34_6043} have shown that such states never violate MABK inequalities for $\sin2\alpha\leq 2^{(1-N)/2}$ with $N>2$. This bound is valid also for WWW\.ZB inequalities for odd $N$. The non-vanishing elements of the $N$-particle correlation tensor are:
\begin{eqnarray}
T_{1...1}=-T_{221...1}=-T_{212...1}=...=T_{22221....1}=...=\sin 2\alpha,\nonumber\\
T_{3...3}=\frac{1+(-1)^N}{2}+\frac{1-(-1)^N}{2}\cos 2\alpha.
\end{eqnarray}
The first line contains all the elements with an even number of subscripts $2$, rather than $1$.
Thus the optimal sums in (\ref{zbc2}) are $\max\{1,2^{N-1}\sin^22\alpha\}$ for $N$ odd and $\max\{1+\sin^22\alpha,2^{N-1}\sin^22\alpha\}$ for $N$ even. In the $N$ odd case a possible choice is to consider the $xz$ part of the correlation tensor, in which we have only two non-vanishing elements, $T_{3...3}=\cos 2\alpha, T_{1...1}=\sin 2\alpha$, thus $\sum_{a,..,n=1,3}T^2_{a...n}=1$. Alternatively, we can consider the $xy$ part of the tensor, which has $2^{N-1}$ non-zero elements, all of modulo $\sin 2\alpha$. Taking into account that $2^{N-1}\sin^22\alpha$ is less than $1$ for sufficiently small $\alpha$, we have shown that there exist pure entangled states of odd $N$ qubits, which never violate any WWW\.ZB inequalities. 

It is also worthy stressing that for $N$ even the optimal sum is always above $1$ (except for a trivial case of a product state). For the purposes of the demonstration of this statement, let us focus on the case of $N=4$ described in \cite{w1} and the $xz$ part of $\hat{T}$, which reads as follows:
\begin{equation}
\hat{T}_{[xz]}=\vec{e}_3\otimes\vec{e}_3\otimes\vec{e}_3\otimes\vec{e}_3+\sin 2\alpha\vec{e}_1\otimes\vec{e}_1\otimes\vec{e}_1\otimes\vec{e}_1.
\end{equation}
First, let three of the observers rotate their coordinate systems by $\pi/4$ around the $y$-axis. Then the correlation subtensor takes a form of
\begin{eqnarray}
\hat{T}'_{[xz]}=&2^{-3/2}\left((\vec{e}\:'_1+\vec{e}\:'_3)\otimes(\vec{e}\:'_1+\vec{e}\:'_3)\otimes(\vec{e}\:'_1+\vec{e}\:'_3)\otimes\vec{e}\:'_3\right.\nonumber\\ &+\left.\sin 2\alpha(\vec{e}\:'_1-\vec{e}\:'_3)\otimes(\vec{e}\:'_1-\vec{e}\:'_3)\otimes(\vec{e}\:'_1-\vec{e}\:'_3)\otimes\vec{e}\:'_1\right).
\end{eqnarray}
 Next, we replace the elements of the tensor with their moduli:
 \begin{eqnarray}
 \hat{T}_{[xz]}'^{mod}=2^{-3/2}(\vec{e}_1\!'+\vec{e}_3\!')\otimes(\vec{e}_1\!'+\vec{e}_3\!')\otimes(\vec{e}_1\!'+\vec{e}_3\!')\otimes(\sin2\alpha\vec{e}_1\!'+\vec{e}_3\!').
 \end{eqnarray}
Finally, first three observers shall perform such transformations that $\vec{e}_1\!''=\frac{1}{\sqrt{2}}(\vec{e}_1\!'+\vec{e}_3\!')$ and the fourth observer should define $\vec{e}_1\!''$ as $\frac{1}{\sqrt{1+\sin^22\alpha}}(\sin 2\alpha\vec{e}_1\!'+\vec{e}_3\!')$. The tensor has now only one non-zero entry:
\begin{equation}
\label{modtensor}
\hat{T}\!''^{mod}_{[xz]}=\sqrt{1+\sin 2\alpha}\vec{e}_1\!''\otimes\vec{e}_1\!''\otimes\vec{e}_1\!''\otimes\vec{e}_1\!''.
\end{equation}
Applying (\ref{zbc3}) to (\ref{modtensor}) we see that at least one of WWW\.ZB inequalities is violated by all such states for $0<\alpha\leq\pi/4$. The same argument can be applied for more qubits. The inequality, which is always violated a generalized CHSH inequality:
\begin{equation}
\left\langle(A_1^{[1]}+A_2^{[1]})\prod_{k=2}^NA_1^{[k]}+(A_1^{[1]}-A_2^{[1]})\prod_{k=2}^NA_1^{[k]}\right\rangle\leq 2.
\end{equation}

The two different behaviors, for $N$ odd and even, are related to an interpretation of $\sum_{a,..,n=1}^2T_{a...n}^2$. Following \cite{brukzukzeilqph0106119} we argue that this sum can be understood as the amount of information stored in some of correlations between the qubits. This interpretation is justified by the fact that when we measure an observable $A$ with two outcomes, ``$\pm 1$", which have equal degeneracies, $1-\Delta^2(A)=\langle A\rangle^2$ is equal to $1$ in the case of one of the results occurring deterministically, and 0 when both results occur with equal probabilities. The first situation would suggest that we have learned one  bit of information about the system\footnote{We need to stress, however, that it is crucial that the degeneracies are equal. Consider an extreme case, in which ``$+1$" appears much more often in the spectrum of $A$ than ``$-1$". For a maximally mixed state, which obviously contains no information, $\langle A\rangle$ would then be close to 1.}.
 Product states, and hence their statistical mixtures, cannot contain more than one bit encoded in these correlations. Thus if the condition (\ref{zbc2}) is not met, the presence of entanglement is immediately implied. In such a case, it is possible, but, as we have mentioned, not necessary, to violate Bell inequalities. 
 
\section{WZLP\.ZB Inequalities}
\hspace*{5mm}

A natural way of extending the set of known Bell expressions is to consider more than two observables per site.  
An example of such an inequality was given for three qubits by Wu and Zong in \cite{wuzong}. The inequality given therein utilizes four alternative measurements for two of three observers and two measurements of the third one.
A generalization of the inequality from \cite{wuzong} was demonstrated by Laskowski, Paterek, \.Zukowski, and Brukner \cite{patlaszukbruk}. They have noticed that for LHV models both the expressions,
\begin{equation}
A_{12,S_1}=\sum_{k_1,l_1=1,2}S_1(k_1,l_1)(A^{[1]}_1+(-1)^{k_1}A^{[1]}_2)(A^{[2]}_1+(-1)^{l_1}A_2^{[2]})
\end{equation}
and
\begin{equation}
A_{34,S_2}=\sum_{k_2,l_2=1,2}S_2(k_2,l_2)(A^{[1]}_3+(-1)^{k_2}A^{[1]}_4)(A^{[2]}_3+(-1)^{l_2}A_4^{[2]}),
\end{equation}
are always equal to $\pm 4$, similarly to $A^{[i]}_j$ being equal to $\pm 1$. Thus a new three-qubit inequality can be derived in a following way:
\begin{eqnarray}
\label{plzb3q}
&|\langle A_{12;12,34}\rangle|&\nonumber\\
=&\left|\left\langle\sum_{k,l=1,2}S(k,l)(A_{12,S_1}+(-1)^kA_{34,S_2})(A_{[1]}^3+(-1)^lA_2^{[3]})\right\rangle\right|&\leq 16.
\end{eqnarray}
Since any of the sign functions $S_1,S_2,S$ can have one of $2^4=16$ forms, (\ref{plzb3q}) stands for $\left(2^4\right)^=2^{12}$ inequalities. If none of $S,S_1,S_2$ is factorisable with respect to $(-1)^k,(-1)^l$,; $(-1)^{k_1},(-1)^{l_1}$, or $(-1)^{k_2},(-1)^{l_2}$, respectively, the inequalities are equivalent to
\begin{eqnarray}
\label{plzb3qa}
&\left|\left\langle (A_1^{[1]}(A^{[2]}_1+A^{[2]}_2)+A_2^{[1]}(A^{[2]}_1-A^{[2]}_2))(A_1^{[3]}+A_2^{[3]})+\right.\right.&\nonumber\\
&\left.\left.(A_3^{[1]}(A^{[2]}_3+A^{[2]}_4)+A_4^{[1]}(A^{[2]}_3-A^{[2]}_4))(A_1^{[3]}-A_2^{[3]})\right\rangle\right|&\leq 16,
\end{eqnarray}
which was found in \cite{wuzong}.
If $S_1$ or $S_2$ is  the only factorisable function under the consideration, we get an inequality equivalent to (\ref{plzb3qa}) with $A_1^{[1]}=A_2^{[1]}$ or $A_3^{[1]}=A_4^{[1]}$. In other cases, as well as if, for example, $A_1^{[1]}=A_3^{[1]},A_2^{[1]}=A_4^{[1]},A_1^{[2]}=A_3^{[2]},A_2^{[2]}=A_4^{[2]}$, the expression already belongs to the set (\ref{zbinequalityspecific}).

It is now possible to extend the derivation to more qubits, for examlpe, for $N=4$:
\begin{eqnarray}
\label{plzb4q}
&|\langle{A_{12;; 12;12,34;; 34;56,78}}\rangle|&\nonumber\\=&\left|\left\langle\sum_{k,l=1}^2S(k,l)(A_{12;12,34}+(-1)^kA_{34;56,78})(A_1^{[4]}+(-1)^lA_2^{[4]})\right\rangle\right|\leq 16.&
\end{eqnarray}

We thus obtain a series of $N$-qubit Bell inequalities, in which the last observer chooses from 2 apparatus settings, and the previous has twice as many as the next one, except the first two, who both have a choice of $2^{N-1}$ settings. 

We recall that for WWW\.ZB inequalities we chose the sum and the difference of the two vectors defining a pair of observables for each observer  to be proportional to $\vec{e}_1\:^{[k]}$ an $\vec{e}_2\:^{[k]}$. The same can be now done for the vectors of the last observer. The mean value of the Bell operator is then expressed as two terms, dependent on disjoint set of observables of all other observers. Thus such a procedure can be applied for the next observer, with independently chosen local Cartesian systems in each term. Repeating the reasoning from Section 3.5 we finally reach the {\em necessary} and {\em sufficient} condition for a state to satisfy the inequality. For the three-qubit inequality (\ref{plzb3qa}) it reads

\begin{equation}
\label{plzbcond3q}
\sum_{a,b=1}^2T_{ab1}^2+\sum_{a,b=1}^2(T'_{ab2})^2\leq 1.
\end{equation}
Carol's coordinate system is the same in both therms. Cartesian bases of Alice and Bob for the second term may be chosen different from those for the first one. The analogous condition for a state to satisfy (\ref{plzb4q}) reads 
\begin{equation}
\label{plzbcond4q}
\sum_{a,b=1,2}T^2_{ab11}+\sum_{a,b=1,2}T'^2_{ab12}+\sum_{a,b=1,2}T''^2_{ab21}+\sum_{a,b=1,2}T'''^2_{ab22}\leq 1.
\end{equation} 
In any of the four terms the two first experimenters are allowed to arbitrarily choose their coordinate systems, individually for each term. The third observer has one Cartesian basis for the first two terms and some other for the other two. The forth works in a fixed coordinate system in all four terms.

Importantly, conditions (\ref{plzbcond3q}), (\ref{plzbcond4q}), and similar ones we would obtain for more qubits are {\em necessary} and {\em sufficient}. Thus the WZLP\.ZB inequalities are violated by noisy W states (\ref{noisyw}) for any $V_N>1/\sqrt{3-\frac{2}{N}}$, unlike in the case of WWW\.ZB inequalities. 

All generalized GHZ states considered in \cite{w1} violate the WZLP\.ZB inequalities for $\alpha\neq 0$, also for $N$ odd. To show that, let the observables of the last observer be $\cos\theta_1\sigma^{[N]}_3\pm\sin\theta_1\sigma^{[N]}_1$. Now, only such elements of the $N$-particle correlation tensor enter the condition for satisfying or violating the inequality, which have the last subscript equal to 1 or 3. In the first case all others should choose observables in such a way that the condition is entered by the elements of the tensor, for which the subscript $``2"$ appears an even number of times, while the remaining subscripts are $``1"$, e.g. $T_{11...1}$ or $T_{2211...1}$. Since the value of the last index is fixed, there is $2^{N-2}$ such terms, each equal to $\pm\sin 2\alpha$. The total contribution of these elements to the sum  is at most $2^{N-2}\sin^22\alpha$. If the last subscript is equal to 3, the rest of observers should ensure that $T_{3...3}^2=\cos^22\alpha$ enters the condition. The total sum would then be $1+(2^{N-2}-1)\sin^2 2\alpha>1$.

\section{Convex Hull Problem}
\hspace*{5mm}

As first considered by Froissart \cite{Froissart}, Bell inequalities define hyperplanes in a statistical space\footnote{By a statistical space we shall mean a real vector space with components given by probabilities of certain events (then the space is referred to as a probability space) or mean values of certain operators (as we will consider products of local observables, this will be called a correlation space or a mean values space).}. In between all these hyperplanes for a given experimental setup there is a convex polytope\footnote{A polytope is a finite region of a hyperspace bounded by a finite number of faces. The word ``polytope" closes a sequence of notions ``point, line segment, polygon, polyhedron,...".}. This polytope contains all possible statistical distributions which can be explained with LHVs. This makes reasonable to ask two following questions about Bell inequalities. 

The first one is ``Is the family, to which a Bell inequality belongs, complete?". That is, if the family completely bounds the polytope of LHV statistics in its most compact form. The answer to this question depends on the dimensionality of  
the statistical space. For example, WWW\.ZB inequalities are complete when each observer can choose between only two measurements, but, as we have shown, there are physical probability distributions, which satisfy the inequalities (in this case, belong to the polytope), but violate WZLP\.ZB inequalities. 

The other problem that can be addressed is whether or not a given inequality is tight\footnote{A tight inequality defines a hyperplane, which actually contains a face of the polytope.}. The original Bell theorem (see Section 2.4) states only that there exist experiments, quantum-mechanical results of which cannot be explained within Local Realism. For the purposes of the falsification of LHV-based theories it is enough to find any hyperplane, which separate any physically accessible point of the statistical space from the polytope of statistics explainable by LHV. However, for the sake of detecting of useful entanglement in as many states as possible, we need to find optimal, or in other words, tight Bell inequalities. The conditions for tightness are that the hyperplane contains at least $D$ extreme points of the polytope and that the whole its interior lies at the same side of the hyperplane. $D$ stands for the dimensionality of the space. This problem is called {\em convex hull problem}. It is thus the usual procedure to find tight Bell inequalities: to find a set of the extreme points of the LHV-permitted polytope, and then to list hyperplanes containing of $D$ of them, which do not intersect the interior. Such lists were given, for instance, by Pitowsky and Svozil \cite{pitowskysvozil}, \'Sliwa \cite{sliwa}, and Collins {\em et al.} \cite{collins}. It also turns out that \cite{bell,clauserhorne,chsh,belinskiiklyshko,zukowskibrukner,patlaszukbruk} are also elegant, yet unconscious, derivations of inequalities optimal in this sense.

Out of many examples from \cite{pitowskysvozil}, let us give the one for two qubits and two events per site, denoted as $A_1,A_2,B_1,B_2$. A point of the statistical space is described by 8 probabilities arranged to a vector:
\begin{eqnarray}
(P(A_1),P(A_2),P(B_1),P(B_2),P(A_1,B_1),P(A_1,B_2),P(A_2,B_1),P(A_2,B_2)).\nonumber
\end{eqnarray}
Local realistic theories describe sixteen points of this space:
\begin{eqnarray}
(1,1,1,1,1,1,1,1),&
(1,1,1,0,1,0,1,0),\nonumber\\
(1,1,0,1,0,1,0,1),&
(1,1,0,0,0,0,0,0),\nonumber\\
(1,0,1,1,1,1,0,0),&
(1,0,1,0,1,0,0,0),\nonumber\\
(1,0,0,1,0,1,0,0),&
(1,0,0,0,0,0,0,0),\nonumber\\
(0,1,1,1,0,0,1,1),&
(0,1,1,0,0,0,1,0),\nonumber\\
(0,1,0,1,0,0,0,1),&
(0,1,0,0,0,0,0,0),\nonumber\\
(0,0,1,1,0,0,0,0),&
(0,0,1,0,0,0,0,0),\nonumber\\
(0,0,0,1,0,0,0,0),&
(0,0,0,0,0,0,0,0),\nonumber
\end{eqnarray}
 and the complete set of inequalities reads
 \begin{eqnarray}
\label{pitowskytrivial}
&0\leq P(A_i,B_j)\leq P(A_i),P(B_j)\leq 1,
\vspace{5pt}
\\
\label{pitowskych1}
&-1\leq P(A_1B_1)+P(A_1B_2)+P(A_2B_1)-P(A_2B_2)-P(A_1)-P(B_1)\leq 0,&\nonumber\\
\\
\label{pitowskych2}
&-1\leq P(A_1B_2)+P(A_1B_1)+P(A_2B_2)-P(A_2B_1)-P(A_1)-P(B_2)\leq 0,&\nonumber\\
\\
\label{pitowskych3}
&-1\leq P(A_2B_1)+P(A_2B_2)+P(A_1B_1)-P(A_1B_2)-P(A_2)-P(B_1)\leq 0,&\nonumber\\
\\
\label{pitowskych4}
&-1\leq P(A_2B_2)+P(A_2B_1)+P(A_1B_2)-P(A_1B_1)-P(A_2)-P(B_2)\leq 0,&\nonumber\\
\end{eqnarray}
$i,j=1,2$. The necessary and sufficient criterion for a point to be in agreement with Local Realism is that it satisfies all these inequalities. (\ref{pitowskytrivial}) are trivial inequalities, which define properties of probabilities, whereas (\ref{pitowskych1}), (\ref{pitowskych2}), (\ref{pitowskych3}), and (\ref{pitowskych4}) are Clauser-Horne inequalities, equivalent to (\ref{chinequality}). However, permitting observers to consider {\em three} events leads to more complicated inequalities given in \cite{pitowskysvozil}.
\section{Tight Bell Inequalities with up to Three Settings per Site}
\hspace*{5mm}

It is important to stress that the approach of convex hull to the problem of finding Bell inequalities is suitable for both the probability space and the correlation space. Derivations of Pitowsky and Svozil \cite{pitowskysvozil} were made for probabilities, whereas \.Zukowski \cite{qph0611086} attempted to find tight inequalities by a direct analysis of the polytope spanned in the correlation space. The subject of his interest was a situation, in which each of $N$ observers chooses between three dichotomic observables with usual outcomes, $``\pm 1"$. 

For the sake of illustrating the problem \.Zukowski considers the case of $N=1$ thus working in the space of vectors built of mean values of observables $A_0^{[1]},A_1^{[1]},$ and $A_2^{[1]}$. The inequalities found for $N=1$ cannot be interpreted as Bell inequalities, since tho whole physically allowed fragment of the space is also per definition accessible with local hidden variables.

An elementary event is that measurements of the three observables give results $a_0^{[1]},a^{[1]}_1,a_2^{[1]}$, respectively. The probability of such an event shall be denoted as $P(a_0^{[1]},a_1^{[1]},a_2^{[1]})$. The mean value vector shall be defined as 
\begin{equation}
\vec{E}=\sum_{a_0^{[1]},a_1^{[1]},a_2^{[1]}=\pm 1}(a_0^{[1]}, a_1^{[1]}, a_2^{[1]}) P(a_0^{[1]},a_1^{[1]},a_2^{[1]}).
\end{equation}
The polytope is a cube, faces of which connect the following quadruples of points ($a,b=\pm 1$):
\begin{eqnarray}
(1,a,b),&(-1,a,b),&(a,1,b),\nonumber\\
(a,-1,b),&(a,b,1),&(a,b,-1).\nonumber
\end{eqnarray}
For a given face any three vertices form a complete linearly independent set in  $\Re^3$, however, vectors of this set are not orthogonal. Any other vector belonging to the space can be uniquely expressed as a linear combination of the three. In particular, the fourth vector of the face can be decomposed to the first three with coefficients of moduli 1 and which add up to 1. For example, 
\begin{equation}
\label{zukowskiface}
\sum_{a,b=\pm 1}ab(1,a,b)=(0,0,0).
\end{equation}

 More generally, we recall the fact from Analytic Geometry that a hyperplane in a $D$-dimensional space which does not cross the origin of the coordinate system, is defined by $D$ linearly independent vectors. All other vectors which point the hyperplane are such linear combinations of the initial vectors, that the expansion coefficients add up to 1. Rephrasing this fact explicitly, we take a complete set of $D$ lineary independent vectors, $\{\vec{r}_i\}_{i=1}^D$. A hyperplane is a defined as a set of all such vectors $\vec{r}=\sum_{i=1}^D\vec{r}_it_i$, that $\sum_{i=1}^Dt_i=1$.
 
Now, to get more intuition on the problem for more complicated cases, \.Zukowski suggests to consider three randomly chosen vertices of the cube. If these three vertices belong to the same face of the cube, the fourth vertex at the plane can be expressed as a linear combination of the three with coefficients adding up to 1. 

In the other case, the three vectors are linearly independent, but do not form a face. Thus there must exist a pair, $\vec{v}$ and $-\vec{v}$, such that its elements lie at two sides of the hyperplane. The hyperplane cuts the polytope through.

The last possibility is that the three vectors do not establish a complete set. Thus there are vector $\vec{v}$ and $-\vec{v}$ which are not expressible by the trio. Moreover, such a trio must consist of a pair $\vec{v}\;'$ and $-\vec{v}\;'$, thus the hyperplane also cuts the cube through. 

Finally, \.Zukowski presents the following corollary. If one builds a non-orthonormalized basis out of three vectors representing the vertices of the polytope, which defines its face, all other vertices are described by convex combinations of basis vectors with coefficients equal to $\pm 1$.

Let us define the analogous problem for two qubits. The correlation space is described by a correlation matrix,
\begin{equation}
\hat{E}=\sum_{a_0^{[1]},a_1^{[1]},a_2^{[1]}=\pm 1}\sum_{a_0^{[2]},a_1^{[2]},a_2^{[2]}=\pm 1}(a_0^{[1]},a_1^{[1]},a_2^{[1]})\otimes(a_0^{[2]},a_1^{[2]},a_2^{[2]})P(a_0^{[1]},a_1^{[1]},a_2^{[1]};a_0^{[2]},a_1^{[2]},a_2^{[2]}).
\end{equation}
\vspace{5pt}
We thus work in a nine-dimensional correlation space and the polytope is spanned between $2^5=32$ vertices\footnote{Having in total six measurements, one might expect to have $2^6=64$ vertices. Please note, however, that $(a_0^{[1]},a_1^{[1]},a_2^{[1]})\otimes(a_0^{[2]},a_1^{[2]},a_2^{[2]})=(-a_0^{[1]},-a_1^{[1]},-a_2^{[1]})\otimes(-a_0^{[2]},-a_1^{[2]},-a_2^{[2]})$, and thus their number is decreased by a factor 2.}. We can always divide all the vectors representing them into two groups in such a way that one element out of each pair $\{\vec{v}\:^i,-\vec{v}\:^i\}$ belongs to one group, and the other to the second. In each such group out of 16 vectors we can choose 9 vector forming a complete set. Which such sets then define a face of the polytope?

\.Zukowski now notices {\em Fact 1}. He considers a randomly chosen complete basis of vectors $\{\vec{v}\;^i\}_{i=1}^9$ and takes an assumption that there exists seven more vectors, which have the expansion coefficients summing up to 1. Thus these sixteen vectors constitute one hyperplane. \.Zukowski now aims to prove that no other vertex belongs to the hyperplane and all of them lie at the same side of it.

{\em Proof:} The necessary and sufficient condition for a vector $\vec{x}$ to be part of the corresponding hyperplane is that the quantity
\begin{equation}
\label{zukowskidcapital}
D[\vec{x},\vec{v}\:^1,...,\vec{v}\:^9]=det\left[\begin{array}{ccccc}1&x_1&x_2&...&x_9\\1&v_1^1&v_2^1&...&v_9^1\\...& & & \\1&v_1^9&v_2^9&...&v_9^9\end{array}\right]
\end{equation}
is equal to 0. Moreover, $D[\vec{x},\vec{v}\:^1,...,\vec{v}\:^9]$ is always positive at one side of the hyperplane an negative the other.

The vertex $\vec{v}_0$, which is from outside the hyperplane is either given by $-\vec{v}\:^k (k=1,...,9)$ or $\sum_{k=1}^9t_k\vec{v}\:^k$ with $\sum_{k=1}^9t_k=-1$. In 
either case, we can add such a linear combinations of rows 2 to 10 to the first row that only its first entry is 2, while the rest of them vanishes. It is then clear that
\begin{equation}
D[\vec{v}\;^0,\vec{v}\;^1,...,\vec{v}\;^9]=2d[\vec{v}\;^1,\vec{v}\;^2,...,\vec{v}\;^9]
\end{equation}
with
\begin{equation}
d[\vec{v}\;^1,...,\vec{v}\;^9]=det\left[\begin{array}{ccc}v_1^1&...&v_9^1\\...& &...\\v_1^9&...&v_9^9\end{array}\right].	
\end{equation}
QED.

{\em Fact 2} states that plane containing both $\vec{v}$ and  $-\vec{v}$ contains also $\vec{0}=(\vec{v}+(-\vec{v}))/2$.

{\em Fact 3} is that coefficients of expansion of any vector in a non-orthonormal basis are integers. In fact, we can show that in our case these coefficients are $\pm 1$ or 0.

{\em Proof:} let us choose 9 basis vectors as $(1,a',b')\otimes(1,c',d')$ where $a',b',c',d'=\pm 1$, but never $(a',b')=(1,1)$ or $(c',d')=(1,1)$. Now, we use (\ref{zukowskiface}) to obtain 6 more vertices:
\begin{eqnarray}
\label{zukowskicombination1}
(1,1,1)\otimes(1,c',d')=\sum_{(a',b')\neq(1,1)}(1,a',b')\otimes(1,c',d'),\\
\label{zukowskicombination2}
(1,1,1)\otimes(1,b',a')=\sum_{(c',d')\neq(1,1)}(1,a',b')\otimes(1,c',d'),
\end{eqnarray}
and the 16th one:
\begin{eqnarray}
\label{zukowskicombination3}
(1,1,1)\otimes(1,1,1)=\sum_{(a',b'),(c',d')\neq(1,1)}(1,a',b')\otimes(1,c',d').
\end{eqnarray}
Sixteen other vectors are obtained by a global sign flip. For a set of nine in which other pairs of $(a',b')$ or $(c',d')$ being excluded, a similar argument can be presented. If a choice of nine is more complicated, yet the set is complete, it is sufficient to use transformations inverse to (\ref{zukowskicombination1}), (\ref{zukowskicombination2}), and (\ref{zukowskicombination3}), coefficients of which are also $\pm 1$ or 0. Applying these transformations once more, we see that all coefficients must be integers. QED.

Finally, assume that we have found a hyperplane containing between 9 and 15 veritces, i. e., there exists a pair of vectors, $\vec{v}\;^0$ and $-\vec{v}\;^0$, neither of which ends at the hyperplane. {\em Fact 4} is that they lie at two opposite sides of the hyperplane and the latter cannot define a face of a polytope.

{\em Proof:} obviously, we can expand $\pm\vec{v}^0=\pm\sum_{k=1}^9t_k\vec{v}\:^k$. By adding an appropriate combination to the first row of the matrix used to compute (\ref{zukowskidcapital}) we get
\begin{equation}
D[\pm\vec{v}\:^0,\vec{v}\:^1,...,\vec{v}\:^9]=1\mp \sum_{k=1}^9t_kd[\vec{v}\:^1,...,\vec{v}\:^9].
\end{equation}
The previous proof showed that all coefficients are integers. It is, moreover, easy to argue that they must add up to an odd integer. Since it is impossible that $\left|\sum_{k=1}^9t_k\right|=1$, $D$ has different signs for $\vec{v}\:^0$ and $-\vec{v}\:^0$. The hyperplane must cut the interior through. We hence conclude that all faces of the polytope contain exactly 16 vertices. QED.

The Bell inequalities will be given by
\begin{equation}
\label{zukowskibell1}
\pm D[\hat{E},\vec{v}\:^1,...,\vec{v}\:^9]\leq 0,
\end{equation}
where the global sign is to be determined\footnote{Certainly, the maximally mixed state, $\frac{1}{2^3}\openone_{2^3\times 2^3}$, which carries no information, satisfies all Bell inequalities. The global sing of the left-hand side of (\ref{zukowskibell1}) shall be hence chosen the opposite as of $d[\vec{v}^1,...,\vec{v}^9]$.}, and $\{\vec{v}\:^k\}_{k=1}^9$ are 9 of 16 vectors defining a  hyperplane, which has the properties emerging from the described facts and contains a face of the polytope.                                                                        

Now, we are only left with the problem of finding valid sets of vertices. One such set is certainly $\{(1,a,b)\otimes(1,c,d)\}_{a,b,c,d=\pm 1}$. Other sets, claims \.Zukowski, are given by $S(a,b,c,d)(1,a,b)\otimes(1,c,d)$, with a sign function $S(a,b,c,d)$ such that $\sum_{a,b}abS(abcd)=\sum_{c,d}cdS(a,b,c,d)=0$. This also guaranties that, e.g., $\sum_{abc}abcS(a,b,c,d)=\sum_{a,b,c,d}abcdS(a,b,c,d)=0$. The explicit form of the sign function is
\begin{equation}
\label{zukowskisg}
S(a,b,c,d)=X+Aa+Bb+Cc+Dd+Eac+Fad+Gbc+Hbd,
\end{equation}
where $X,...,H$ are constants.

{\em Proof:} let us choose the basis as $S(a',b',c',d)(1,a',b')\otimes(1,c',d')$, with cases $(a',b')=(1,1)$ and $(c',d')=(1,1)$ excluded.

For the beginning, let us consider a constant sign function. The remaining 7 vertices belonging to the face are obtained from (\ref{zukowskiface}), and therefore the coefficients of expanding $(1,1,1)\otimes(1,c',d')$ and $(1,a',b')\otimes(1,1,1)$ add up to $-\sum_{(a',b')\neq(1,1)}a'b'=-\sum_{(c',d')\neq(1,1)}c'd'=1$. Eventually, $(1,1,1)\otimes(1,1,1)=\sum_{(a',b')\neq (1,1)}\sum_{(c',d')\neq(1,1)}a'b'c'd'(1,a',b')\otimes(1,c',d')=1$.\\
We will now generalize these considerations for an arbitrary sign function. When $S(a,b,c,d)$ is not constant, we cancel its action by putting it into the coefficients:
\begin{eqnarray}
&S(1,1,c',d')(S(1,1,c',d')(1,1,1)\otimes(1,c',d'))&\nonumber\\
\nonumber\\
&=\sum_{(a',b')\neq(1,1)}a'b'S(a',b',c',d')(S(a',b',c'd')(1,a',b')\otimes(1,c',d'),&
\\
&S(a',b',1,1)(S(a',b',1,1)(1,a',b')\otimes(1,1,1))&\nonumber\\
\nonumber\\
&=\sum_{(c',d')\neq(1,1)}c'd'S(a',b',c',d')(S(a',b',c'd')(1,a',b')\otimes(1,c',d'),&
\\
&S(1,1,1,1)(S(1,1,1,1)(1,1,1)\otimes(1,1,1))&\nonumber\\
\nonumber\\
&=\sum_{(a',b')\neq (1,1)}\sum_{(c',d')\neq(1,1)}a'b'c'd'S(a',b',c',d')(S(a',b',c'd')(1,a',b')\otimes(1,c',d').&\nonumber\\
\end{eqnarray}
Thus the sums of the coefficients of the expansion are given by
\begin{eqnarray}
&-S(1,1,c',d')\sum_{(a',b')\neq(1,1)}a'b'S(a',b',c',d')=S(1,1,c',d')^2=1,&\\
&-S(a',b',1,1)\sum_{(c',d')\neq(1,1)}c'd'S(a',b',c',d')=S(a',b',c,d)^2=1,&\\
&-S(1,1,1,1)\sum_{(a',b')\neq (1,1)}\sum_{(c',d')\neq(1,1)} a'b'c'd'S(a',b',c',d')=S(1,1,1,1)^2=1,&\nonumber\\
\end{eqnarray}
where these identities follow from (\ref{zukowskiface}) and (\ref{zukowskisg}).

\.Zukowski next considers a sign function $\sigma(a',b',c',d')$ which does not have the form of (\ref{zukowskisg}). In particular, let us focus on the case of $\sum_{(c',d')\neq(1,1)}c'd'\sigma(a',b',c',d')=x\sigma(a',b',1,1),x\neq -1$. This means that either $x=1$ or $x=\pm 3$. As the relation
\begin{eqnarray}
&\sum_{(c'd')\neq(1,1)}c'd'\sigma(a',b',c',d')(\sigma(a',b',c',d')(1,a',b')\otimes(1,c',d'))&\nonumber\\
&=-\sigma(a',b',1,1)(\sigma(a',b',1,1)(1,a',b')\otimes(1,1,1))&
\end{eqnarray}
holds, the sum of the expansion coefficients is
\begin{equation}
-\sum_{(c',d')\neq(1,1)}c'd'\sigma(a',b',c',d')\sigma(1,1,c',d')=-x\sigma(a',b',1,1)^2=-x\neq 1,
\end{equation}
i. e., the hyperplane does not contain the vertex. QED.

The Bell inequalities in the form of (\ref{zukowskibell1}) look complicated, not suitable for direct applications. The equivalent form for the case of two qubits and three observables per qubit is
\begin{equation}
\label{zukowskibell2}
\sum_{a,b,c,d=\pm 1}S(a,b,c,d)\hat{E}\cdot(1,a,b)\otimes(1,c,d)\leq 2^4,
\end{equation}
or, more explicitly,
\begin{eqnarray}
\label{zukowskibell2a}
&\sum_{a,b,c,d=\pm 1}S(a,b,c,d)&\nonumber\\
&\times(E_{00}+aE_{10}+bE_{20}+cE_{01}+dE_{02}+acE_{11}+adE_{12}+bcE_{21}+bdE_{22})&\nonumber\\
&\leq 2^4,&
\end{eqnarray}
or, using (\ref{zukowskisg}),
\begin{eqnarray}
\label{zukowksibell2b}
&XE_{00}+CE_{01}+DE_{02}&\nonumber\\
+&AE_{10}+EE_{11}+FE_{12}&\nonumber\\
+&BE_{20}+GE_{21}+HE_{22}&\leq 1,
\end{eqnarray}
where $E_{ij}=\langle A_i^{[1]}A_j^{[2]}\rangle$.

Such a form of the inequality can be straight-forwardly generalized for more complicated cases, like $N$ qubits, or more observables per site.
\section{Explicit Form of Inequalities \cite{w8}}
\hspace*{5mm}

\.Zukowski \cite{qph0611086} has presented a general method to find tight Bell inequalities, but never gave the inequalities in their explicit form. The problem was thus brought under further analysis by Wie\'sniak, Badzi\k{a}g, and \.Zukowski in \cite{w8}. The paper aims to show a method of building a sign function.

We shall start with the case of $N=2$ For completeness of the analysis we first need to redefine the sign function. In the discussed case, instead of two, let $S$ be dependent on three signs per observer, so that (\ref{zukowskibell2a}) has the form
\begin{equation}
\label{wiesniakbell1}
\frac{1}{2^6}\sum_{s_0^{[1]},s_1^{[1]},s_2^{[1]}=\pm 1}\sum_{s_0^{[2]},s_1^{[2]},s_2^{[2]}=\pm 1}
S(s_0^{[1]},s_1^{[1]},s_2^{[1]};s_0^{[2]},s_1^{[2]},s_2^{[2]})\left(\sum_{i,j=0}^2s_i^{[1]}s_j^{[2]}E_{ij}\right)\leq 1,
\end{equation}
and the discrete Fourier transform (up to a multiplicative constant) of the sign function is equal to
\begin{equation}
\label{wiesniaksignfunction}
S(s_0^{[1]},s_1^{[1]},s_2^{[1]};s_0^{[2]},s_1^{[2]},s_2^{[2]})=\sum_{i,j=0}^2g_{ij}s_i^{[1]}s_j^{[2]}.
\end{equation}
Coefficients $g_{ij}$ are equal to
\begin{equation}
\label{gij}
g_{ij}=\frac{1}{2^6}\sum_{s_0^{[1]},s_1^{[1]},s_2^{[1]}=\pm 1}\sum_{s_0^{[2]},s_1^{[2]},s_2^{[2]}=\pm 1}S(s_0^{[1]},s_1^{[1]},s_2^{[1]};s_0^{[2]},s_1^{[2]},s_2^{[2]})s_i^{[1]}s_j^{[2]}
\end{equation}
so that (\ref{wiesniakbell1}) can by shortly written as 
\begin{equation}
\label{wiesniakbell2}
\sum_{i,j=0}^2g_{ij}E_{ij}\leq 1.
\end{equation}

Let us concentrate on the properties of the coefficients. We start with defining functions, which we shall call deltas of some order. First order delta with respect to $s_i^{[1]}$ is given by
\begin{equation}
\label{delta1}
\Delta_{s_i^{[1]}}=\frac{S(s_i^{[1]}=1)-S(s_i^{[1]}=-1)}{2}.
\end{equation}
Since for any given combination of all other signs the sign function may, or may not flip, $\Delta_{s_i^{[1]}}$ takes values of $\pm 1$ or 0. Similarly, we define second order delta with respect to $s_i^{[1]}$ and $s_j^{[2]}$:
\begin{equation}
\label{delta2}
\Delta_{s_i^{[1]}s_j^{[2]}}=\frac{\Delta_{s_i^{[1]}}(s_j^{[2]}=1)-\Delta_{s_i^{[1]}}(s_j^{[2]}=-1)}{2}.
\end{equation} 
Under the flip of $s_j^{[2]}$ $\Delta_{s_i^{[1]}}$ can remain unchanged, or change between $\pm 1$ and $0$ or between $1$ and $-1$. Thus possible values of a second-order deltas are $0,\pm\frac{1}{2},\pm 1$. In general $k$th order deltas can take values between $-1$ and $1$, which are multiples of $2^{1-k}$.

The other obvious thing about deltas is that for $N$ qubits $N$th order deltas are coefficients of the Fourier transform of the sign function. The higher order deltas can be thus reconstructed by
\begin{equation}
\Delta_{s_i^{[1]}}=\Delta_{s_i^{[1]}s_0^{[2]}}s_0^{[2]}+\Delta_{s_i^{[1]}s_1^{[2]}}s_1^{[2]}+\Delta_{s_i^{[1]}s_2^{[2]}}s_2^{[2]}=g_{i0}s_0^{[2]}+g_{i1}s_1^{[2]}+g_{i2}s_2^{[2]}
\end{equation}
and
\begin{equation}
S(s_0^{[1]},s_1^{[1]},s_2^{[1]};s_0^{[2]},s_1^{[2]},s_2^{[2]})=\Delta_{s_0^{[1]}}s_0^{[1]}+\Delta_{s_1^{[1]}}s_1^{[1]}+\Delta_{s_2^{[1]}}s_2^{[1]}.
\end{equation}
Other facts, which would be helpful in constructing valid sign functions, are
\begin{equation}
\label{consg}
\left|\sum_{i,j=0}^2g_{ij}\right|=1=\sum_{i,j=0}^2g_{ij}^2.
\end{equation}
The first equality follows from Fact 4 from \cite{qph0611086} that each face of the polytope contains 16 vertices. The second is a consequence of Parseval's theorem, which states that the Fourier transform preserves the Euclidean norm of the transformed object. Since the norm of any sign function is 
\begin{equation}
\sum_{s_0^{[1]},s_1^{[1]},s_2^{[1]}=\pm 1}\sum_{s_0^{[2]},s_1^{[2]},s_2^{[2]}=\pm 1}S(s_0^{[1]};s_1^{[1]},s_2^{[1]},s_0^{[2]};s_1^{[2]},s_2^{[2]})^2=64, 
\end{equation}
also its norm in the Fourier picture must be constant. We have dropped the factor guarantying normalization of the Fourier transform, however. 

It is easy to confirm that the only non-trivial possibility for first order delta is $\frac{1}{2}(s^{[2]}_i\pm s^{[2]}_{i'})$ with $i\neq i'$. For simplicity let us take $i=0$ and $i'=1$. Any more terms would allow the delta to exceed 1 for a certain choice of signs. Moreover, the norm of such a delta $|\Delta_{s^{[1]}_k}|^2$, that is the sum of squares of all coefficients entering the delta, is $\frac{1}{2}$. Thus from (\ref{consg}) we conclude that two such deltas establish a sign function, one with a $``+"$ sing, the other with $``-"$. The only non-trivial sign function for two qubits reads
\begin{eqnarray}
S(s_0^{[1]},s_1^{[1]};s_0^{[2]},s_1^{[2]})=\frac{1}{2}(s_0^{[1]}(s_0^{[2]}+s_1^{[2]})+s_1^{[1]}(s_0^{[2]}-s_1^{[2]})),
\end{eqnarray}
which represents the CHSH inequality, (\ref{chshinequality}). Thus we conclude that for correlations of two qubits the polytope is bounded by trivial inequalities $\left|\langle A_i^{[1]} A_j^{[2]}\rangle\right|\leq 1$ and CHSH inequalities.

Let us apply the same method for the case of three qubits and three observables per site. Coefficients $g_{ijk}$ satisfy conditions similar to (\ref{consg}) and take values $0,\pm\frac{1}{4},\pm\frac{1}{2},\pm\frac{3}{4},\pm 1$. The last case, of course, corresponds to trivial inequalities and thus it will be skipped in further discussions. The Bell inequalities take a form 
\begin{equation}
\label{wiesniakbell3qubits}
\sum_{ijk}g_{ijk}E_{ijk}\leq 1,
\end{equation}
where $E_{ijk}=\langle A^{[1]}_iA^{[2]}_jA^{[3]}_k\rangle$. 

The following table gives all possible forms of $\Delta_{s_k^{[1]}s_l^{[2]}}$. It also gives a list of other $\Delta_{s_i^{[1]}s_j^{[2]}}$s, a certain one can go with in order to create a valid delta of the first order (obvious repetitions are avoided):
\vspace{5mm}
\begin{center}
\begin{tabular}{|c|c|}
\hline
$\Delta_{s_i^{[1]}s_j^{[2]}}$&goes with:\\
\hline
$\frac{1}{4}(3(-1)^ms^{[3]}_x+(-1)^ns^{[3]}_y)$&$\pm\frac{1}{4}((-1)^ms^{[3]}_x-(-1)^ns^{[3]}_y)$\\
&\\
\hline
$\frac{1}{4}(2(-1)^ms^{[3]}_x$&$\pm\frac{1}{4}((-1)^ms^{[3]}_x-(-1)^ns^{[3]}_y)$\\
$+(-1)^ns^{[3]}_y+(-1)^os^{[3]}_z)$&or\\
& $\pm\frac{1}{4}(2(-1)^ms^{[3]}_x-(-1)^n s^{[3]}y-(-1)^os^{[3]}_z)$ \\
&or\\
&$\pm\frac{1}{4}(-1)^ms^{[3]}_x-(-1)^ns^{[3]}_y)$ \\& and\\& $\pm\frac{1}{4}((-1)^ms^{[3]}_x-(-1)^{n} s^{[3]}_y)$ \\
&\\
\hline
$\frac{1}{2}((-1)^ms^{[3]}_x+(-1)^ns^{[3]}_y)$&alone or with\\ &$\frac{1}{2}((-1)^ms^{[3]}_x-(-1)^ns^{[3]}_y)$\\ 
& or\\
& two $\pm\frac{1}{4}((-1)^ms^{[3]}_x-(-1)^ns^{[3]}_y)$s\\
&\\
\hline
$\frac{1}{2}(-1)^ms^{[3]}_x$&any one of\\& $\pm\frac{1}{2}s^{[3]}_x,\pm\frac{1}{2}s^{[3]}_y,\pm\frac{1}{2}s^{[3]}_z$\\& or\\
& $\pm\frac{1}{4}((-1)^ms^{[3]}_x+(-1)^ns^{[3]}_y)$\\& and\\
&$\pm\frac{1}{4}((-1)^ms^{[3]}_x-(-1)^ns^{[3]}_y)$\\& or\\
& $\pm\frac{1}{4}((-1)^ns^{[3]}_y+(-1)^os^{[3]}_z)$\\& and
\\& $\pm\frac{1}{4}((-1)^ns^{[3]}_y-(-1)^os^{[3]}_z)$\\
&\\
\hline
$\frac{1}{4}((-1)^ms^{[3]}_x+(-1)^ns^{[3]}_y)$&$\pm\frac{1}{4}((-1)^ms^{[3]}_x+(-1)^ns^{[3]}_y)$\\& or\\ 
& $\pm\frac{1}{4}((-1)^ns^{[3]}_y-(-1)^os^{[3]}_z)$\\& and\\ & $\pm\frac{1}{4}((-1)^ms^{[3]}_x+(-1)^os^{[3]}_x)$\\
&\\
\hline
\end{tabular}
\vspace{5mm}
\end{center}
In the table $(x,y,z)$ is a permutation of $(0,1,2)$ and $m,n,o=\pm 1$.

Thus any first order delta belongs (after local transformations, i. e., permutations of observables and sign flips) to one of families listed below (we take $s_0^{[1]}=s_0^{[2]}=s_0^{[3]}=1$ for a time):

\vspace{5mm}
\begin{center}
\begin{tabular}{|c|c|c|}
\hline
&$\Delta_{s_i^{[1]}}$&$|\Delta_{s_i^{[1]}}|^2$\\
\hline
$\Delta_{0}$&$\frac{1}{2}(1+s_1^{[3]}+s^{[2]}_1(1-s^{[3]}_1)$&$\frac{16}{16}$\\
\hline
$\Delta_{I}$&$\frac{1}{4}(-3+s^{[3]}_1+s^{[2]}_1(1+s^{[3]}_1))$&$\frac{12}{16}$\\
\hline
$\Delta_{II}$&$\frac{1}{2}(1+s^{[3]}_1)$&$\frac{8}{16}$\\
\hline
$\Delta_{III}$&$\frac{1}{2}(1+s^{[2]}_1)$&$\frac{8}{16}$\\
\hline
$\Delta_{IV}$&$\frac{1}{2}(1+s^{[2]}_1s^{[3]}_1)$&$\frac{8}{16}$\\
\hline
$\Delta_V$&$\frac{1}{4}(2+s^{[3]}_1+s^{[3]}_2+s^{[2]}_1(s^{[3]}_1-s^{[3]}_2))$&$\frac{8}{16}$\\
\hline
$\Delta_{VI}$&$\frac{1}{4}(1+s^{[2]}_1)(1+s^{[3]}_1)$&$\frac{4}{16}$\\
\hline
$\Delta_{VII}$&$\frac{1}{4}((s^{[3]}_1-s^{[3]}_2)+s^{[2]}_1(1-s^{[3]}_1)+s^{[2]}_2(1-s^{[3]}_2))$&$\frac{6}{16}$\\
\hline
$\Delta_{VIII}$&$\frac{1}{4}(2+s^{[3]}_1+s^{[3]}_2+s^{[2]}_1(1-s^{[3]}_1)+s^{[2]}_2(1-s^{[3]}_2))$&$\frac{10}{16}$\\
\hline
$\Delta_{IX}$&$\frac{1}{4}(2+s^{[3]}_1+s^{[3]}_2+s^{[2]}_1(2-s^{[3]}_1-s^{[3]}_2))$&$\frac{12}{16}$\\
\hline
$\Delta_{X}$&$\frac{1}{4}(2+2s^{[3]}_1+(s^{[2]}_1+s^{[2]}_2)(1-s^{[3]}_1))$&$\frac{12}{16}$\\ 
\hline
$\Delta_{XI}$&$\frac{1}{4}(2+s^{[3]}_1+s^{[3]}_2+s^{[2]}_1(2-s^{[3]}_1-s^{[3]}_2))$&$\frac{12}{16}$\\
\hline
$\Delta_{XII}$&$\frac{1}{4}(2+s^{[2]}_1(s^{[3]}_1+s^{[3]}_2)+s^{[2]}_2(s^{[3]}_1-s^{[3]}_2))$&$\frac{8}{16}$\\
\hline
\end{tabular}
\end{center}
\vspace{5mm}

Now, it is a necessary but not a sufficient condition that such a set of first order deltas enter the sign function, norms of of which add up to 1. Note that $\Delta_0$ is already a sign function, which represents a CHSH-like inequality:
\begin{equation}
\label{wiesniakchsh1}
\langle E_{000}+E_{001}+E_{010}-E_{011}\rangle\leq 2.
\end{equation}

In other cases we need perform such local actions (observable permutations and sign flips) on $\Delta_{s_0^{[1]}}$, $\Delta_{s_1^{[1]}}$, and possibly $\Delta_{s_2^{[1]}}$, that together they constitute a sign function.

As we have mentioned, WWW\.ZB inequalities, i. e. those, which utilize no more than two observables per site, are tight. Thus it should be possible to obtain them from the analysis of the sign function. For example, by using $\Delta_I$ and $\Delta_{IV}$ we can obtain the following sign function:
\begin{eqnarray}
S&=&\frac{1}{4}(-3+s^{[3]}_1+s^{[2]}_1(1+s^{[3]}_1)\nonumber\\
&+&s^{[1]}_1(1+s^{[2]}_1)(1+s^{[3]}_1)),
\end{eqnarray} 
and by putting it into (\ref{wiesniakbell3qubits}) we have 
\begin{eqnarray}
\frac{1}{4}(-3E_{000}+E_{001}+E_{010}+E_{011}&\nonumber\\
+E_{100}+E_{101}+E_{110}+E_{111})&\leq 1.
\end{eqnarray}
Taking two $\Delta_{II}$s or two $\Delta_{III}$ we construct
\begin{eqnarray}
S&=&\frac{1}{2}(1+s^{[3]}_1\nonumber\\
&+&s^{[1]}_1(1-s^{[3]}_1)),\\
\nonumber\\
S&=&\frac{1}{2}(1+s^{[2]}_1\nonumber\\
&+&s_1^{[1]}(1-s^{[2]}_1),
\end{eqnarray}
which lead to CHSH-like inequalities (\ref{wiesniakchsh1}). Another possibility given by these deltas is
\begin{eqnarray}
S&=&\frac{1}{2}(1+s^{[3]}_1\nonumber\\
&+&s_1^{[1]}s^{[2]}_1(1-s^{[3]}_1),
\end{eqnarray}
which implies 
\begin{eqnarray}
\frac{1}{2}(E_{000}+E_{001}&\nonumber\\
+E_{110}-E_{111})&\leq 1.
\end{eqnarray}
A combination of $\Delta_{II}$ and $\Delta_{III}$ leads to one of Mermin inequalities:
\begin{eqnarray}
S=&\frac{1}{2}(1+s^{[2]}_1s^{[3]}_1\nonumber\\
+&s_1^{[1]}(s^{[2]}_1-s^{[3]}_1)),\\
\nonumber\\
&\frac{1}{2}(E_{000}+E_{011}&\nonumber\\
&+E_{101}-E_{110})&\leq 1.
\end{eqnarray}
This closes the set of WWW\.ZB inequalities, at least one of each kind, for three qubits. Now we ready to show Bell inequalities for which at least one observer has a choice of three observables. For example, two $\Delta_V$s can be arranged into
\begin{eqnarray}
S&=&\frac{1}{4}(2+s^{[3]}_1+s^{[3]}_2+s^{[2]}_1(s^{[3]}_1-s^{[3]}_2)\nonumber\\
&+&s^{[1]}_1(2-s^{[3]}_1-s^{[3]}_2+s^{[2]}_1(s^{[3]}_2-s^{[3]}_1))),
\end{eqnarray}
what expresses the inequality
\begin{eqnarray}
\label{wiesniakineq1}
\frac{1}{4}&(2E_{000}+E_{001}+E_{002}+E_{011}-E_{012}&\nonumber\\
+&2E_{100}-E_{101}-E_{102}-E_{111}+E_{112})&\leq 1.
\end{eqnarray}
This sign function can also be a combination of two $\Delta_{XI}$s or $\Delta_{VI}$ and two $\Delta_{IX}$s. Also two $\Delta_{XII}$s can be used to construct
\begin{eqnarray}
S&=&\frac{1}4(2+s^{[2]}_1(s^{[3]}_1+s^{[3]}_2)+s^{[2]}_2(s^{[3]}_1-s^{[3]}_2)\nonumber\\
&+&s^{[1]}_1(2-s^{[2]}_1(s^{[3]}_1+s^{[3]}_2)-s^{[2]}_2(s^{[3]}_1-s^{[3]}_2)),
\end{eqnarray}
which can also be obtained from $\Delta_{III}$ and two $\Delta_{VI}$s. After putting into (\ref{wiesniakbell3qubits}) it gives
\begin{eqnarray}
\label{wiesniakineq2}
\frac{1}{4}&(2E_{000}+E_{011}+E_{012}+E_{021}-E_{022}&\nonumber\\
+&2E_{100}-E_{111}-E_{112}-E_{121}+E_{122})&\leq 1.
\end{eqnarray}
Let us also consider a pair of $\Delta_{VII}$ and $\Delta_{X}$:
\begin{eqnarray}
S&=&\frac{1}{4}(2+s^{[3]}_1+s^{[3]}_2+s^{[2]}_1(1-s^{[3]}_1)+s^{[2]}_2(1-s^{[3]}_2)\nonumber\\
&+&s^1_1((s^{[3]}_1-s^{[3]}_2)+s^{[2]}_1(1-s^{[3]}_1)-s^{[2]}_2(1-s^{[3]}_2))).
\end{eqnarray}
With this sign function, which can also be obtained from $\Delta_V$ and two $\Delta_{VI}$s, we can construct an inequality of a form
\begin{eqnarray}
\label{wiesniakineq3}
\frac{1}{4}&(2E_{000}+E_{001}+E_{002}+E_{010}-E_{011}+E_{020}-E_{022}\nonumber\\
+&E_{101}-E_{102}+E_{110}-E_{111}-E_{120}+E_{122})&\leq 1.
\end{eqnarray}
The last possibility of constructing a sign function is with two $\Delta_{VII}$s and a $\Delta_{V}$:
\begin{eqnarray}
S&=&\frac{1}{4}(1+s_1^{[3]}+s_1^{[2]}(1+s_2^{[3]})+s_2^{[2]}(s_1^{[3]}-s_2^{[3]})\nonumber\\
&+&s_1^{[1]}(1-s_2^{[3]}+s_1^{[2]}(1-s_1^{[3]})-s_2^{[2]}(s_1^{[3]}-s_2^{[3]}))\nonumber\\
&+&s_2^{[1]}(s_1^{[3]}+s_2^{[3]}-s_1^{[2]}(s_1^{[3]}+s_2^{[3]}))),
\end{eqnarray}
which leads to
\begin{eqnarray}
\label{wiesniakineq4}
\frac{1}{4}&(E_{000}+E_{001}+E_{010}+E_{012}+E_{021}-E_{022}&\nonumber\\
+&E_{100}-E_{102}+E_{110}-E_{111}-E_{121}+E_{122}&\nonumber\\
+&E_{201}+E_{202}-E_{211}-E_{222})&\leq 1.
\end{eqnarray}

Interestingly, (\ref{wiesniakineq1}), (\ref{wiesniakineq2}), and (\ref{wiesniakineq3}) are special forms of (\ref{plzb3qa}) derived in \cite{wuzong}. Namely by choosing $A_0^{[3]}=A_1^{[3]}$ we obtain (\ref{wiesniakineq2}) and subsequently (\ref{wiesniakineq1}) with $A_0^{[2]}=A_2^{[2]}$. Finally, we obtain (\ref{wiesniakineq3}) by choosing $A_0^{[2]}=A_2^{[2]}$ and $A_0^{[3]}=A_1^{[3]}$. Further such simplifications lead to WWW\.ZB inequalities.

(\ref{wiesniakineq1}), (\ref{wiesniakineq2}), and (\ref{wiesniakineq3}) are also special cases of the $3\times 3\times 3$ inequality, (\ref{wiesniakineq4}). (\ref{wiesniakineq1}) is obtained by e.g. $A_0^{[1]}=\pm A_2^{[1]}$. With e.g. $A_0^{[1]}=\pm A_1^{[1]}$, we obtain (\ref{wiesniakineq2}). By putting e.g. $A_2^{[1]}=\pm A_3^{[1]}$ we get (\ref{wiesniakineq3}).

It is now interesting to find conditions on states to satisfy these inequalities. This can be, of course, done in the same fashion as for WWW\.ZB and WZLP\.ZB inequalities and basing on the fact explained in the previous paragraph. The sum and the difference of two normalized vectors are a pair of observables are two orthogonal vector. It important to stress that a fixed direction of one of the original vectors does not fully determine the plane spanned by the pair. However, as observers choose fewer observables than in (\ref{plzb3qa}), the conditions presented below are only sufficient, but not necessary. Repeating the argumentation we obtain that (\ref{plzbcond3q}) is a sufficient condition for (\ref{wiesniakineq3}) to be satisfied.

Now, if one of the sign functions $S_1,S_2$ in (\ref{plzb3q}) is factorisable, for example $S_1(k_1,l_1)$ with respect to $(-1)^{k_1}$ and $(-1)^{l_1}$, we get (\ref{wiesniakineq2}). In such a case the condition reads
\begin{equation}
\label{wiesniakcond2}
T_{111}^2+\sum_{i,j=1,2}(T'_{2ij})^2\leq 1, 
\end{equation}
where, again, Bob and Charlie can choose different coordinate systems in both terms.

As for (\ref{wiesniakineq1}), we can take $\vec{a}_0\!^{[1]}+\vec{a}_1\!^{[2]}=2\cos\alpha\vec{e}_1\!^{[3]},\vec{a}_0\!^{[1]}-\vec{a}_1\!^{[1]}=2\sin\alpha\vec{e}_2\!^{[1]},\vec{a}_1\!^{[3]}+\vec{a}_2\!^{[3]}=2\cos\gamma\vec{e}_1\!^{[3]}$ and $\vec{a}_1\!^{[3]}-\vec{a}_2\!^{[3]}=2\sin\gamma\vec{e}_2\!^{[3]}$ ($\vec{e}_1=(1,0,0),\vec{e}_2=(0,1,0)$). After performing necessary calculations similar to ones from Section 3.5 we obtain the first form of the condition:
\begin{equation}
\label{wiesniakcond1a}
(T_{111}^{(c)})^2+T_{211}^2+(T_{212}^{(b)})^2\leq 1.
\end{equation}
Superscripts $(c)$ and $(b)$ denote that in these terms Carlie and Bob, respectively, are allowed to perform arbitrary rotations of coordinate systems in which the second term is expressed. Please note that (\ref{wiesniakcond1a}) contains only 3 elements of the correlation tensor, and thus (\ref{wiesniakineq1}) can be violated at most by a factor $\sqrt{3}$. The three terms can be brought to a common basis, since $\max_{(c)}(T_{111}^{(c)})^2=T_{111}^3+T_{112}^2+T_{113}^2$, and similarly, $\max_{(b)}(T_{212}^{(b)})^2=T_{212}^3+T_{222}^2+T_{232}^2$:
\begin{equation}
\label{wiesniakcond2a}
T_{211}^2+\sum_{i=1}^3(T_{11i}^2+T_{2i2}^2)\leq 1.
\end{equation}

\section{Additional Constraint on Local Realism \cite{w3}}
\hspace*{5mm}

As we have mentioned in Chapter 2 the theories that are excluded by Quantum Mechanics \cite{bell} are based on two main assumptions; {\em Realism}, which states that physical systems posses properties, regardless if already measured, and {\em Locality}, which forbids superluminal effects. Nagata, Laskowski, Wie\'sniak, and \.Zukowski \cite{w3} have considered a case, in which a local realist is allowed to construct his theories only under a condition that the correlation function is a rotationally invariant function of unit vectors defining local measurements, i.e.,
\begin{equation}
\label{rotationinv}
E(\vec{a}^1,\vec{a}^2,...,\vec{a}^N)=\hat{T}\cdot(\vec{a}^1\otimes\vec{a}^2\otimes...\otimes\vec{a}^N).
\end{equation}
This form is independent of particular choices of local coordinate systems, but dependent on mutual relations between the measuring apparata and the physical system

Let us first focus on equatorial observables, i. e., those, vectors of which lie in the $xy$-plane. The correlation function is now a function of angles, which determine positions of the vectors and can be seen as a vector living in the $\Re^N$ space of real square-integrable functions. The measure is then  given by $d\Omega=\prod_{i=1}^Nd\phi_i$, and let the scalar product be 
\begin{equation}
\label{rotinvsp}
(\vec{G},\vec{H})=\int_0^{2\pi}d\phi_1...\int_0^{2\pi}...d\phi_NG(\vec{a}^1,...,\vec{a}^N)H(\vec{a}^1,...,\vec{a}^N).
\end{equation}
We can now utilize facts from Analytical Geometry, especially the one that if for two vectors $\vec{G},\vec{H}$ there exist a vector $\vec{F}$, such that
\begin{equation}
(\vec{F},\vec{G})>(\vec{F},\vec{H}),
\label{rotinvbasis}
\end{equation}
it is certain that $\vec{G}\neq \vec{H}$. Now, a Bell inequality can be constructed by finding such a functional, for which, for a given state, $(\vec{F},\vec{E}_{QM})>(\vec{F},\vec{E}_{LHV})$, where $\vec{E}_{QM}$ and $\vec{E}_{LHV}$ are correlation functions permitted by Quantum Mechanics and Local Realism, respectively. For some state we can take the quantum-mechanical correlation function as $\vec{F}$. Thus we cannot falsify Local Realism if the inequality
\begin{eqnarray}
\label{rotinvineq}
&\int_0^{2\pi}...\int_0^{2\pi}E_{QM}^2(\phi_1,...,\phi_N)d\phi_1...d\phi_N&\nonumber\\
&\leq\int_0^{2\pi}...\int_0^{2\pi}E_{QM}(\phi_1,...,\phi_N)E_{LHV}(\phi_1,...,\phi_N)d\phi_1...d\phi_N&
\end{eqnarray}
is always saturated, independently of the chosen LHV model.


We will now find the necessary and sufficient condition for the violation of the inequality. The quantum-mechanical correlation function reads
\begin{eqnarray}
\vec{E}_{QM}=&
\sum_{i_1,..,i_N=1}^2T_{i_1...i_N}\prod_{k=1}^Nc^{i_k}_k,
\end{eqnarray}
where  $c^{i_k}_k=\cos(\phi_k-(i-1)\pi/2)$. $\left\{\prod_{k=1}^Nc^{i_k}_k\right\}_{i_1,...,i_N=1}^2$ is a set of mutually orthogonal vectors of norm $\pi^{N/2}$. Thus the left-hand side of (\ref{rotinvineq}) is $(\vec{E}_{QM},\vec{E}_{QM})=\pi^N\sum_{a,...n=1}^2T_{a..n}^2$

In Local Realism it is sufficient to consider functions in the form
\begin{equation}
\vec{E}_{LHV}(\phi_1,...\phi_N)=\prod_{i=1}^NI^{[i]}(\phi_i).
\end{equation}
The right-hand side is bounded by a specific number dependent on the correlation tensor, namely,
\begin{eqnarray}
\label{copiedw41}
&(\vec{E}_{QM},\vec{E}_{LHV}=\int_0^{2\pi}d\phi_1...\int_0^{2\pi}d\phi_N E_{LHV}\sum_{i_1,...,i_N=1,2}T_{i_1...i_N}\prod_{k=1}^Nc^{i_k}_k&\nonumber\\
&\leq 4^NT_{max},
\end{eqnarray}
where $T_{max}$ is the maximal possible value of $E_{QM}$ in the equatorial planes,
\begin{equation}
\label{emax}
T_{max}=\max_{\phi_1,...,\phi_N}E_{QM}(\phi_0,...,\phi_N)
\end{equation}
Let us now argue for this bound. Notice that in (\ref{copiedw41}) we deal with a sum of products of the integrals
$\int_{0}^{2\pi}d\phi_iI^{[i]}\cos\phi_{i}$ and
$\int_{0}^{2\pi}d\phi_iI^{[i]}\sin\phi_{i}$ with coefficients $T_{i_1...i_N}$. Thus only a projection of $I^{[i]}$ onto the subspace of normalized vectors $\frac{1}{\sqrt{\pi}}\cos\phi_i$ and $\frac{1}{\sqrt{\pi}}\sin\phi_i$ is relevant. Any normalized function in that subspace can be written as
\begin{equation}
\cos\phi_{0,i}\frac{1}{\sqrt{\pi}}\cos\phi_i+\sin\phi_{0,i}\frac{1}{\sqrt{\pi}}\sin\phi_i=\frac{1}{\sqrt{\pi}}\cos(\phi_i-\phi_{0,i}).
\end{equation}
As $|I^{[i]}(\phi_{i})|=1$, one has
\begin{equation}
\label{imax}
||I^{[i]||}||=\max_{\phi_{0,i}}=\int_0^{2\pi}d\phi_iI^{[i]}\frac{1}{\sqrt{\pi}}\cos(\phi_{0,i}-\phi_{i})\leq 4/\sqrt{\pi}.
\end{equation}
Now, since $\frac{1}{\sqrt{\pi}}\cos\phi_i$ and $\frac{1}{\sqrt{\pi}}\sin\phi_i$ are orthogonal we have
\begin{equation}
\int_0^{2\pi}d\phi_iI^{[i]}\cos\phi_i=\cos\varphi_i||I^{[i]||}||
\end{equation}
and
\begin{equation}
\int_0^{2\pi}d\phi_iI^{[i]}\sin\phi_i=\sin\varphi_i||I^{[i]||}||,
\end{equation}
where $\varphi_i$ is some angle. We now can express $(\vec{E}_{QM},\vec{E}_{LHV})$ as
\begin{equation}
(\vec{E}_{QM},\vec{E}_{LHV})=\pi^{N/2}\left(\prod_{i=1}^N||I^{[i]||}||\right)\sum_{i_1,...,i_N=1}^2T_{i_1...i_N}\prod_{j=1}^Nd_{j}^{i_j},
\end{equation}
where $d^{i_j}_j=\cos(\varphi_j-(i_j-1)\pi/2)$. The last expression can be now bounded with (\ref{emax}) and
(\ref{imax}):
\begin{equation}
(\vec{E}_{QM},\vec{E}_{LHV})\leq 4^NT_{max}.
\end{equation}
Thus the necessary and sufficient condition for a violation of (\ref{rotinvineq}) is

\begin{equation}
\label{rotinvcond}
\left(\frac{\pi}{4}\right)^N\frac{\sum_{a,...,n=1}^2 T_{a..n}^2}{T_{max}}> 1.
\end{equation}

It is now easy to show that the falsification of Local Realism using (\ref{rotinvineq}) by noisy GHZ states (\ref{ghzstate}), $\rho=V|GHZ_N\rangle\langle GHZ_N|+(1-V)\frac{I_{2^N\times 2^N}}{2^N}$ ($0\leq V\leq 1$), is more robust against the admixture of the white noise than for WWW\.ZB inequalities. As we have explained in this Chapter, the latter are satisfied if, and in case of noisy GHZ states only if, $\sum_{a,...,n=1,2}T_{a...n}^2\leq 1$. In $xy$-planes a noisy GHZ state has $2^{N-1}$ non-zero entries of the correlation tensor, all equal to $\pm V$. The sum is equal to $V^22^{N-2}$. From (\ref{rotinvcond2}) and (\ref{zbc2}) we obtain that for $2(2/\pi)^N<V\leq 2^{-(N-1)/2}$ WWW\.ZB inequalities are satisfied, but (\ref{rotinvineq}) is violated. Thus the region of $V$, in which Local Realism is still possible, shrinks exponentially with $N$. This decay is faster for (\ref{rotinvineq}) than in case of MAKB, WWW\.ZB or WZLP\.ZB inequalities.

Such a derivation can generalized to inequalities, which utilize {\em all} possible apparata settings, when we allow $\theta_i\neq\pi/2$. In such a case, if all LHV-based models satisfy
\begin{eqnarray}
\label{rotinvineq2}
&\int_0^\pi\sin\theta_1d\theta_1\int_0^{2\pi}d\phi_1...\int_0^\pi\sin\theta_Nd\theta_N\int_0^{2\pi}d\phi_N&\nonumber\\
&\times E_{QM}^2(\theta_{1},\phi_{1},...,\theta_N,\phi_N)&\nonumber\\
\leq &\int_0^\pi\sin\theta_1d\theta_1\int_0^{2\pi}d\phi_1...\int_0^\pi\sin\theta_Nd\theta_N\int_0^{2\pi}d\phi_N&\nonumber\\
&\times E_{QM}(\theta_{1},\phi_{1},...,\theta_N,\phi_N)E_{LHV}(\theta_{1},\phi_{1},...,\theta_N,\phi_N)&,
\end{eqnarray}
the local realistic description cannot be falsified. Similarly to the previous case of the monoplanar inequalities (\ref{rotinvineq}) ($\theta_i=\pi/2$), we introduce local bases which this time are the spherical harmonics, in particular the three, tensor products of which span the subspace of $\vec{E}_{QM}$ are $\sqrt{\frac{3}{2\pi}}\cos\theta_i,\sqrt{\frac{3}{2\pi}}\sin\theta_i\cos\phi_i,\sqrt{\frac{3}{2\pi}}\sin\theta_i\sin\phi_i$. By an argument similar to the one described above they lead to a necessary and sufficient condition for (\ref{rotinvineq2}) to be violated, similar to (\ref{rotinvineq}):
\begin{equation}
\label{rotinvcond2}
\left(\frac{2}{3}\right)^N\frac{\sum_{a,...,n=1}^3T_{a...n}^2}{T_{max}}>1.
\end{equation}

This type of inequalities was first discussed by \.Zukowski \cite{zukowski93} and Kaszlikowski and \.Zukowski \cite{kaszlikowskizukowski} in papers entitled {\em Bell Theorem Involving all Settings of Measuring Apparatus} and {\em Bell Inequalities Involving All Possible Measurements}, respectively. Indeed, if one strictly follows (\ref{rotinvineq}) or (\ref{rotinvineq2}), observers must perform infinitively many experimental runs, at least having their apparata set in $xy$-planes. However, a correlation function can be reconstructed with $2^N$ (or $3^N$) measurement settings. Such data are enough to for (\ref{rotinvcond}) (or (\ref{rotinvcond2})). 

An important thing to be noticed is that additional constraints on Local Realism lead to a stronger version of the Bell theorem. In \cite{w3} we demand that the correlation function is invariant under all rotations of the Cartesian frame. This assumption leads to Bell inequalities with an exponential advantage over WWW\.ZB and WZLP\.ZB inequalities. This advantage is expressed in terms of the highest maximal violation ratio. One may thus ask what other constrains are helpful in falsifying Local Realism.

A direct application of such inequalities can be more feasible experimentally if the scalar product (\ref{rotinvsp}) is defined not as an integral, but on several measurement settings per observer. Some applications of such inequalities are described in, e.g., \cite{laskowski}. 
\chapter{Critical Parameters in BJSS scheme}
\section{Proposal of Tan, Walls, and Collett}
\hspace*{5mm}

Most of proposed and performed Einstein-Podolsky-Rosen experiment were based on coherent correlations between two (or more) quantum systems. It is worthy to recall that the existence of entanglement emerges directly from the principle of superposition, which in general does not require two separate subsystems. It is thus natural to ask whether coherences of a state of an individual quantum system can be understood as a sing of entanglement in any situation. Can such states violate Bell inequalities in physically feasible experiments? The first difficulty is to ensure that the superposition is realized in spatial degrees of freedom. Local superpositions, like an elliptical polarization of a photon, obviously cannot provide space-like separation. After having this problem solved by an appropriate unitary transformation, we must deal with super-selection rules, which do not allow us to know a relative phase between components of the the states with different numbers of particles. In the most straight-forward attempt of a realization of a Bell-like experiment with a single particle the observers can at most detect the particle in one of the localizations, but they cannot realize any other observables, which do not commute with the first measurement.

\begin{figure}
\centering
\includegraphics[width=4cm]{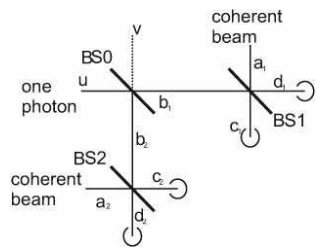}
\caption{The scheme of Tan, Walls, and Collett.}
\label{tanwallscollettpicture}
\end{figure}

One of the first articles concerning this problem is due to Tan, Walls, and Collett \cite{tanwallscollett}. They proposed a setup presented in Figure 4.1. It consists of three balanced beam splitters\footnote{A balanced beam splitter transmits and reflects every photon with equal probabilities, regardless of the polarization.}, BS0, BS1, and BS2, as well as four detectors, which are able to determine number of photons detected within the exposure time. The action of BS0 is described by
\begin{equation}
\left(\begin{array}{c}\hat{b}_1\\ \hat{b}_2\end{array}\right)=\frac{1}{\sqrt{2}}\left(\begin{array}{cc}1&i\\ i&1\end{array}\right)\left(\begin{array}{c}\hat{u}\\ \hat{v}\end{array}\right),
\end{equation}
and respectively for BS1 and BS2:
\begin{equation}
\left(\begin{array}{c}\hat{c}_k\\ \hat{d}_k\end{array}\right)=\frac{1}{\sqrt{2}}\left(\begin{array}{cc}1&i\\ i&1\end{array}\right)\left(\begin{array}{c}\hat{a}_k\\ \hat{b}_k\end{array}\right),
\end{equation}
with $k=1,2$. $\hat{x}$ is an annihilation operator related to a spatial mode $x$. $u$ and $v$ are inputs of BS0, and $b_1, b_2$ are its outputs and also inputs of BS1 and BS2. Their other inputs are $a_1$ and $a_2$. The rest of the symbols denote outputs of BS1 and BS2. Mentioning only the spatial modes we have assumed that all the signals will have the same frequency an polarization. The total transformation realized by the beam splitters can be written as
\begin{equation}
\left(\begin{array}{c}\hat{c}_1\\ \hat{d}_1\\ \hat{c}_2\\ \hat{d}_2\end{array}\right)=\left(\begin{array}{cccc}\frac{1}{\sqrt{2}}&\frac{i}{2}&0&-\frac{1}{2}\\ \frac{i}{\sqrt{2}}&\frac{1}{2}&0&\frac{i}{2}\\ 0&-\frac{1}{2}&\frac{1}{\sqrt{2}}&\frac{i}{2}\\0&\frac{i}{2}&\frac{i}{\sqrt{2}}&\frac{1}{2}\end{array}\right)\left(\begin{array}{c}\hat{a}_1\\ \hat{v}\\ \hat{a}_2\\ \hat{u}\end{array}\right).
\end{equation}
Tan, Walls, and Collett first suggest the coherent states of light,
\begin{equation}
|\alpha e^{i\theta_k}\rangle=e^{-\alpha^2/2}\sum_{n=0}^{\infty}\frac{(\alpha e^{i\theta_k})^n}{\sqrt{n!}}|n\rangle,
\end{equation}
(with $\alpha$ real and $|n\rangle$ being a normalized Fock state of $n$ photons) to be sent to input channels $a_1$ and $a_2$ with the vacuum in channels $u$ an $v$. In such a case the mean number of photons registered by any of the detectors is $\langle \hat{c}^\dagger_k\hat{c}_k\rangle=\langle\hat{d}^\dagger_k\hat{d}_k\rangle=\frac{1}{2}\alpha^2$, and the mean product of the results yielded by an arbitrary pair of detectors equals to $\frac{1}{4}\alpha^4$. As expected, there are no correlations, since the two signals never interfere with each other.

In the next step we introduce a single-photon signal to input $u$. It is easy to argue that the mean number of photons registered by a single detector will be increased by $\frac{1}{4}$, as the single photon has equal chances to propagate to any of the four detectors. Moreover, it interferes at BS1 and BS2 with the coherent beams, which implies the following coincidences of light intensities between pairs of detectors:
\begin{eqnarray}
\langle I_{c_1}I_{c_2}\rangle=\langle I_{d_1}I_{d_2}\rangle\propto\langle\hat{c}_1^\dagger\hat{c}_1\hat{c}_2^\dagger\hat{c}_2\rangle=\langle\hat{d}_1^\dagger\hat{d}_1\hat{d}_2^\dagger\hat{d}_2\rangle=\frac{1}{4}(\alpha^2+\alpha(1+\sin(\theta_1-\theta_2))),\nonumber\\
\\
\langle I_{c_1}I_{d_2}\rangle=\langle I_{d_1}I_{c_2}\rangle\propto\langle\hat{c}_1^\dagger\hat{c}_1\hat{d}_2^\dagger\hat{d}_2\rangle=\langle\hat{d}_1^\dagger\hat{d}_1\hat{c}_2^\dagger\hat{c}_2\rangle=\frac{1}{4}(\alpha^2+\alpha(1-\sin(\theta_1-\theta_2))).\nonumber\\
\end{eqnarray}
The above relations suggest that the phases of the coherent beams can play a role similar to apparatus settings in a two-particle Bell experiments. We can now calculate the interference visibility according to the Michelson formula:
\begin{equation}
\label{visibilitydefinition}
V=\frac{\max \langle I_{c_1}I_{d_2}\rangle-\min\langle I_{c_1}I_{d_2}\rangle}{\max \langle I_{c_1}I_{d_2}\rangle+\min\langle I_{c_1}I_{d_2}\rangle},
\end{equation}
with maxima and minima taken over $\theta_1$ and $\theta_2$. In the case of a single photon as the $u$-input, we get 
\begin{equation}
\label{visibilityquantum}
V_{QM}=\frac{1}{1+\alpha^2}.
\end{equation}

The classical analogue of this experiment is a situation, in which the $u$-input is not a Fock state, but also a coherent state $|\beta\rangle$. Then the coincidences are
\begin{eqnarray}
\langle I_{c_1}I_{c_2}\rangle=\langle I_{d_1}I_{d_2}\rangle\propto\frac{1}{4}(\alpha^4+\alpha^2\beta^2(1+\sin(\theta_1-\theta_2))+\frac{1}{4}\beta^4)\\
\langle I_{c_1}I_{d_2}\rangle=\langle I_{d_1}I_{c_2}\rangle\propto\frac{1}{4}(\alpha^4+\alpha^2\beta^2(1-\sin(\theta_1-\theta_2))+\frac{1}{4}\beta^4).
\end{eqnarray}
These put into (\ref{visibilitydefinition}) give
\begin{equation}
\label{visibilityclassical}
V_{cl}=\frac{\left(\frac{\alpha}{\beta}\right)^2}{\left(\frac{\alpha}{\beta}\right)^4+\left(\frac{\alpha}{\beta}\right)^2+\frac{1}{4}}.
\end{equation}

The relations (\ref{visibilityquantum}) and (\ref{visibilityclassical}) are enough to argue the non-classicality of the first situation. For very low values of $\alpha$ the term $\left(\frac{\alpha}{\beta}\right)^4$ in the coincidences is neglible and thus $V_{QM}$ tends to the unity. On the other hand, in the second case coincidences have terms $\frac{1}{16}\beta^4$ in the denominator, which are fixed with respect to $\alpha$ and thus the maximum of $V_{cl}=\frac{1}{2}$ is achieved for $\alpha^2=\frac{1}{2}\beta^2$. Any value of the visibility above $\frac{1}{2}$ thus implies that the field has no classical description. This is not a question, however, of a possibility of local hidden variables, which would allow to model probabilities of results of measurements, but rather whether can the state be reconstructed using only coherent states, which are here seen as classical.

The analysis is pushed further by defining a correlation function,
\begin{equation}
-1\leq E(\theta_1,\theta_2)=\frac{\langle(\hat{d}_1^\dagger\hat{d}_1-\hat{c}_1^\dagger\hat{c}_1)(\hat{d}_2^\dagger\hat{d}_2-\hat{c}_2^\dagger\hat{c}_2)\rangle}{\langle(\hat{d}_1^\dagger\hat{d}_1+\hat{c}_1^\dagger\hat{c}_1)(\hat{d}_2^\dagger\hat{d}_2+\hat{c}_2^\dagger\hat{c}_2)\rangle}\leq 1, 
\end{equation}
which can be used in (\ref{chshinequality}). After taking into account the coherent states of modes $a_1$ and $a_2$, as well as the vacuum in $v$ we get
\begin{equation}
\label{twccorr}
E(\theta_1,\theta_2)=-\frac{\alpha^2(\langle \hat{u}^\dagger\hat{u}\rangle\sin(\theta_2-\theta_1)+|\langle\hat{u}^2\rangle|\sin(\theta_1+\theta_2-\xi))}{\alpha^4+\langle\hat{u}^\dagger\hat{u}\rangle\alpha^2+\frac{1}{4}\langle\hat{u}^{\dagger 2}\hat{u}^2\rangle},
\end{equation}
where $\langle \hat{u}^2\rangle=|\langle \hat{u}^2\rangle|e^{i\xi}$. When we assume that there is one photon in $u$ the last terms in the numerator and the denominator vanish, and the correlation function reads $E(\theta_1,\theta_2)=\frac{\sin(\theta_1-\theta_2)}{1+\alpha^2}$. The CHSH inequality can be violated if $\alpha\leq\sqrt{\sqrt{2}-1}$. 
If, however, the signal sent to the channel $u$ is a coherent state $|\beta e^{i\xi/2}\rangle$, we have $\langle\hat{u}^\dagger\hat{u}\rangle=|\langle\hat{u}^2\rangle|=\beta^2$ and $\langle\hat{u}^{\dagger 2}\hat{u}^2\rangle=\beta^4$. (\ref{twccorr}) function can be then factorized to $E(\theta_1,\theta_2)=8\alpha^2\beta^2\cos(\theta_1-\xi/2)\sin(\theta_2+\xi/2)/(\alpha^2+2\beta^2)^2$ and hence no violation of a Bell inequality is possible.
\section{Scheme of Bj\"ork, Jonsson, and S\'anchez-Soto} 
\hspace*{5mm}

The idea of  Tan, Walls, and Collett started a discussion on the non-classicality of a single photon \cite{hardy,czachor,peressinglephoton,gerry}. 
An important variant of the original proposal is due to Bj\"ork, Jonsson and S\'anchez-Soto (BJSS) \cite{bjorkjonssonsanchezsoto}. In their scheme one is interested in probabilities of certain events, rather than mean values. This allows to use the CH inequality (\ref{chinequality}), which turns out to be inequivalent to the CHSH inequality (\ref{chshinequality}).

\begin{figure}
\label{bjorkfigure}
\centering
\includegraphics[width=7cm]{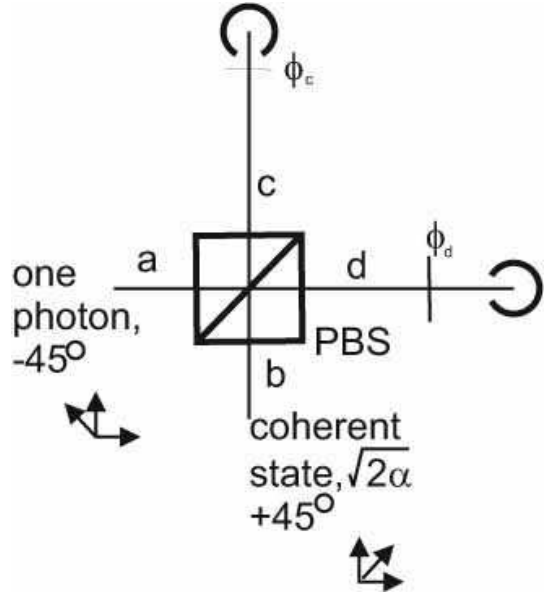}
\caption{The scheme of Bj\"ork, Jonsson, and S\'anchez-Soto.}
\end{figure}

Let us start with a description of the experimental setup. In its center we have a polarizing beam splitter (PBS), the inputs of which are denoted by $a, b$ and outputs by $c,d$. It reflects horizontally polarized ($H$) light and transmits vertically polarized ($V$). The single photon, sent to input $a$, has a $-45^\circ$ polarization. Its state is thus $(\hat{a}_H^\dagger-\hat{a}_V^\dagger)|\Omega\rangle/\sqrt{2}$ in front of PBS and $(\hat{c}_H^\dagger-\hat{d}_V^\dagger)|\Omega\rangle/\sqrt{2}$. The photon is superposed in two modes, which are distinguishable spatially, as well as with respect to the polarization. The other input is used to inject the coherent beam from the local oscillator with a $+45^\circ$ polarization, and the mean photon number $2\alpha^2$. Behind PBS the state of all four modes can be written as 
\begin{equation}
|\psi\rangle=\left(
e^{i\omega\tau_d}|0,\alpha e^{i\omega\tau_c},1,\alpha e^{i\omega\tau_d}\rangle-
e^{i\omega\tau_c}|1,\alpha e^{i\omega\tau_c},0,\alpha e^{i\omega\tau_d}\rangle\right)/\sqrt{2}
\end{equation}
with mode ordering $c_H,c_V,d_V,d_H$. In the single photon modes we have used the Fock formalism, whereas the coherent state formalism is applied to the other two modes. The complex phases appeared due to a propagation between PBS and the measuring devices for times $\tau_c$ and $\tau_d$, but they can be neglected without any loss of generality.

Now, the observers shall try to detect their modes in one of the states 
\begin{equation}
|+,n_k,\phi_k\rangle=\frac{1}{\sqrt{1+\frac{n_k}{\alpha^2}}}\left(\sqrt{\frac{n_k}{\alpha^2}}|0,n_k\rangle+e^{i\phi_k}|1,n_k-1\rangle\right).
\end{equation}
It is worthy noting that the local vacuum in both modes, $|0,0\rangle$, is not covered by these states. The probability of having such a state at one side is $e^{-\alpha^2}/2$. This is one of main reasons to have $\alpha$ large, in order to minimize the occurrence of 0-photon states.

We shall begin with demonstrating that in the perfectly realized experiment we can violate (\ref{chinequality}). A local projection onto $|+,n_k,\phi_k\rangle\langle +,n_k,\phi_k|$ occurs with a probability
\begin{equation}
P_+(n_k,\phi_k)=\frac{1}{1+\frac{n_k}{\alpha^2}}e^{-\alpha^2}\frac{(\alpha)^{(n_k-1)}}{(n_k-1)!}.
\end{equation}
These probabilities sum up to $P_+(\phi_k)=\sum_{n_k=1}^{\infty}P_{+}(n_k,\phi_k)$, and an interferometric relation
\begin{equation}
\label{bjorkinterference}
P_{++}(\phi_c,\phi_d)=P_+(\phi_c)P_+(\phi_d)(1-\cos(\phi_c-\phi_d))
\end{equation}
holds.

We need to argue in what way these probabilities can be added. $P_+(\phi_k)$ contains a function of a random variable, $n_k-1$, and a Poisson probability describing the distribution of this variable. The only parameter of this distribution is $\alpha^2$, equal to the mean value, the variance, and all higher cumulants of $n_k-1$. We can easily find these quantities by treating $\alpha^2$ as a thermodynamical parameter and taking the partition function $Z=e^{\alpha^2}$:
\begin{eqnarray}
\label{bjorkmean}
&\langle n_k-1\rangle=\frac{\partial\log Z}{\partial(\alpha^2)}=\alpha^2,&\\
\label{bjorkvariance}
&\langle (n_k-1)^2\rangle-\langle (n_k-1)\rangle^2=\frac{\partial^2 \log Z}{\partial(\alpha^2)^2}=\alpha^2,&\\
\label{bjorkcumul}
&\langle (n_k-1)^3\rangle-3\langle n_k-1\rangle\langle (n_k-1)^2\rangle+2\langle n_k-1\rangle^3=\frac{\partial^3\log Z}{\partial(\alpha^2)^3}=\alpha^2,&\\
&...&\nonumber
\end{eqnarray}
Let us first consider the variance. If $\alpha^2$ is much larger than the unity, it can be neglected against $\alpha^4$ appearing in the two terms on the left side of (\ref{bjorkvariance}), $\langle n_k^2\rangle$ and $\langle n_k\rangle^2$ (hereafter, we take $n_k\approx n_k-1$ appearing in averages). It is thus reasonable to assume $\langle n_k^2\rangle\approx\langle n_k\rangle^2$. Subsequently, we can draw a similar argument for the third cumulant and show that $\langle n_k^3\rangle\approx \langle n_k\rangle^3$, etc.. Thus for any sufficiently smooth function we can assume $\langle f(n_k)\rangle\approx f(\langle n_k\rangle)$. In particular, let us introduce approximations useful in further consideration:
\begin{equation}
\label{approx1}
\sum_{n_k=1}^\infty\frac{e^{-\alpha^2 x}}{1+\frac{n_k}{\alpha^2}}\frac{(\alpha^2x)^{n_k-1}}{(n_k-1)!}\approx \frac{1}{1+x}
\end{equation}
\begin{equation}
\label{approx2}
\sum_{n_k=1}^\infty\frac{e^{-\alpha^2 x}}{1+\frac{n_k}{\alpha^2}}\frac{n_k(\alpha^2x)^{n_k}}{\alpha^2n_k!}\approx \frac{x}{1+x}
\end{equation}
\begin{equation}
\label{approx3}
\sum_{n_k=1}^\infty\frac{e^{-\alpha^2 x}}{1+\frac{n_k}{\alpha^2}}\frac{n_k(\alpha^2x)^{n_k-1}}{\alpha^2(n_k-1)!}\approx \frac{x}{1+x}
\end{equation}
\begin{equation}
\label{approx4}
\sum_{n_k=1}^\infty\frac{e^{-\alpha^2 x}}{1+\frac{n_k}{\alpha^2}}\frac{(\alpha^2x)^{n_k}}{n_k!}\approx \frac{1}{1+x}
\end{equation}
\begin{figure}
\centering
\subfigure{\includegraphics[width=6cm]{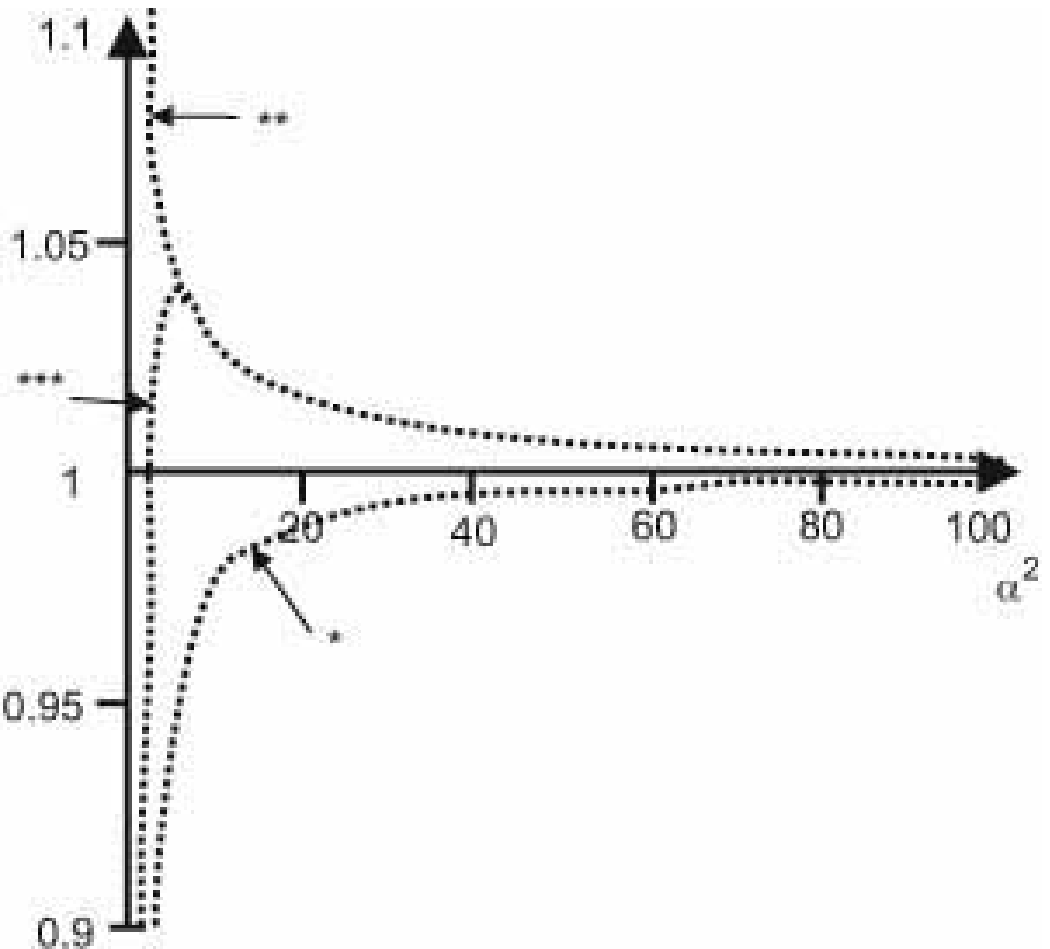}}
\subfigure{\includegraphics[width=6cm]{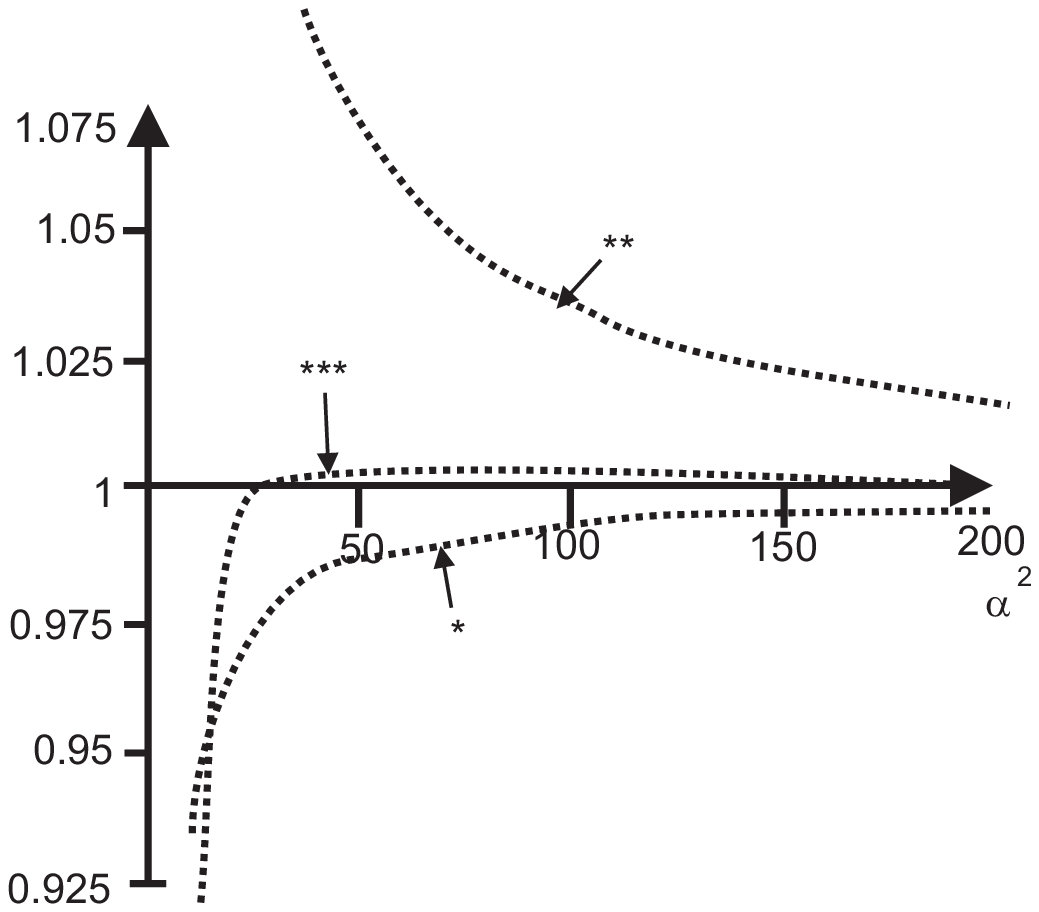}}
\caption{Accuracies of approximations (\ref{approx1}),(\ref{approx2}),(\ref{approx3}),(\ref{approx4}) for $x=1$ (left) and $x=0.2$ (right). * stands for $(1+x)\sum_{n=1}^{1000}\frac{e^{-\alpha^2x}}{1+\frac{n}{\alpha^2}}\frac{(\alpha^2x)^{n-1}}{(n-1)!}$, ** represents $\frac{1+x}{x}\sum_{n=1}^{1000}\frac{e^{-\alpha^2x}}{(1+\frac{n}{\alpha^2})}\frac{n(\alpha^2x)^{n-1}}{\alpha^2(n-1)!}$, and *** reads $(1+x)\sum_{n=1}^{1000}\frac{e^{-\alpha^2x}}{1+\frac{n}{\alpha^2}}\frac{(\alpha^2x)^n}{n!}$.}
\end{figure}
The role of $x$ will be clarified below. Strictly speaking, in (\ref{approx1}), (\ref{approx2}), (\ref{approx3}), and (\ref{approx4}) we demand that $\alpha^2x$, rather than $\alpha^2$, is much larger than 1. We are interested in the range $0\leq x\leq 1$. For the perfect case of $x=1$ the local probabilities obviously add up to $\frac{1}{2}$, and due to the relation (\ref{bjorkinterference}) the inequality (\ref{chinequality}) can be maximally violated, if the chosen phases are $\phi_c=0,\phi'_c=\pi/2,\phi_d=\pi/4,\phi'_d=-\pi/4$. The expression equals $\frac{1}{2}(\sqrt{2}-1)>0$.
\section{Some Possible Imperfections \cite{w7}}
\hspace*{5mm}

Let us now reproduce the results from \cite{w7}. Having shown that the experiment might be considered a valid test against Local Realism, we can introduce some imperfections to our considerations. The first one is the coherence loss. However, since the coherent state is a superposition of infinitively many Fock states, for simplicity we make a conjecture that only the single photon part is affected:
\begin{eqnarray}
&\frac{1}{2}(|0,\alpha,1,\alpha\rangle-|1,\alpha,0,\alpha\rangle)(\langle 0,\alpha,1,\alpha|-\langle 1,\alpha,0,\alpha|)&\nonumber\\
\rightarrow&\frac{1}{2}(|0,\alpha,1,\alpha\rangle\langle 0,\alpha,1,\alpha|+|1,\alpha,0,\alpha\rangle\langle 1,\alpha,0,\alpha|)&\nonumber\\
&-\frac{l}{2}(|0,\alpha,1,\alpha\rangle\langle 1,\alpha,0,\alpha|+|1,\alpha,0,\alpha\rangle\langle 0,\alpha,1,\alpha|).&
\end{eqnarray}
We call $0\leq l\leq 1$ the decoherence parameter. Such an action can be justified by classicality (in some sense) of a coherent beam, and strong non-classicality of a single photon. 

Another imperfection we may introduce is the loss of photons during the propagation. Let us assume that the probability that a photon will reach the measuring devices is $\eta$. This effect can be modeled by putting beam splitters with transmittivities $\eta$ in channels $c$ and $d$ and tracing out the reflected modes. Was it possible to design a device realizing projections $|+,n_k,\phi_k\rangle\langle +,n_k,\phi_k|$ with only one photosensitive element, the transparency of the channel would be equivalent to the detection efficiency of this photosensor, provided that the propagation channels are perfectly transparent. The single photon state is stochastically transformed into the vacuum in respective modes, whereas the amplitude of the coherent part is decreased by a factor $\sqrt{\eta}$. The whole transition due to both imperfections reads
\begin{eqnarray}
\label{bjorkstateleta}
 &\frac{1}{2}(|0,\alpha,1,\alpha\rangle-|1,\alpha,0,\alpha\rangle)(\langle 0,\alpha,1,\alpha|-\langle 1,\alpha,0,\alpha|)\rightarrow\rho(l,\eta)&\nonumber\\
=&(1-\eta)|0,\alpha\sqrt{\eta},0,\alpha\sqrt{\eta}\rangle\langle 0,\alpha\sqrt{\eta},0,\alpha\sqrt{\eta}|&\nonumber\\
+&\frac{\eta}{2}(|1,\alpha\sqrt{\eta},0,\alpha\sqrt{\eta}\rangle\langle 1,\alpha\sqrt{\eta},0\alpha\sqrt{\eta}|+|0,\alpha\sqrt{\eta},1,\alpha\sqrt{\eta}\rangle\langle 0,\alpha\sqrt{\eta},1,\alpha\sqrt{\eta}|)&\nonumber\\
-&\frac{l\eta}{2}(|1,\alpha\sqrt{\eta},0,\alpha\sqrt{\eta}\rangle\langle 0,\alpha\sqrt{\eta},1,\alpha\sqrt{\eta}|+|0,\alpha\sqrt{\eta},1,\alpha\sqrt{\eta}\rangle\langle 1,\alpha\sqrt{\eta},0,\alpha\sqrt{\eta}|).&\nonumber\\
\end{eqnarray}
The first line of the right-hand side is the term in which the single photon is lost, the second are diagonal elemenst of the density matrix with single photon not lost, the last are off-diagonal elements.

Using formulae (\ref{approx1}) to the vacuum part and (\ref{approx1}),(\ref{approx2}) to the other part, we obtain
\begin{eqnarray}
&P_+(\phi_k)\approx\frac{\eta(3-\eta)}{2(1+\eta)},&\\
&P_{++}(\phi_c,\phi_d)\approx\left(\frac{\eta}{1+\eta}\right)^2(2-\eta-l\cos(\phi_c-\phi_d)).&
\end{eqnarray} 
We now choose the optimal angles to obtain $-\cos(\phi_c-\phi_d)-\cos(\phi'_c-\phi_d)-\cos(\phi_c-\phi'_d)+\cos(\phi_c-\phi_d)=2\sqrt{2}$. To compute the dependence between the critical decoherence parameter and the critical channel transparency, above which Local Realism can be falsified, we set the left-hand side of (\ref{chinequality}) equal to 0 and put in the computed probabilities. The obtained equation,
\begin{equation}
\frac{-\eta^3_{CRIT}+2\eta^2_{CRIT}(1+l_{CRIT}\sqrt{2})-3\eta_{CRIT}}{(1+\eta_{CRIT})^2}=0,
\end{equation}
apart from a trivial solution $\eta_{CRIT}=0$ is satisfied for
\begin{equation}
\label{bjorkcritpp}
l_{CRIT}=\frac{3-2\eta_{CRIT}+\eta^2_{CRIT}}{2\sqrt{2}\eta_{CRIT}}.
\end{equation}

Equation (\ref{bjorkcritpp}) implies that if coherence is perfectly preserved, the required channel transparency is $\eta_{CRIT}=1+\sqrt{2}-2^{3/4}\approx 73.4\%$. This about 0.096 less than the required efficiency in two-photon Bell-type experiments with the maximally entangled state.

The next step of the analysis of the BJSS scheme presented in \cite{w7} is using the CHSH inequality (\ref{chshinequality}). For this purpose we need to define a correlation function, which, depending on local phases, would take values between $-1$ and $+1$. We can naively do it by associating the states $|+,n_k,\phi_k\rangle$ proposed by Bj\"ork, Jonsson, and S\'anchez-Soto with outcomes ``$+1$" of local observables, whereas the states $|-,n_k,\phi_k\rangle=\frac{1}{1+\frac{n_k}{\alpha^2}}\left(|0,n_k\rangle-e^{i\phi_k}\sqrt{\frac{n_k}{\alpha}}|1,n_k-1\rangle\right)$ will correspond to outcomes ``$-1$". Then the correlation function is taken as the mean value of the product of local outcomes, and after applying approximations (\ref{approx1}), (\ref{approx2}),( \ref{approx3}), and (\ref{approx4}) we get
\begin{equation}
\label{bjorke}
E(\phi_c,\phi_d)=-\eta\frac{3-5\eta+2\eta^2+4\eta l\cos(\phi_c-\phi_d)}{(1+\eta)^2}.
\end{equation}
After putting (\ref{bjorke}) into (\ref{chshinequality}) and optimizing over the angles we obtain the necessary condition for the violation,
\begin{equation}
\label{bjorkcritchsh}
l>-\frac{2\eta^3-6\eta^2+\eta-1}{4\sqrt{2}\eta^2},
\end{equation}
which implies that the inequality can be violated only for $\eta\geq 63.2\%$, significantly less than the value obtained from (\ref{bjorkcritpp}). Both results are correct according to the approximation. However, together they suggest that the analysis is not complete. The CHSH inequality is based on the correlation function, whereas in the CH inequality we utilize local and global probabilities. This fact reveals a greater generality of the latter. Thus if there exist a region of an advantage of a correlation function-based expression over a certain probability-based one, there must be a set of CH inequalities, in which at least one is violated in the region of the violation of the CHSH inequality.

The CH inequality, the violation of which turns out to be more robust against the signal attenuation of the propagation channels than in the case shown above, is constructed with the states $|+,n_k,\phi_k\rangle$ being associated with the first events at each side, which enter (\ref{chinequality}) and $|-,n_k,\phi_k\rangle$ with the second events. In such a case, the CH inequality reads
\begin{equation}
\label{chineqbjork}
P_{++}(\phi_c,\phi_d)+P_{+-}(\phi_c,\phi'_d)+P_{-+}(\phi'_c,\phi_d)-P_{--}(\phi'_c,\phi'_d)-P_+(\phi_c)-P_+(\phi_d)\leq 0,
\end{equation}
and probabilities entering (\ref{chineqbjork}), that have not yet been computed, are
\begin{eqnarray}
&P_{--}(\phi_c,\phi_d)&\nonumber\\&=\sum_{n_c,n_d=1}^{\infty}\langle -,n_c,\phi_c|^{[c]}\langle -,n_d,\phi_d|^{[d]}\rho(\eta,l)|-,n_c,\phi_c\rangle^{[c]}|-,n_d,\phi_d\rangle^{[d]}&\nonumber\\
&\approx\left(\frac{\eta}{1+\eta}\right)^2(1-\cos(\phi_c-\phi_d)),&\\
\nonumber\\
&P_{+-}(\phi_c,\phi_d)=P_{-+}(\phi_c,\phi_d)&\nonumber\\&=\sum_{n_c,n_d=1}^{\infty}\langle +,n_c,\phi_c|^{[c]}\langle -,n_d,\phi_d|^{[d]}\rho(\eta,l)|+,n_c,\phi_c\rangle^{[c]}|-,n_d,\phi_d\rangle^{[d]}&\nonumber\\&=\sum_{n_c,n_d=1}^{\infty}\langle -,n_c,\phi_c|^{[c]}\langle +,n_d,\phi_d|^{[d]}\rho(\eta,l)|-,n_c,\phi_c\rangle^{[c]}|+,n_d,\phi_d\rangle^{[d]}&\nonumber\\
&\approx\frac{\eta}{2(1+\eta)^2}(3-2\eta+\eta^2+2l\eta\cos(\phi_c-\phi_d))&.
\end{eqnarray}
Altogether, after optimizing over the apparata settings, (\ref{chinequality}) simplifies to
\begin{equation}
\label{bjorkcritpm}
l>\frac{3-\eta}{2\sqrt{2}}.
\end{equation}
In such a case, the critical channel transparency with the perfect coherence preservation is $3-2\sqrt{2}\approx 17.2\%$.

In \cite{w7} we have thus shown that a Bell test inspired by the original idea of Tan, Walls, and Collett \cite{tanwallscollett} provides a much lower threshold on the channel transparency than the critical value of the similar parameter, the detection efficiency in two-qubit Bell experiments. A possible explanation is that the non-classicality of the state in the first case is brought in by the single photon. Coherent states are seen as quantum realizations of classically allowed states of the field. They are eigenstates of annihilation operators, which corresponds to multiplying by the intensity of the field in classical or semi-classical Electrodynamics. The single photon state is purely quantum, as the notion of a photon does not exist in classical physics. Another argument to justify this robustness is the following. When we loose the photon, we admix the vacuum state in the modes referring to it. The density matrix reduced to the modes of the single photon, in the basis $\{|0,0\rangle,|0,1\rangle,|1,0\rangle,|1,1\rangle\}$, experiences the following transition:
\begin{equation}
\left(\begin{array}{cccc}0&0&0&0\\ 0&\frac{1}{2}&-\frac{1}{2}&0\\0&-\frac{1}{2}&\frac{1}{2}&0\\0&0&0&0\end{array}\right)\rightarrow\left(\begin{array}{cccc}1-\eta&0&0&0\\0&\frac{1}{2}\eta&-\frac{1}{2}\eta l&0\\0&-\frac{1}{2}\eta l&\frac{1}{2}\eta&0\\0&0&0&0\end{array}\right).
\end{equation}
This matrix has one negative eigenvalue ofter the partial transposition, $\frac{1}{2}(1-\eta-\sqrt{(1-\eta)^2+l^2\eta^2})$, for any non-zero value of the product of $l$ and $\eta$ and thus basing on the Peres-Horodecki criterion \cite{peresppt,horodeckippt} we conclude that, except for trivial cases of $\eta=0$ and $l=0$, the state is always entangled. We thus expect that this entanglement can be revealed by Bell inequalities in a large range of the parameters. 

All the inequalities in the BBJS scheme are never violated if $l\leq \frac{1}{\sqrt{2}}$. This greatly resembles the admixture of the white noise (see Section 2.5). However, in this case, rather than the maximally mixed state, we admix a classically correlated mixture $\frac{1}{2}\left(|01\rangle\langle 01|+|10\rangle\langle 01|\right)$.  These correlations have no importance in the bases the observers perform measurements. 

One can also consider the case, in which the observers are able to predict the opaqueness of the channels and set the initial intensity of the coherent beam to be $2\alpha^2/\eta$. Then the vacuum state would be still admixed with a weight $1-\eta$, but the approximated sums (\ref{approx1}), (\ref{approx2}), (\ref{approx3}), and (\ref{approx4}) are equal to $\frac{1}{2}$. Thus the probabilities we use in the inequalities are equal to
\begin{eqnarray}
&P_{+}(\phi_k)=P_-(\phi_k)=\frac{1}{2},&\\
&P_{++}(\phi_c,\phi_d)=P_{--}(\phi_c,\phi_d)=\frac{1}{4}(1-l\eta\cos(\phi_c-\phi_d)),&\\
&P_{+-}(\phi_c,\phi_d)=P_{-+}(\phi_c,\phi_d)=\frac{1}{4}(1-l\eta\cos(\phi_c-\phi_d)),&
\end{eqnarray}
which trivially gives a condition $l_{CRIT}\eta_{CRIT}=1/\sqrt{2}$. The photon loss has now the same effect as the decoherence. The inequalities cannot be violated for $\eta\leq 1/\sqrt{2}$.

\begin{figure}
\centering
\includegraphics[width=5cm]{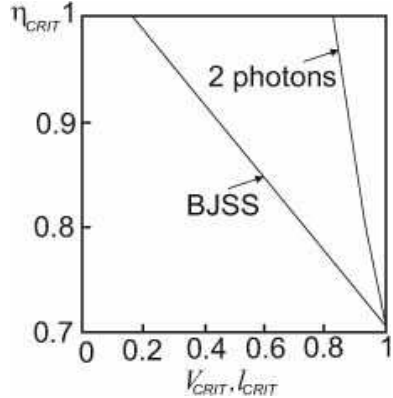}
\caption{$\eta_{CRIT}$ versus $V_{CRIT}$ for the two-photon experiment with a maximally entangled state, and versus $l_{CRIT}$ in the scheme of Bj\"ork, Jonsson, and S\'anchez-Soto. The Clauser-Horne inequality is violated in the upper right region of the plot.}
\end{figure}

\section{Experiment of Hessmo {\em et al.}}
\hspace*{5mm}

At the end of the discussion of possible non-classicality of Fock states we will consider an experiment performed by Hessmo {\em et al.} \cite{hessmo}. The setup consists of a non-polarizing beam splitter, which receives a single photon and a coherent beam of a small intensity as input signals. The polarizations of the two signals are mutually orthogonal, and the state of the field behind the beam splitter can be written similarly to the one in the proposal of Bj\"ork, Jonsson, and S\'anchez-Soto, in \cite{hessmo} it differs only by a sign between the components. Each measuring device is built of a birefringent wave-plate to control the relative phases $\phi_c$ and $\phi_d$ between the single photon and the coherent beam polarizations. Behind every wave-plate there is a polarizing beam splitter, which transmits the single photon with a probability $t^2$ and reflects it with a probability $r^2$ ($t$ and $r$ are taken real), thus the transmittivity and the reflectivity for the coherent beam are $r^2$ and $t^2$, respectively. In the experiment $t^2\approx\cos^22^\circ$. At the transmitted outputs the observers had single-photon  detectors, whereas the reflected outputs were left unobserved.

\begin{figure}
\centering
\includegraphics[width=5cm]{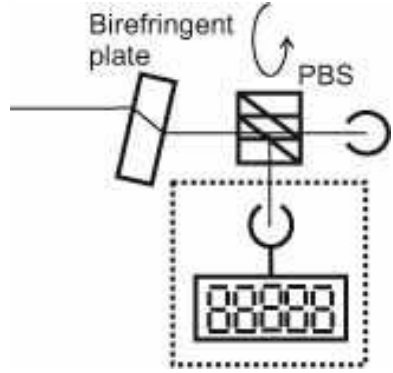}
\caption{A scheme of a measuring device in the Hessmo {\em et al.} experiment. If we add a photon counter in a reflected mode (in a dashed box) and replace detector $D$ with one able to detect {\em exactly} one photon, we would be able to perform a BJSS experiment.}
\end{figure}

The probabilities of the detections are
\begin{eqnarray}
P_c=P_d=\frac{1}{2}e^{-\alpha^2r^2}(1+r^2+ \alpha^2 r^2 t^2),
\end{eqnarray}
the probability that the both detectors register a photon reads
\begin{equation}
P_{cd}=e^{-2\alpha^2r^2}\left(r^2+r^2t^2(1+cos(\phi_c-\phi_d))\right),
\end{equation}
which allows to calculate the coincidence probability as
\begin{eqnarray}
&P_{coinc}=1-P_c-P_d+P_{cd}&\nonumber\\&=1-e^{-\alpha^2r^2}(1+r^2+\alpha^2r^2t^2)&\nonumber\\
&+e^{-2\alpha^2r^2}\left(r^2+2r^2t^2(1+cos(\phi_c-\phi_d))\right).&
\end{eqnarray}
Moreover, Hessmo {\em et al.} consider also the cases when the single photon does not reach any of the detectors and the two detections are caused by two photons from the coherent beam. The total probability of the coincidence is $P_{coinc}^{total}=\eta P_{coinc}+(1-\eta)(1-e^{-\alpha^2r^2})^2$. One can minimize the effect of false coincidences by choosing $r^2<<t^2$ and $\alpha^2r^2<<1$. The second-order interference visibility found in \cite{hessmo} is $(66\pm 2)\%$ and $(91\pm 3)\%$ after the background correlation correction. As shown by Tan, Walls and Collett, both results are enough to agrue for the non-classicality of a singe photon. Moreover, after the background correlation correction the second-order interference visibility is high enough for violation of the Bell inequalities.

Obviously, the Hessmo {\em et al.} experiment is not equivalent to the the one proposed in \cite{bjorkjonssonsanchezsoto}, The non-classicality is demonstrated not with violation of the CH inequality, but only with high values of the second-order interference visibility. On the other hand, to conduct the BJSS experiment one needs to enrich the measuring devices with detectors, which are able to reveal exact numbers of photons detected in reflected modes. The other necessary modification is to replace  detectors that detect {\em at least} one photon with such that detect exactly one. As we show below, such detectors can be used to realize the projections utilized in the BJSS scheme under the condition that the second signal apart from the coherent beam contains at most one photon.  Behind the PBS the detected state is 
\begin{equation}
|det,n_k\rangle=\frac{1}{\sqrt{(n_k-1)!}}\hat{e}^{\dagger}_k(\hat{f}^\dagger_k)^{n_k-1}|0,0\rangle,
\end{equation}
 where $e_k$, $f_k$ are spatial modes of the PBS outputs. Since the signal enters each polarizing beam splitter trough only one input channel, the polarizations in spatial modes in outputs are already fixed. Now, let us denote the modes of the single photon and the coherent beam at each side as $k_{SP}$ and $k_{CB}$, respectively. Then the action of the PBS is
\begin{eqnarray}
\left(\begin{array}{c}\hat{e}^\dagger_k\\ \hat{f}^\dagger_k\end{array}\right)=\left(\begin{array}{cc}t&r\\
-r& t\end{array}\right)\left(\begin{array}{c}\hat{k}^\dagger_{SP}\\ \hat{k}^\dagger_{CB}\end{array}\right), 
\end{eqnarray}
so that the detectors are sensitive to states $\frac{1}{\sqrt{(n_k-1)!}}(t\hat{k}_{SP}^\dagger+r\hat{k}_{CB}^\dagger)(-r\hat{k}_{SP}+t\hat{k}^\dagger_{CB})^{n_k-1}|0,0\rangle$. Finally, we include the birefringent plate and the states detected by the devices are
\begin{equation}
|det_k,n_k,\phi_k\rangle\propto(e^{i\phi_k}t\hat{k}_{SP}^\dagger+r\hat{k}_{CB})(-e^{i\phi_k}\sqrt{r}\hat{k}^\dagger_{SF}+\sqrt{t}\hat{k}_{CB}^\dagger)^{n_k-1}|0,0\rangle.
\end{equation}
When we plug in our assumption of having at most one photon apart from the coherent beam, we get
\begin{equation}
|det_k,n_k,\phi_k\rangle\propto -rt^{(n_k-1)}\sqrt{n_k}|0,n_k\rangle+e^{i\phi_k}(t^{n_k}-r^2t^{n_k-2})|1,n_k-1\rangle,
\end{equation}
where the notation from the last section was adopted. 
The properly chosen position of the PBS in the measuring device for each $n_k$ individually would allow to conduct the BJSS experiment.

\chapter{Entanglement in Bulk Systems}
\section{Introduction}
\hspace*{5mm}
In previous Chapters we have studied entanglement between only a few quantum subsystems. We have assumed that the state of the system is known and can be arbitrarily manipulated.

However, the presented methods of detecting entanglement do not seem to appropriate for macroscopic systems. For ensembles of the order of $10^{23}$ spins one may not refer to individual properties of every constituent of a solid sample or correlations between all of them. Instead, one should base one's arguments for non-classical correlations in the system on a statistical description of the whole ensemble. 

This is possible when the system is in a thermal equilibrium at some temperature $T$. Given the Hamiltonian $H$ describing the system, the state is  $\rho=\exp(-H/\kappa T)/Z$, where $\kappa$ is the Boltzmann's constant, taken equal to 1 for all numerical purposes in this dissertation. The collective thermodynamical quantities are computed by differentiating the logarithm of the partition function $Z=\textbf{Tr}\exp(-H/\kappa T)$ with respect to external parameters, like temperature or the magnitude of the magnetic field $B$. For example, the internal energy is equal to $U=-\frac{\partial \log Z}{\partial(1/\kappa T)}$, the heat capacity $C=\frac{\partial U}{\partial T}$, and the magnetic susceptibility $\chi=\frac{1}{\kappa T}\frac{\partial^2 \log Z}{\partial B^2}$. In this Chapter we will show that under some conditions these quantities are entanglement witnesses \cite{witness}, that is their low-temperature values can reveal correlations of purely quantum nature.

The problem of the detection of entanglement in a thermal equilibrium is an appealing challenge of the Quantum Information Theory. First, it allows to see quantum correlations as a natural feature of quantum systems at low temperatures. Second, thermal entanglement present in a large scale in solids might be important for future Quantum Computation Technology.

\section{Concept of Entanglement Witness}
\hspace*{5mm}

Before we give a definition of an entanglement witness, let us recall in details the Peres-Horodecki criterion \cite{peresppt,horodeckippt} mentioned in the Sections 3.5 and 4.3. It states that under a partial transposition separable states always remain physical, while entangled ones can afterwards have a negative eigenvalue. For non-entangled states the statement is easily proved. The transposition changes any state $\rho$ into another density operator. Thus, for any two states $\rho^{[1]}$ and $\rho^{[2]}$, $\rho^{[1]}\otimes(\rho^{[2]})^T$ must also be positive. To finish the proof we notice that also a statistical mixture, $\sum_i P_i\rho_i^{[1]}\otimes(\rho_i^{[2]})^T$, is also a non-negative operator of trace 1, provided that $P_i\geq 0$ and $\sum_iP_i=1$.

To show that the positivity of a density operator under a partial transposition does is not preserved in general, it is enough to consider an entangled state of a system of dimensionality $2\times 2$ or $2\times 3$. For these systems the positivity of a partially transposed state was shown to be a {\em necessary} and {\em sufficient} condition for its separability \cite{3horodecki99}. Let us, for example, consider the case of a two-qubit singlet state:
\begin{equation}
\left(\begin{array}{cccc}0&0&0&0\\0&\frac{1}{2}&-\frac{1}{2}&0\\ 0&-\frac{1}{2}&\frac{1}{2}&0\\0&0&0&0 \end{array}\right)^{T_1}=\left(\begin{array}{cccc}0&0&0&-\frac{1}{2}\\0&\frac{1}{2}&0&0\\ 0&0&\frac{1}{2}&0\\-\frac{1}{2}&0&0&0 \end{array}\right),
\end{equation}
The $T_1$ index stands for the partial transposition with respect to the first qubit. The same effect is obtained by transposing with respect to the second subsystem. The partially transposed matrix has eigenvalues $\{\frac{1}{2},\frac{1}{2},\frac{1}{2},-\frac{1}{2}\}$. However, there are known states, which, despite of their entanglement, have a positive partial transpose \cite{wernerwolf}.

The partial transposition is an example of a linear, positive, but not completely positive map\footnote{A positive map transforms one non-negative operator into another. A completely positive map $\epsilon$ is such that $\openone\otimes\epsilon$ is positive, regardless of the dimensionality of the first Hilbert space.}. Such an operation, mapping a state from Hilbert space $\mathcal{H}^{[2]}$ onto $\mathcal{H}^{[3]}$ can be generally written as
\begin{equation}
\label{linearmap}
\epsilon(\rho^{[12]})^{[13]}=\sum_{k_1,k_2,l_1,l_2}W_{k_1,l_1,k_2,l_2}|l_1\rangle^{[3]}\langle k_1|^{[2]}\rho^{[12]}|k_2\rangle^{[2]}\langle l_2|^{[3]}.
\end{equation}
From this relation we see that with every linear map we can associate a self-adjoint matrix $W_{k_1,k_2,l_1,l_2}=\langle k_1l_1|^{[23]}W^{[23]}|k_2l_2\rangle^{[23]}$. In a short form the action of the map on the second subsystem can be written as
\begin{equation}
\epsilon(\rho^{[12]})^{[13]}=\textbf{Tr}_2(\rho^{[12]T_2}W^{[23]}).
\end{equation}
If we assume that dimensionalities of $\mathcal{H}^{[2]}$ and $\mathcal{H}^{[3]}$ are equal, the reverse relation can be also obtained with the maximally entangled state $|\psi^+\rangle^{[22']}=\frac{1}{\sqrt{D_2}}\sum_{i=1}^{D_2}|i\rangle^{[2]}|i\rangle^{[2']}$,where $D_2$ is the dimensionality of $\mathcal{H}^{[2]}$. Then
\begin{equation}
(\openone^{[2]}\otimes\epsilon^{[2'3]})(|\psi^+\rangle^{[22']}\langle\psi^+|^{[22']})=W^{[23]}.
\end{equation}
We stress that here $\openone^{[2]}$ denotes the trivial map on $\mathcal{H}^{[2]}$.

{\em Proof:}
\begin{eqnarray}
&(\openone^{[1]}\otimes\epsilon)(|\psi^+\rangle^{[22']}\langle\psi^+|^{[22']})=(\openone^{[2]}\otimes\epsilon)\left(\frac{1}{D_2}\sum_{i,i'=1}^{D_2}|i\rangle^{[2]}|i\rangle^{[2']}\langle i'|^{[2]}\langle i'|^{[2']}\right)&\nonumber\\
&=\frac{1}{D_2}\sum_{i,i'=1}^{D_2}|i\rangle^{[2]}\langle i'|^{[2]}&\nonumber\\
&\otimes\left(\sum_{k_1,l_1,k_2,l_2=1}^{D_2}\langle k_1l_1|^{[2'3]}W^{[2'3]}|k_2l_2\rangle^{[2'3]}(\langle i'|^{[2']}|k_1l_1\rangle^{[2'3]})(\langle k_2l_2|^{[2'3]}|i\rangle^{[2']})\right)&\nonumber\\
&=\frac{1}{D_2}\sum_{i,i=1}^{D_2}|i\rangle^{[2]}\langle i'|^{[2]}\otimes\sum_{l_1,l_2=1}^{D_2}\langle i l_1|^{[2'3]}W^{[2'3]}|i'l_2\rangle^{[2'3]}|l_1\rangle^{[3]}\langle l_2|^{[3]}&\nonumber\\
&=\frac{1}{D_2}\left(\sum_{i,l_1=1}^{D_2}|il_1\rangle^{[23]}\langle il_1|^{[23]}\right)W^{[23]}\left(\sum_{i',l_2=1}^{D_2}|i'l_2\rangle^{[23]}\langle i'l_2|^{[23]}\right)=\frac{1}{D_2}W^{[23]}.&
\end{eqnarray}
QED.

The mutual relation between the linear maps and the operators is known as the Jamio{\l}kowski isomorphism \cite{jamiolkowski}. It was also shown that
\begin{itemize}
\item{if $\epsilon$ is a completely positive map, matrix $W$ (upper indices of $W$ are hereafter dropped) is a positive operator,}
\item{if $\epsilon$ is a positive, but not completely positive, map, $W$ is an entanglement witness.}
\end{itemize}
In other words, in the second case the mean value of $W$ in any separable state is 0 or positive, but there exists an entangled state, for which the expected value of $W$ is negative. Of course, the threshold, above or below which entanglement is demonstrated, can be arbitrarily redefined, as in case of the \cite{chsh} operator. The Horodecki Family has shown \cite{3horodecki99} that any type of entanglement, bound \cite{boundentanglement}, distillable \cite{distillation}, or directly violating a Bell inequality \cite{bell}, can be detected with a proper positive map, thus by a proper entanglement witness.
\section{Internal Energy as Entanglement Witness}
\hspace*{5mm}

One of the first significant papers concerning entanglement in solid state models is due to Wang and Zanardi \cite{pla301}, who have considered nearest-neighbor entanglement in a spin-$\frac{1}{2}$ Heisenberg ring. The Hamiltonian of this system is given by
\begin{equation}
\label{wzhamiltonian}
H_{xxx}=J\left(\sum_{i=1}^N\vec{\sigma}\:^{[i]}\cdot\vec{\sigma}\:^{[i+1]}\right).
\end{equation} 
Periodic boundary conditions are guarantied by an identification $N+1\equiv 1$. $J$ is a coupling constant, positive for the antiferromagnetic case (AFM), and negative for ferromagnetic (FM) systems. The two symmetries of the system, the invariance under an arbitrary collective rotation of Cartesian frames of all qubits and a cyclic permutation of them require that the state of any two neighboring spins, $\rho^{[12]}=\textbf{Tr}_{3....N}\rho$, is
\begin{equation}
\label{wzstate}
\rho^{[12]}=\left(\begin{array}{cccc}u_1&0&0&0\\0&u_2&z^*&0\\0&z&u_3&0\\0&0&0&u_4\end{array}\right).
\end{equation} 
The first noticeable fact is that due to isotropy of the system the sample is not magnetized, i.e., the Bloch vector of any qubit is vanishing, $\langle \vec{\sigma}\;^{[1]}\rangle=\langle \vec{\sigma}\;^{[2]}\rangle=\vec{0}$. Then the entries of the density matrix are found as
\begin{eqnarray}
u_1=u_4=\frac{1}{4}(1+T_{33}),\nonumber\\
u_2=u_3=\frac{1}{4}(1-T_{33}),\nonumber\\
z=\frac{1}{4}(T_{11}+T_{22}+iT_{12}-iT_{21}),
\end{eqnarray}
where $T_{ii'}$s are elements of the correlation tensor (see Section 2.4). The second observation to be made is that, again, because of the rotational symmetry, $T_{12}=T_{21}=0$, thus $z$ is real.

Wang and Zanardi have then computed the two-qubit entanglement measure introduced by Wootters \cite{concurrence} and known as the concurrence. It is defined in a following way: let $\rho^{[12]*}$ be a complex conjugate of a two-qubit state $\rho^{[12]}$. Define a matrix $R=\sqrt{\rho^{[12]}\sigma^{[1]}_2\sigma^{[2]}_2\rho^{[12]*}\sigma^{[1]}_2\sigma^{[2]}_2}$ with non-negative eigenvalues $\lambda_1,\lambda_2,\lambda_3,\lambda_4$ in the decreasing order. The concurrence is defined as $Conc=\max\{0,\lambda_1-\lambda_2-\lambda_3-\lambda_4\}$. In this specific case $Conc=\frac{1}2\max\{0,|T_{11}+T_{22}|-T_{33}-1\}$. Notice that $T_{11}=T_{22}=T_{33}=U/3NJ$. The internal energy increases with temperature and is equal 0 in the limit of $T\rightarrow\infty$, as the thermal state is then the maximally mixed state, and the Hamiltonian is traceless. Therefore we conclude that the internal energy is always negative. Hence we have $Conc=\frac{1}{2}\max\left\{0,\left(-\frac{U}{NJ}-1\right)\right\}$ for AFM and $\max\left\{0,\frac{1}{2}\left(\frac{U}{3NJ}-1\right)\right\}$ for FM. In particular, the nearest-neighbor concurrence at $T\rightarrow 0$ is determined by the ground-state energy $E_0$ per spin. Hult\'en \cite{hulten} found it numerically to be $E_0/NJ\approx -1.773$ for $J>0$ and $N\rightarrow\infty$, whereas for FM systems it is known to be $-1$. In the latter case it we do not observe thermal entanglement at any temperature, whereas for the antiferromagnetic case the criterion of Wang and Zanardi fails above the critical temperature $T_c$ at which $U(T_c)/NJ=-1$. This temperature would be different for different $N$. The only case, in which an AFM ring reveals no thermal entanglement, is $N=3$.

Wang and Zanardi have linked the internal energy with entanglement using the concurrence. A similar, though more direct argument has been drawn by Brukner and Vedral \cite{qph0406040}. They have considered  xx and xxx Heisenberg rings described by a Hamiltonian
\begin{equation}
H_{xxz}(B)=J\left(\sum_{i=1}^N\sigma_1^{[i]}\sigma_1^{[i+1]}+\sigma_2^{[i]}\sigma_2^{[i+1]}+\frac{J_3}{J}\sigma_3^{[i]}\sigma_3^{[i+1]}\right)+B\underbrace{\sum_{i=1}^N\sigma_3^{[i]}}_{=M_3},
\end{equation}
with $B$ being the external magnetic field\footnote{We assume that $B$ contains all the physical constants. For experimental purposes, we would need  to include a of product of the Bohr magnetron $\mu_B$ and the gyromagnetic factor $g$. Then it is also necessary to remember about the Planck constant being a part of the spin operator.} and $J_3=0$ for an xx ring, and $J_3=J$ for an xxx model. Brukner and Vedral have defined the internal energy and the magnetization per site, $\bar{U}=U/N, \bar{M}_3=\langle M_3\rangle/N=\kappa T\frac{\partial \ln Z}{N\partial B}$. They argue that in both cases the quantity
$\frac{\bar{U}+B\bar{M}_3}{J}$ is an entanglement witness. 

Let us first focus on the case of $J_3=J$. Then the explicit form of the Brukner-Vedral reads
\begin{equation}
\label{bvwitness}
\frac{\bar{U}+B\bar{M}_3}{J}=\left\langle\frac{1}{N}\sum_{i=1}^N\vec{\sigma}\;^{[i]}\cdot\vec{\sigma}\;^{[i+1]}\right\rangle.
\end{equation}
The proof of the witnessing properties was first presented by T\'oth, G\"uhne, and Cirac \cite{tothposter}. Notice that for product states, $\rho=\bigotimes_{i=1}^N \rho^{[i]}$, each element of the sum can be bounded by
\begin{eqnarray}
\label{bvbound}
&\left|\frac{1}N\sum_{i=1}^N\langle\vec{\sigma}\:^{[i]}\cdot\vec{\sigma}\:^{[i+1]}\rangle_{PROD}\right|&\nonumber\\
=&\left|\frac{1}{N}\sum_{i=1}^N\langle\vec{\sigma}\:^{[i]}\rangle_{PROD}\cdot\langle\cdot\vec{\sigma}\:^{[i+1]}\rangle_{PROD}\right|&\nonumber\\
\leq&\frac{1}{N}\sum_{i=1}^N\sqrt{\langle\vec{\sigma}\:^{[i]}\rangle_{PROD}^2\langle\vec{\sigma}\:^{[i+1]}\rangle_{PROD}^2}&\leq 1,
\end{eqnarray} 
where the first inequality comes from the Cauchy inequality. By convexity, the same bound holds for separable states.

Thus any value of (\ref{bvwitness}), the modulo of which exceeds 1, immediately implies thermal entanglement. As we have said, the lowest possible value in the thermodynamical limit was found to be about $-1.773$. In fact, for the xxx models without the magnetic field the result of Brukner and Vedral fully agree with the calculation of the concurrence for nearest neighbors by Wang and Zanardi. 

The argument can be extended to more general cases of $J_3\neq J$ and $B$ not equal to zero. In particular, Brukner and Vedral consider an infinite xx ring, the analitycal solution for which was presented by Katsura \cite{katsura}. In order to present his results we introduce a short-hand notation $L=\frac{B}{\kappa T}$ and $K=\frac{J}{2\kappa T}$, and a function
\begin{equation}
f(K,L,\omega)=\sqrt{2K^2+2K^2\cos 2\omega-4KL\cos\omega+L^2}.
\end{equation}
Now, the internal energy and the magnetization are found to be 
\begin{eqnarray}
\bar{U}=\frac{2\kappa T}{\pi}\int_0^\pi f(K,L,\omega)\tanh f(K,L,\omega)d\omega,\\
\langle\bar{M}_3\rangle=-\frac{1}\pi\int_0^\pi\frac{L-2K\cos\omega}{f(K,L,\omega)}\tanh f(K,:,\omega)d\omega.
\end{eqnarray}
These two integrals allow to determine a region of $\frac{B}{J}$ and $\frac{\kappa T}{J}$ (for $J>0$) in which thermal entanglement is revealed by (\ref{bvwitness}). The region is presented in Figure 5.1.

\begin{figure}
\label{bvfigure}
\centering
\includegraphics[width=10cm]{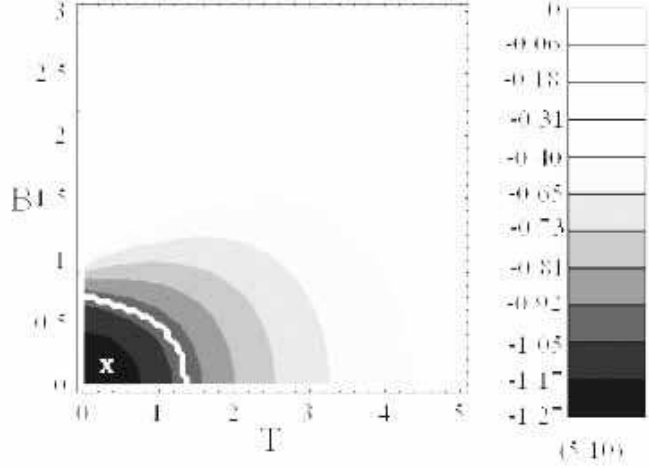}
\caption{(\ref{bvwitness}) for an infinte xx spin$-\frac{1}{2}$ ring (for $J>0$) as a function of the magnetic field and temperature. Only in the region marked with a white ,,x" the values of (\ref{bvwitness}) reveal entanglement in the thermal state.}
\end{figure}

Other examples of using the internal energy as an entanglement witness can be found
in e.g. \cite{dowling,wu,tothinternalenergy}.

\section{Detecting Multipartite Entanglement with Internal Energy}
\hspace*{5mm}

A bound for presence of multipartite entanglement in state of a Heisenberg chain was presented in \cite{guehnetothbriegel}. The Authors have notices that the Hamiltonian of an xxx open chain with an odd number of sites can be written as
\begin{equation}
\label{gtbhamiltonian}
H_{xxx}=J\sum_{i=1}^{(N-1)/2}W_{2i-1,2i,2i+1},
\end{equation}
with $W_{i,j,k}=\vec{\sigma}\!^{[i]}\cdot\vec{\sigma}\;^{[j]}+\vec{\sigma}\;^{[j]}\cdot\vec{\sigma}\;^{[k]}$.  

First let us show that genuine tripartite entanglement must exist if
\begin{equation}
\label{gtbbound1}
\frac{\bar{U}}{J}<-\frac{1+\sqrt{5}}{2}\approx -1.618.
\end{equation}
{\em Proof:} The bipartite entanglement bound for $H_{xxx}$ can be found by summing over such bound for three-qubit hermitian operators $W_{2l-1,2l,2l+1}$, even if the do not mutually commute. We are interested in mean values possible only for states with at least tripartite entanglement, we allow two qubits to be in an entangled state.  However, $W_{2l-1,2l,2l+1}$ is not sensitive to correlations between non-nearest neighbours, hence we shall assume that the state is pure and such, that $|\psi^{[2l-1\: 2l\: 2l+1]}\rangle=|\psi^{[2l-1]}\rangle|\psi^{[2k+1\: 2l+2]}\rangle$. In the language of the correlation tensor this means that $T_{ijk}=T_{i00}T_{0jk}$ for any $i=1,2,3$ and $j,k=0,1,2,3$. Furthermore, we have $\sum_{i=1}^3T^2_{i00}=1$. Such a form of the state implies
\begin{eqnarray}
|\langle W_{123}\rangle|=|T_{100}T_{010}+T_{200}T_{020}+T_{300}T_{030}+T_{011}+T_{022}+T_{033}|.
\end{eqnarray}
In the next step we choose $(T_{100},T_{200},T_{300})$ to be parallel to $(T_{010},T_{020},T_{030})$ and $|\psi^{[23]}\rangle$ to be Schmidt-decomposed, $|\psi^{[23]}\rangle=\cos\alpha|10\rangle-\sin\alpha|01\rangle$. The non-vanishing elements of the correlation tensor are $T_{03}=-T_{30}=\cos 2\alpha$, $T_{11}=T_{22}=-\sin 2\alpha$, and $T_{33}=-1$. $|\langle W_{123} \rangle|$ simplifies to
\begin{equation}
|\langle W_{123}\rangle|=|\cos 2\alpha|+|1+2\sin 2\alpha|,
\end{equation}
which optimized over $\alpha$ gives $\sqrt{5}+1$. The Hamiltonian consists of $(N-1)/2$ $W_{ijk}$ elements. Since we are interested in the limit of large $N$, the bound obtained above is divided by 2. 

It is easy to see, however, that the bound (\ref{gtbbound1}) is not optimal. The argument is not self-consistent, as we first assume, say, the third qubit to be entangled with the second, then we want it to be in a pure state.

To find a better threshold, the Authors assume that the state of the ring of $N=4M$ qubits is bi-factorisable, $|\psi\rangle=|\psi^{[12]}\rangle|\psi^{[34]}\rangle...|\psi^{[N-1\!N]}\rangle$, and define two $6N$-dimensional vectors
\begin{eqnarray}
\label{gtbvector1}
\vec{v}_1=([1],[1:2],[2],[0],[5],[5:6],[6],...,[0]),\\
\label{gtbvector2}
\vec{v}_2=([N],[0],[3],[3:4],[4],[0],[7],....,[N-1:N]).
\end{eqnarray}
Here $[i]=(\langle\sigma_1^{[i]}\rangle,\langle\sigma_2^{[i]}\rangle,\langle\sigma_3^{[i]}\rangle)$, $[i:j]=(\langle\sigma_1^{[i]}\sigma_1^{[j]}\rangle,\langle\sigma_2^{[i]}\sigma_2^{[j]}\rangle,\langle\sigma_3^{[i]}\sigma_3^{[j]}\rangle)$ and $[0]=(1,1,1)$. Then the mean value of the Hamiltonian is given by $\langle H_{xxx}\rangle=\frac{1}{2}J\vec{v}_1\cdot\vec{v}_2$. It is important to notice that for any two-qubit state $\sum_{i=1}^3(T^2_{i0}+T^2_{0i}+T^2_{ii})\leq 3$. Hence the norms  (\ref{gtbvector1}) and (\ref{gtbvector2}) are bounded by $2|\langle H_{xxx}\rangle|\leq J\sqrt{\vec{v}_1\:^2},J\sqrt{\vec{v}_2\:^2}\leq J\sqrt{3N}$. Applying the Cauchy inequality we get that if
\begin{equation}
\label{gtbbound2}
\bar{U}<-\frac{3}{2}J,
\end{equation}
the state must consist of genuine three-partite entanglement.

In a similar fashion G\"uhne, T\'oth, and Briegel argue for the bound on the energy for bi-factorisable states in case of the xx ring. 

\section{Non-linear Entanglement Witnesses}
\hspace*{5mm}

In most of considerations presented up to this point we have discussed detecting entanglement with linear functions of the state, mean values of some operators. However, unlike the set of statistics allowed by LHV, the set of separable state is not a hull spanned between a number of vertices. It is rather a convex object with infinitively many extreme points (factorisable states). Thus we expect that its curved boundary would be better approximated by a non-linear function of the state than by a hyperplane. In other words, an optimal non-linear witness could  detect entanglement in more states than a linear operator, which bounds the set of separable states (see Figure 5.2). A non-linear criterion of entanglement can be drawn for the gedankenexperiment of Einstein, Podolsky, and Rosen \cite{epr}. In that situation it can be suprising that for an entangled state uncertainties of a difference of positions of two particles and a sum of their momenta can simultaneously vanish. For product states one has $\Delta^2(p^{[1]}+p^{[2]})=\Delta^2(p^{[1]})+\Delta^2(p^{[2]})$ and $\Delta^2(q^{[1]}-q^{[2]})=\Delta^2(q^{[1]})+\Delta^2(q^{[2]})$. Now, since $\Delta^2(q^{[i]})\Delta^2(p^{[i]})\geq\frac{\hbar^2}{4}$, it is obvious that the both uncertainties cannot be equal to 0 at the same time. More explicitly this argument was drawn for continuous systems by Reid \cite{pra40913} and Duan {\em et al.} \cite{prl842722}. 
\begin{figure}
\label{nonlinear}
\centering
\includegraphics[width=5cm]{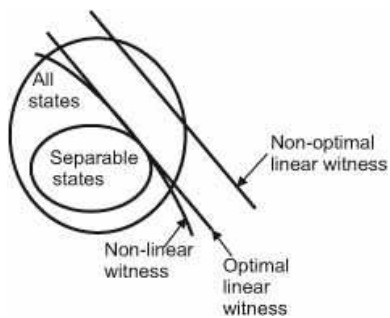}
\caption{A non-optimal linear witness defines a hyperplane far from the convex set of separable states and thus can detect only strong entanglement. A hyperplane of an optimal witness has common point with the boundary of the set, but a non-linear witness better describes the curvature of of the boundary.}
\end{figure}

We, however, want to focus on a version of this criterion formulated for discrete systems by Hofmann and Takeuchi \cite{hofmanntakeuchi}. Let us consider a spin of arbitrary magnitude $l$ and the variances of its three components, first in a pure state $|\psi\rangle$ ($\hbar=1)$:
\begin{eqnarray}
\label{htlocal}
&\Delta^2(S_1)+\Delta^2(S_2)+\Delta^2(S_3)&\nonumber\\&=\langle\psi|S_1^2+S_2^2+S_3^2|\psi\rangle-\langle\psi|S_1|\psi\rangle^2-\langle\psi|S_2|\psi\rangle^2-\langle\psi|S_3|\psi\rangle^2&\nonumber\\
&\geq l(l+1)-l^2=l.&
\end{eqnarray}

It is also necessary to show that a mixedness of a state $\rho=\sum_i P_i|\psi_i\rangle\langle\psi_i|$ can only increase the variance of any Hermitian operator $A$:
\begin{eqnarray}
\label{htmixed}
&\Delta^2(A)=\sum_{i}P_i\langle(A-\langle A\rangle)^2\rangle_i\nonumber\\
&=\sum_iP_i\left(\underbrace{\langle A^2\rangle_i-\langle A\rangle^2_i}_{=\Delta^2(A)_i}+\underbrace{(\langle A\rangle_i-\langle A\rangle)^2}_{\geq 0}\right)\nonumber\\
&\geq \sum_i P_i\Delta^2(A)_i,
\end{eqnarray}
with $\langle A\rangle_i=\langle\psi_i|A|\psi_i\rangle$. Hence, if we have two local spin operators $\vec{S}^{[1]}$ and $\vec{S}^{[2]}$ (the two spins do not need to have the same magnitude, let here us assume, however, that $l^{[1]}=l^{[2]}=l$), for separable states
\begin{equation}
\label{htcondition}
\Delta^2(\vec{S}^{[1]}+\vec{S}^{[2]})\geq 2l
\end{equation}
holds. On the other hand, one can have the two spins in a singlet state $|S=l\rangle$, for which
\begin{equation}
\Delta^2(\vec{S}^{[1]}+\vec{S}^{[2]})=\langle (\vec{S}^{[1]}+\vec{S}^{[2]})^2\rangle-\langle \vec{S}^{[1]}+\vec{S}^{[2]}\rangle^2=0.
\end{equation}
The argument can be straight-forward generalized for more spins. This will be the main theme of the next Section.	
\section{Magnetic Susceptibility as Entanglement Witness \cite{w4}}
\hspace*{5mm}

The criterion of Hofmann and Takeuchi can be verified for a wide class of large spin lattices with a thermodynamic quantity, i. e., the magnetic susceptibility. The use of this function of the state as a witness was first demonstrated for a specific material, copper nitrate \cite{bruknervedralzeil}, and then shown in its full generality by Wie\'sniak, Vedral, and Brukner \cite{w4}. The arguments from \cite{bruknervedralzeil} will be recalled in the next Section.

The appropriate systems to study their magnetic susceptibility from the theoretical point of view are those, Hamiltonians of which are invariant under arbitrary collective rotations. Let the system be described in the zero magnetic field by Hamiltonian $H_0$. In presence of the external field $\vec{B}$ it becomes $H=H_0+\vec{B}\cdot\vec{M}=H_0+\vec{B}\cdot\sum_{i=1}^N\vec{S}^{[i]}$. Let us consider the magnetization in the direction of the field $\vec{B}$, $\langle M_{\vec{B}}\rangle=-\kappa T\frac{\partial \ln Z}{\partial |\vec{B}|}$. Using the Trotter identity, 
\begin{equation}
\label{wiesniakapprox}
e^{A+B}=\lim_{n\rightarrow\infty}\left(e^{\frac{A}{n}}e^{\frac{B}{n}}\right)^n,
\end{equation}
which is true even if $[A,B]\neq 0$, and the fact that operators can be cyclically permuted under the trace, $\textbf{Tr} ABC=\textbf{Tr} BCA$, we see that independently of the Hamiltonian the magnetization along any direction pointed by a unit vector $\vec{n}$ is $\langle M_{\vec{n}}\rangle=\textbf{Tr}\rho\:\vec{n}\cdot\sum_{i=1}^N\vec{S}\:^{[i]}$. However, from the same two facts we see that the magnetic susceptibility, $\chi_{\vec{n}}=\frac{\partial^2 \log Z}{\partial B^2_{\vec{n}}}$, is in general equal to
\begin{equation}
\label{wiesniakintegrate}
\chi_{\vec{n}}=\frac{1}{\kappa T}\left(\int_0^1\textbf{Tr}(M_{\vec{n}}\rho^x M_{\vec{n}}\rho^{1-x}) dx-\langle M_{\vec{n}}\rangle^2\right),
\end{equation}
where $B_{\vec{N}}$ is the magnitude of the magnetic field projection onto the unit vector $\vec{n}$. Only in cases, in which the Hamiltonian, and thus the thermal state commutes with the magnetization operator, the magnetic susceptibility in a direction of vector $\vec{n}$ equals to $\chi_{\vec{n}}=\frac{1}{\kappa T}\Delta^2(M_{\vec{n}})$. Since we need zero-field magnetic susceptibilities with respect to three orthogonal directions to apply the Hofmann-Takeuchi criterion, the Hamiltonian must commute with the magnetization operators in all orthogonal directions. The interaction must be isotropic and no magnetic field should act on any spin. On the other hand, we do not have any other requirements about the lattice. It is irrelevant what are magnitudes of spins seated in sites of the lattice, what is its shape, and what are exact values of coupling constants. The lattice could be one-, two- or three-dimensional, with or without non-nearest-neighbor interactions, or inhomogeneous. In any case we argue that
\begin{equation}
\label{wvbwitness1}
\left.\chi'\right|_{\vec{B}=\vec{0}}=\left.\chi_1\right|_{\vec{B}=\vec{0}}+\left.\chi_2\right|_{\vec{B}=\vec{0}}+\left.\chi_3\right|_{\vec{B}=0}<\frac{\sum_{i=1}^{N}l^{[i]}}{\kappa T}
\end{equation}
cannot be explained without entanglement. The values are summed over three orthogonal directions. When all spins in the lattice have a common magnitude $l$, (\ref{wvbwitness1}) simplifies to
\begin{equation}
\label{wvbwitness2}
\left.\chi'\right|_{\vec{B}=\vec{0}}< \frac{Nl}{\kappa T},
\end{equation}
and due to the rotational symmetry it is sufficient to consider only one direction:
\begin{equation}
\label{wvbwitness3}
\left.\chi_{\vec{n}}\right|_{\vec{B}=0}<\frac{\sum_{i=1}^{N}l^{[i]}}{3\kappa T}.
\end{equation}
In particular, the zero-field magnetic susceptibility at $T\rightarrow 0$ must tend to be infinite if one does not allow the concept of entanglement. Experimental studies of various materials give evidences, however, that this not the case, that is also zero-field magnetic susceptibilities of ferromagnets tend to 0 with temperature. This might be due to a fact that for these materials at certain temperature we deal with a significant symmetry breaking, either spontaneous, or due to weak magnetic fields in the environment, e.g. the geomagnetic field or fields originating from the measuring apparata. In presence of a distinguished direction, our criterion cannot be applied.

Let us assume that $\forall_i l^{[i]}=l$. In \cite{w4} we have also argued that the magnetization and its variance, which in specific situation can be associated with the magnetic susceptibility, together satisfy a certain complementarity relation:
\begin{equation}
\label{wvbcomplementarity}
\underbrace{1-\frac{\Delta^2(\vec{M})}{Nl}}_{\textbf{Non-local properties}}+\underbrace{\frac{\langle\vec{M}\rangle^2}{N^2 l^2}}_{\textbf{Local properties}}\leq 1.
\end{equation}

{\em Proof:} we start with showing that
\begin{equation}
\label{wiesniaklemma1}
\langle\vec{M}^2\rangle\geq\left(\frac{Nl+1}{Nl}\right)\langle \vec{M}\rangle^2.
\end{equation}
For this purpose we will work in the total angular momentum basis $\{|L,m,i\rangle\}_{L,m,i}$, where $\vec{M}^2|L,m,i\rangle=L(L+1)|L,m,i\rangle$, the magnetization in the $z$-direction $M_3|L,m,i\rangle=m|L,m,i\rangle$, and index $i$ is due to possible degeneracies. Since both sides of (\ref{wiesniaklemma1}) are rotationally invariant, we can choose the $z$-axis as the direction of the magnetization, $\langle M_1\rangle=\langle M_2\rangle=0, \langle\vec{M}\rangle^2=\langle M_3\rangle^2$. Let us now define define operator $\hat{K}$, such that $\hat{K}|J,m,i\rangle=J|J,m,i\rangle$. Given probabilities $P_{J,m}=\sum_iP_{J,m,i}$ to find the system with the total angular momentum $J$ and the magnetization $m$ we notice that $\langle M\rangle=\sum_{J,m}P_{J,m}m\leq\sum_{J,m}P_{J,m}J=\leq\langle \hat{K}\rangle$. The proof is being completed by facts that $\langle \vec{M}\!^2\rangle=\langle \hat{K}(\hat{K}+1)\rangle$ and that $Nl\geq j$ implies $Nl\langle \hat{K}\rangle\geq\langle \hat{K}^2\rangle$. Thus we have
\begin{equation}
\label{wiesniaklemma2}
\langle\vec{M}^2\rangle-\frac{Nl+1}{Nl}\langle\vec{M}\rangle^2\geq\langle\hat{K}(\hat{K}+1)\rangle-\frac{Nl+1}{Nl}\geq\frac{Nl+1}{Nl}\Delta^2(\hat{K})\geq 0.
\end{equation}
QED.

The left-hand side of (\ref{wvbcomplementarity}) has been divided into two parts. The first, as shown by Hofmann and Takeuchi, is strictly related to non-local properties of the state and can be positive only in presence of entanglement. The second part, the square of the mean magnetization depends on properties of individual spins. In particular, if our sample is in a singlet state, both fractions vanish and the inequality is saturated by the constant in the front. On the opposite, when the magnetization is maximal, the second part is equal to 1, and the variance of the magnetization is equal to $Nl$. The first part of the left-hand side of (\ref{wvbcomplementarity}) cancels itself out and the inequality is again saturated. In that sense (\ref{wvbcomplementarity}) describes macroscopic quantum information sharing. Information can be stored in either individual properties of solid constituents, or relations between them, or, finally, in a form, which does not saturate (\ref{wvbcomplementarity}).

Some interesting examples of applications of (\ref{wvbwitness1}) are given in the following Section.
\section{Examples}
\hspace*{5mm}

Let us start illustrating the usefulness of the magnetic susceptibility as an macroscopic entanglement witness with repeating the argument by Brukner, Vedral, and Zeilinger \cite{bruknervedralzeil}. It is drawn in a rather different fashion than in \cite{w4}, but also correctly demonstrates the principle. The substance under consideration is copper nitrate, Cu(NO$_3$)$_2\cdot$ 2.5D$_2$O, in a form of a crystal. The detailed structure of the crystal is given in \cite{xu}. What is the most important for this thesis is that the system is a dimerized spin-$\frac{1}{2}$ chain, that is
\begin{equation}
\label{dimer}
H_{CN}=\sum_{j}J_1\vec{S}\;^{[2j]}\cdot\vec{S}\;^{[2j+1]}+J_2\vec{S}\;^{[2j+1]}\cdot\vec{S}\;^{[2j+2]}.
\end{equation}
The inner-dimer coupling constant is $J_1=0.44meV$ and the inter-dimer constant was found to be $J_2=0.11meV$. The Authors assume at the beginning of the proof that, since the system is isotropic, the part of the magnetic susceptibility dependent on the square of the magnetization vanishes. Having neglected the non-nearest-neighbor contribution, they are left with the zero-field magnetic susceptibility in any direction being\footnote{The aim of this paragraph is to confront the entanglement witness with the experimental data published in \cite{cn63}. Thus, following \cite{bruknervedralzeil} we for the moment include all physical constants. Brukner, Vedral, and Zeilinger themselves have not included $\hbar$ in spin magnitudes.}
\begin{equation}
\chi=\frac{g^2\mu_B^2 N}{2\kappa T}\left(\frac{1}{4}\hbar^2+\langle\vec{S}^{o}\cdot\vec{S}^{e}\rangle/3\right).
\end{equation}
$\vec{S}^o$ and $\vec{S}^e$ denote an odd and an even spin (as in (\ref{dimer}) within the same dimer, respectively. Since both spins have magnitudes $\frac{1}{2}\hbar$, the scalar product between them is not larger in modulo than $\frac{1}{4}\hbar^2$, provided that they are not entangled. Thus if the zero-field magnetic susceptibility implies presence of entanglement if
\begin{equation}
\label{bvzwitness}
\chi<\frac{g^2\mu_0^2\hbar^2}{6\kappa T}.
\end{equation}

The reasoning of Brukner, Vedral, and Zeilinger is obviously different from presented in \cite{w4}. On one hand, they have neglected long-distance correlations, whereas we treat contributions from all pairs equally. On the other, they have a priori used isotropy of the system, what has allowed to lower the product state bounds. It would not be suitable for our argument, which is no more than a thermodynamical implementation of the criterion of Hofmann and Takeuchi. The latter is in line just a mathematical criterion, which shall be valid for all states, not only rotationally invariant. The other argument for not taking this assumption initially in \cite{w4} is that we then loose the extensibility of the magnetic susceptibility. The extensibility is here understood as the fact that for a collection of not interacting, identical systems (for example, monocrystals or molecules) the quantity shall be proportional to their number. This is a key feature for studying bulk objects, as we are unable to precisely determine the population of individual systems. 

Interestingly, the bound obtained by Brukner, Vedral and Zeilinger is the same as  in (\ref{wvbwitness3}), despite of the described differences in the derivations. This result was subsequently confronted with the experimental data published in \cite{cn63}.
\begin{figure}
\centering
\label{bvzplot}
\includegraphics[width=10cm]{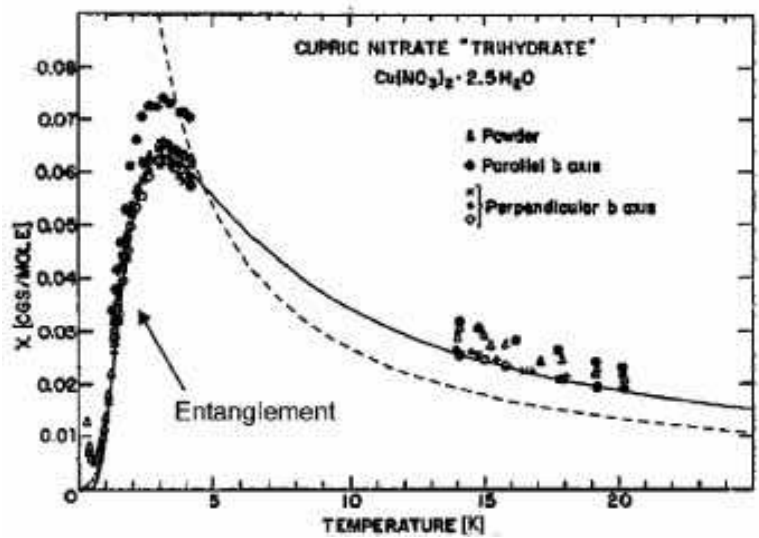}
\caption{The temperature dependence of magnetic susceptibility
of powder cupric nitrate (triangles) and a single-crystal cupric nitrate measured at low-
field parallel (open squares) and perpendicular (open circles,
crosses, filled circles) to the monoclinic b axis. The data and the
figure are from Ref. \cite{cn63}. The solid curve is the theoretical curve for
a dimer rescaled for the amount of noise estimated from the experiment.
This noise is computed as the ratio of the maximal experimental
value (averaged over crystal data) and the maximal theoretical
value. The dashed curve represents the macroscopic
entanglement witness (\ref{bvzwitness}) and is rescaled in exactly the same way.
The intersection point of this curve and the experimental one de-
fines the temperature range (left from the intersection point) with
entanglement in the substance. The critical temperature is around $T_c^{
exp}\approx 5 K$.
Note that the entanglement witness will cut the experimental curve
(and hence yield a critical temperature) independently of a particular
rescaling procedure. (The figure and the caption taken from \cite{bruknervedralzeil}).}
\end{figure}
 
Figure 5.3 presents the measured molar magnetic susceptibility of cupric nitrate measured at various temperatures and with crystal powdering. The solid line is an interpolation of the measured susceptibility and the dashed curve represents (\ref{bvzwitness}). The lines cross at about $5K$, below which the presence entanglement is manifested. It is particularly interesting that the experiment was performed in 1963, much before the discussion on quantum information became intensive. 

Now, let us pass to theoretical applications of the criterion. We begin with analyzing results obtained by Xiang in \cite{xiang}. Therein he computes the thermodynamical quantities of long, efficiently infinite, spin$-\frac{1}{2}$ and spin$-1$ (as well as spin-$\frac{3}{2}$) chains described by the Hamiltonian 
\begin{equation}
\label{hxxxb}
H_{xxx}(B)=\sum_{i=-\infty}^\infty \vec{S}^{[i]}\cdot\vec{S}^{[i+1]}+B\sum_{i=-\infty}^\infty \vec{S}^{[i]}_3.
\end{equation}
The used numerical procedure, called the transfer matrix renolmalization group, is described in details in \cite{wangxiang}. Figure 5.4 presents computed zero-field magnetic susceptibilities for spin-$\frac{1}2$ and spin-1 chains, respectively. Dotted lines are our entanglement criteria given by (\ref{wvbwitness3}). The lines cross in 5.4 at $T\approx 1.4$ and at $T=2$. These results are to be compared with the critical temperatures found using the internal energy argument, (\ref{bvbound}). The internal energies for spin-$\frac{1}{2}$ and spin-1 chains \cite{xiang} are presented in Figure 5.5. If one uses the Brukner-Vedral witness rather than the magnetic susceptibility, one finds that entanglement is certainly present below $T\approx 0.8$ for spins$-\frac{1}{2}$ and $T\approx 1.4$ in the case of spins-1. The advantage of (\ref{wvbwitness3}) over (\ref{bvbound}) shall be explained with the fact that, unlike (\ref{bvbound}), where only the nearest-neighbor pairs are taken into account, in (\ref{wvbwitness3}) we also include the correlations between not neighboring sites. 

\begin{figure}
\centering
\includegraphics[width=8cm]{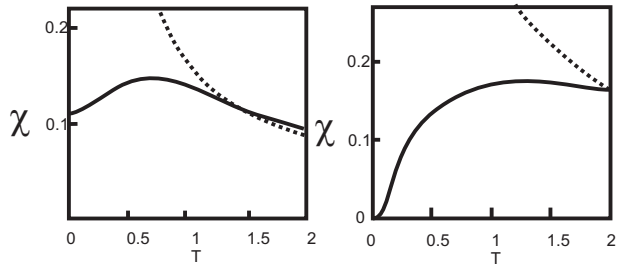}
\label{suscxiang}
\caption{The zero-field magnetic susceptibility of an infinite Heisenberg chain of spins$-\frac{1}{2}$ (left) and spins-1 (right) computed in \cite{xiang}. The dotted line is our entanglement criterion (\ref{wvbwitness3})}.
\end{figure}

\begin{figure}
\centering
\includegraphics[width=8cm]{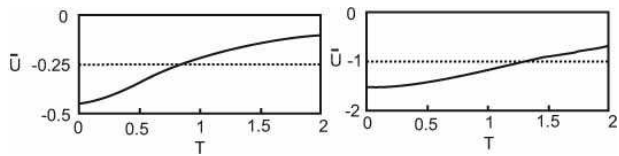}
\label{energyxiang}
\caption{The internal energy of an infinite Heisenberg chain of spins$-\frac{1}{2}$ (left) and spins-1 (right) computed in \cite{xiang}. The dotted line is the entanglement criterion (\ref{bvwitness})}.
\end{figure}

Let us study lattices of spins-1 in more details. For spins$-\frac{1}{2}$ the interaction, which would be biquadratic, rather than bilinear against spin operators, would be trivial. However, for spins of higher magnitudes it is an interesting isotropic interaction. We have numerically computed the zero-field magnetic susceptibility for the ring of six spins-1 with adjustable bilinear and biquadratic interaction constants. This system is described by the following Hamiltonian:
\begin{equation}
H(a)=\cos a\sum_{i=1}^6\vec{S}^{[i]}\cdot\vec{S}^{[i+1]}+\sin a\sum_{i=1}^6\left(\vec{S}^{[i]}\cdot\vec{S}^{[i+1]}\right)^2,
\end{equation}
with the periodicity condition $[7]\equiv[1]$. The magnetic susceptibility per spin is presented in Figures 5.6 and 5.7. From the latter we see that entanglement is implied by (\ref{wvbwitness2}) between $a=-\frac{3}4\pi$ and $a=\frac{1}{2}\pi$, and only below $T_c\approx 2.66$. Quantum correlations are more temperature-persistent for $-\frac{3}{4}\pi\leq a\leq 0$ than for $0\leq a\leq \frac{1}{2}\pi$. This might  be due to a fact that the biquadratic term with a negative factor favors states with small total angular momenta over those with large angular momenta. For example, the two-spin-1 operator $\left(\vec{S}^{[1]}\cdot\vec{S}^{[2]}\right)^2$ has the singlet state as an eigenstates with the corresponding eigenvalue 36, while the states of the total angular momentum 6(=2(2+1)) have the eigenvalue 1. In the region of the entangled ground state and positive $a$, the bilinear interaction is dominant over the biquadratic term. For $a$ close to $-\frac{3}{4}\pi$, on the other hand, the latter overcomes the former.

\begin{figure}
\centering
\includegraphics[height=5cm]{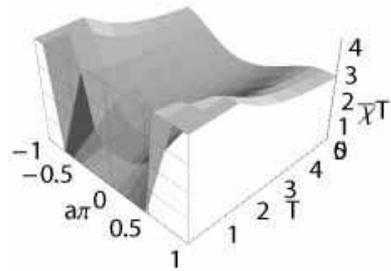}
\label{spin1suscfull}
\caption{The zero-field magnetic susceptibility (summed over three orthogonal directions and  multiplied by temperature) per spin of a Heisenberg ring of 6 spins-1 as a function of temperature and $a$.}
\end{figure}

\begin{figure}
\centering
\includegraphics[height=5cm]{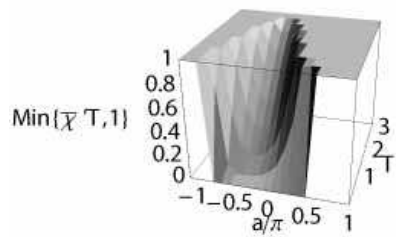}
\label{spin1susc}
\caption{The zero-field magnetic susceptibility (summed over three orthogonal directions and  multiplied by temperature) per spin of a Heisenberg ring of 6 spins-1 as a function of temperature and $a$ plotted up to the entanglement threshold.}
\end{figure}

Another model we would like to consider is an example of a 3-dimensional lattice. Let us consider the system of nine spins$-\frac{1}{2}$, eight of which are placed in vertices of a cube and interact with one another only along edges. The ninth is placed in the center of the cube and interacts with all others equally strongly. With the free parameter $a$ we will this time vary the proportion between the constant of the coupling along edges and with the central qubit:
\begin{eqnarray}
H=&\cos a\left(\vec{S}^{[1]}\cdot\vec{S}^{[2]}+\vec{S}^{[2]}\cdot\vec{S}^{[3]}+\vec{S}^{[3]}\cdot\vec{S}^{[4]}+\vec{S}^{[4]}\cdot\vec{S}^{[1]}\right.\\
+&\vec{S}^{[5]}\cdot\vec{S}^{[6]}+\vec{S}^{[6]}\cdot\vec{S}^{[7]}+\vec{S}^{[7]}\cdot\vec{S}^{[8]}+\vec{S}^{[8]}\cdot\vec{S}^{[5]}\\
+&\left.\vec{S}^{[1]}\cdot\vec{S}^{[5]}+\vec{S}^{[2]}\cdot\vec{S}^{[6]}+\vec{S}^{[3]}\cdot\vec{S}^{[7]}+\vec{S}^{[4]}\cdot\vec{S}^{[8]}\right)\\
+&\sin a \left(\vec{S}^{[9]}\cdot\sum_{i=1}^8\vec{S}^{[i]}\right).
\end{eqnarray} 
\begin{figure}
\centering
\label{cubesystem}
\includegraphics[height=1.5cm]{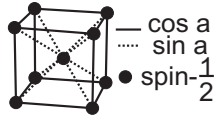}
\caption{8 spins-$\frac{1}{2}$ forming a cube and surrounding another spin-$\frac{1}{2}$.}
\end{figure}

In Figure 5.9 we present the computed susceptibility per spin multiplied by temperature up to the entanglement threshold 0.5. It can be seen that the thermal state is entangled for $-0.4\leq a\leq 0.43$ at sufficiently low temperatures. However, the quantity presented in Figure 5.9 does not vanish at $T=0$, hence the magnetic susceptibility would, as it seems, diverge with $T\rightarrow 0$. This is for the reason that given an odd number of odd-half-integer spins it is not possible to construct a singlet state. The ground state shall possess the rotation symmetry of the Hamiltonian at the expense of being mixed. However, in realistic situations this symmetry shall be broken and the probe shall then gain the magnetization.

\begin{figure}
\centering
\includegraphics[height=7cm]{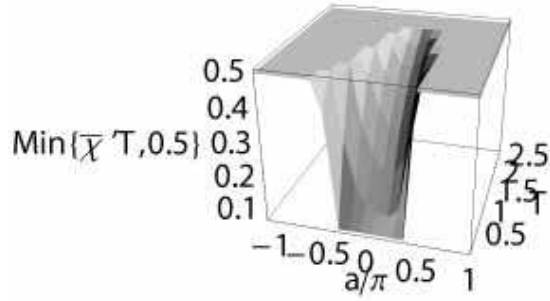}
\label{susccube}
\caption{The zero-field magnetic susceptibility per spin (summed over three orthogonal directions and  multiplied by temperature) of a cube of spins-$\frac{1}{2}$ with another spin-$\frac{1}{2}$ in its center and plotted up to the entanglement threshold, 0.5}\end{figure}

\begin{figure}
\centering
\includegraphics[width=7cm]{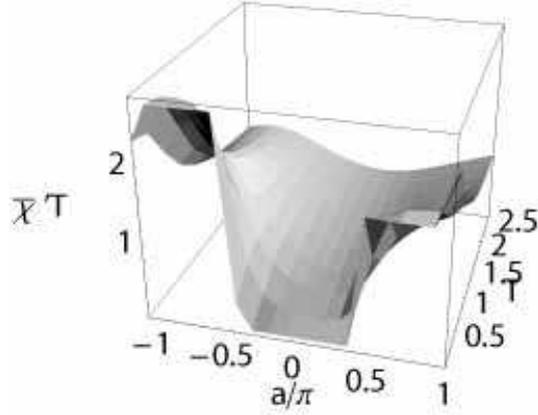}
\label{susccubefull}
\caption{The zero-field magnetic susceptibility per spin (summed over three orthogonal directions and  multiplied by temperature) of a cube of spins-$\frac{1}{2}$ with another spin-$\frac{1}{2}$ in its center multiplied by temperature.}
\end{figure}

This interesting result motivated us to find a model, which in various phases has all three ways of the low-temperature behavior of the zero-field susceptibility at $T\rightarrow 0$:
\begin{itemize}
\item{$\chi'T$ tends to the constant above the entanglement threshold,}
\item{$\chi'T$ tends to the non-zero constant below the threshold,}
\item{$\chi'T$ tends to zero.}
\end{itemize}
The system which satisfies this requirement is a simple modification of the previous one. Instead of a spin$-\frac{1}{2}$, we shall place a spin-1 in the middle. To save the computational resources, we shall also decrease the number of surrounding spins and arrange them into an octahedron (o) and a tetrahedron (t). The systems are drawn in Figure 5.11, and the corresponding Hamiltonians shall be given by
\begin{eqnarray}
H_{(o)}=&\cos a\left(\left(\vec{S}^{[1]}+\vec{S}^{[6]}\right)\cdot\left(\sum_{i=2}^5\vec{S}^{[i]}\right)\right.\nonumber\\
+&\left.\vec{S}^{[2]}\cdot\vec{S}^{[5]}+\sum_{i=2}^4\vec{S}^{[i]}\cdot\vec{S}^{[i+1]}\right)\nonumber\\
+&\sin a\vec{S}^{[7]}\cdot\sum_{i=1}^6\vec{S}^{[i]}
\end{eqnarray}
for the octahedron, where $l^{[i]}=\frac{1}{2}$ for $i$ between 1 and 6 and $l^{[7]}=1$, and
\begin{eqnarray}
H_{(t)}=&\cos a\left(\left(\sum_{i=1}^4\vec{S}^{[i]}\right)^2-3\right)/2\nonumber\\
+&\sin a\vec{S}^{[5]}\cdot\sum_{i=1}^4\vec{S}^{[i]}
\end{eqnarray}
for the tetrahedron, with $l^{[5]}=1$ and $\frac{1}{2}$ being the magnitude of all other spins. 
\begin{figure}
\centering
\label{otsystem}
\includegraphics[width=6cm]{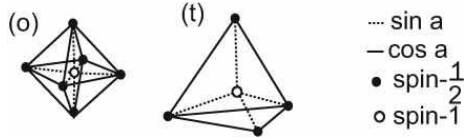}
\caption{6 (left) and 4 (right) spins-$\frac{1}{2}$ surrounding a spin-1.}
\end{figure}
These systems have at least 8 and 6, respectively, distinct phases with respect to $a$. Transitions between these phases can be seen in Figure 5.12 as rapid changes of values of low-temperature $\chi'$. In their nature lies an interchange of the ground state and one of the excited states. 

\begin{figure}[htp]
\centering
\subfigure{
\includegraphics[width=7cm]{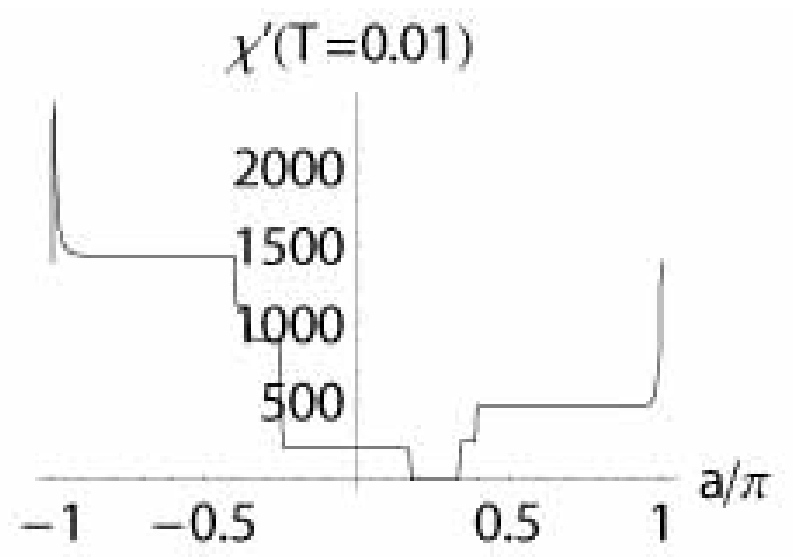}}
\subfigure{
\includegraphics[width=7cm]{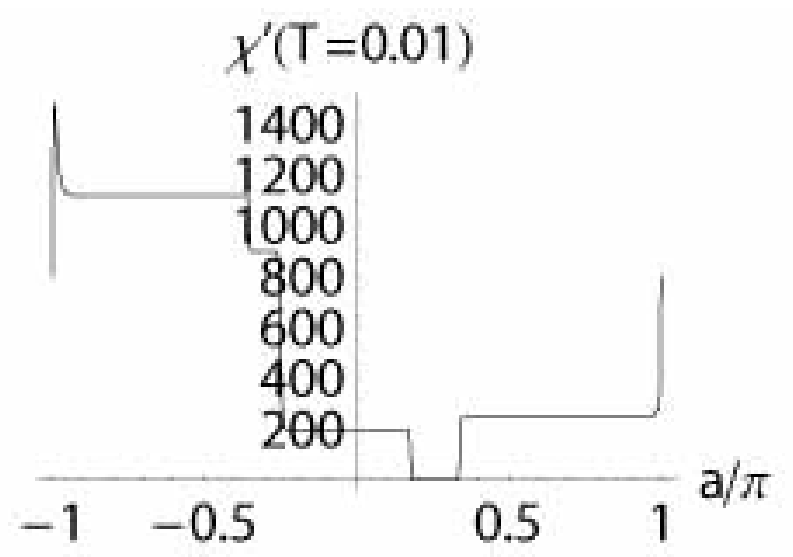}}
\label{groundstates}
\caption{$\chi'$ at $T=0.01$ for an octahedron (left) and a tetrahedron (right) of spins-$\frac{1}{2}$ with a spin-1 in the center. Every rapid change of the value corresponds to a quantum phase transition of the first kind, a change of ground system.}
\end{figure}

The susceptibilities multiplied by temperature are presented in Figures 5.13 and 5.14 (up to entanglement thresholds, 4 for (o) and 3 for (t), respectively). Both systems have interesting features. The ground state of (t) is recognized as entangled whenever the interaction between the spin-1 and surrounding spins is antiferromagnetic, and for (o) there exists an interval of $a$ in which the thermal is satisfies our separability criterion at $T$ close to 0, but then the magnetic susceptibility goes under the threshold as temperature grows and low excited states significantly contribute to the mixture.

\begin{figure}
\centering
\includegraphics[width=6cm]{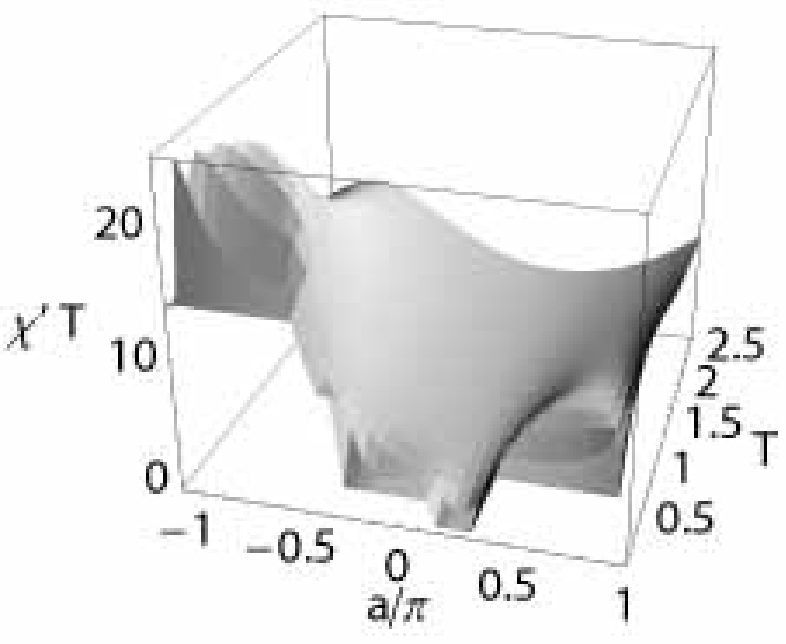}
\includegraphics[width=6cm]{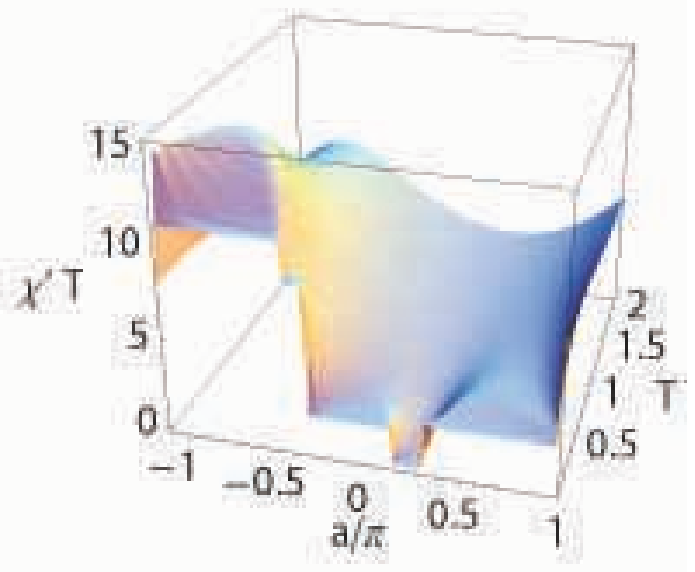}
\label{otsuscfull}
\caption{$\chi'T$ plotted for an octahedron (left) and a tetrahedron (right) of spins-$\frac{1}{2}$ with a spin-1 in the center.}
\end{figure}

\begin{figure}
\centering
\includegraphics[height=7cm]{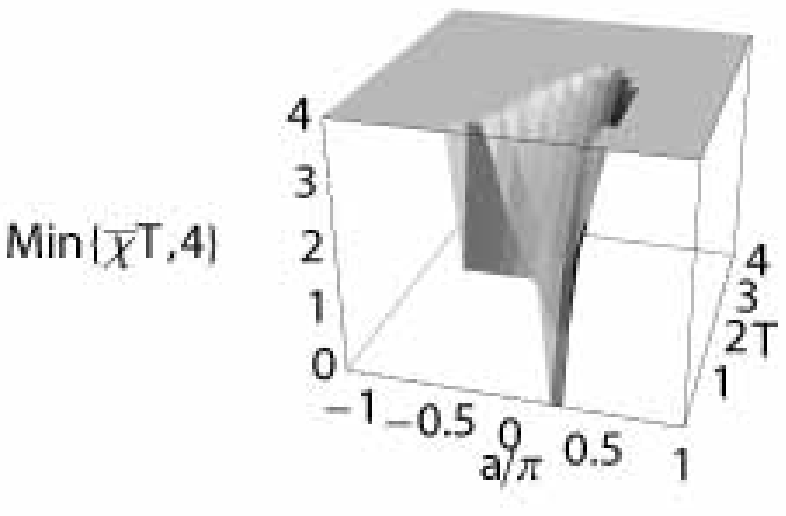}
\includegraphics[height=7cm]{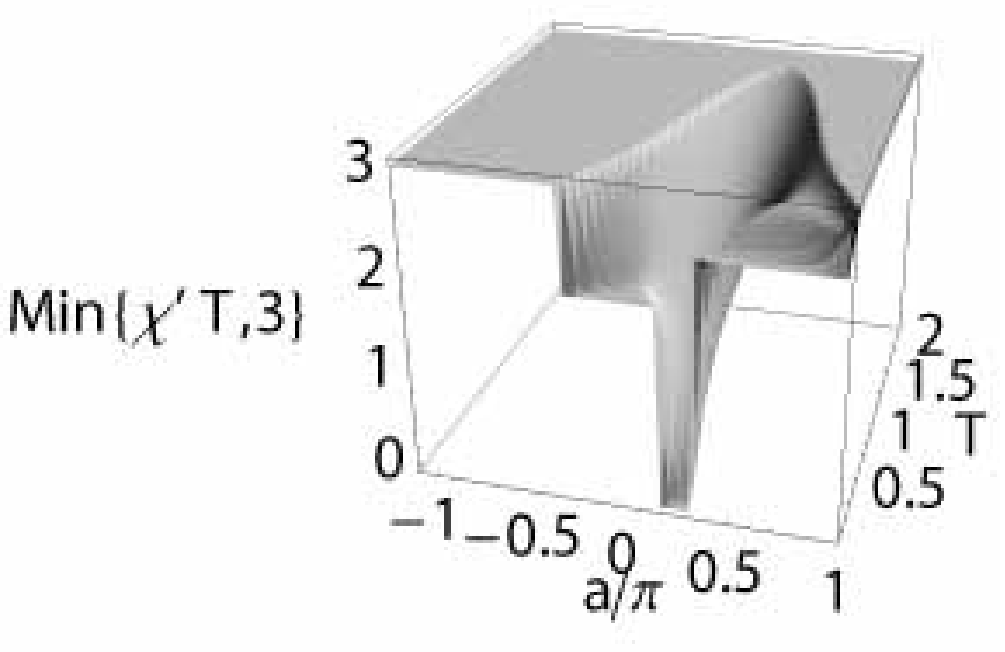}
\label{otsusc}
\caption{$\chi'T$ plotted for an octahedron (left) and a tetrahedron (right) of spins-frac{1}{2} with a spin-1 in the center, plotted to entanglement thresholds, 4 and 3, respectively.}
\end{figure}

\section{Heat Capacity as Entanglement Witness \cite{w5}}
\hspace*{5mm}

In Section 5.3 we have argued that the presence of thermal entanglement can be deduced from the internal energy. However, it is physically difficult, or even impossible to precisely measure the internal energy of an object, as it is always given up to an irrelevant additive constant. One can gauge the internal energy, for example, by bringing the sample close to the absolute zero or the infinite temperature, and then by a careful energetic balance of bringing it to a desired temperature. As we have already mentioned, for spin lattices and other systems described by traceless Hamiltonians the energy at $T\rightarrow\infty$ is $0$, whereas the ground-state energy can be calculated from $H$. Neither of the limits is attainable physically, but both can be reached sufficiently close to make arbitrarily good approximations. To use the zero-temperature energetic balance we need to know the precise form of the Hamiltonian. While heating the sample and subsequent cooling it back to the temperature of our interest, our sample can overcome structural phase transitions, like melting or evaporating.

From the experimental point of view, it is much easier determine the heat capacity, the change of internal energy per an infinitely small change of temperature, $C=\frac{\partial U}{\partial T}=\frac{1}{\kappa T^2}\Delta^2(H)$ (at a constant volume). In this section we will show that low-temperature values of $C$ can witness entanglement. In particular, we will argue that in cases, in which the ground state is entangled, the Third Law of Thermodynamics is related to non-separability of the state. The Law was first given by Nernst \cite{nernst} as a statement that the entropy of the ground state is a constant dependent only on its mixedness. We are more interested in an alternative formulation, namely that the absolute zero temperature is unattainable in any finite physical process, as the heat capacity then tends to zero. The equivalence of these two statements is still a subject to discussions, for example by Langsberg \cite{amjphys65296refti}. A convincing argument was given by Guggenheim \cite{guggenheim}, and based on fact that for finite systems, the entropy at a finite temperature, $S(T)=\int_0^{T}\frac{C(T')}{T'}dT'$ must also be finite. This puts a strong bound on the possible low-temperature behavior of $C$. The converse implication is discussed in e.g. \cite{munster}.

The first way to demonstrate the heat capacity as an entanglement witness is direct. We will show that the variance of some Hamiltonian cannot vanish if one allows only product states. By (\ref{htmixed}) it must also be true for separable states. Hence either there must appear thermal entanglement, or the heat capacity diverges with $T\rightarrow 0$.

An example of a model with non-zero energy variance for separable states is an Ising ring in a transverse magnetic field, the Hamiltonian of which reads
\begin{equation}
\label{isinghamiltonian}
H_{Ising}=\pm\sum_{i=1}^N\sigma_3^{[i]}\sigma_3^{[i+1]}+B\sum_{i=1}^N\sigma_1^{[i]},
\end{equation}
with the periodicity $N+1\equiv 1$. 

Here follows the proof that no factorisable state is an eigenstate of the Hamiltonian for $B\neq 0$. Assume, that this is not the case and the state $|\psi\rangle=\bigotimes_{i=1}^N(a_i|0\rangle+b_i|1\rangle)$ is an element of the eigenbasis of $H_{Ising}$, with some energy $E$. We first need to notice that this is necessary that $\forall_i a_i,b_i\neq 0$  because of the action of the magnetic field. Then one should have
\begin{equation}
E=\frac{\langle 00..0|H_{Ising}|\psi\rangle}{\langle 00..0|\psi\rangle}=\frac{\langle 10..0|H_{Ising}|\psi\rangle}{\langle 10..0|\psi\rangle},
\end{equation}
which written explicitly is
\begin{equation}
\frac{\left(B\Sigma+N\right)\prod_{i=1}^Na_i}{\prod_{i=1}^Na_i}=\frac{\left(B+\frac{b_1}{a_1}\left(B\Sigma-B\frac{b_1}{a_1}+N-4\right)\right)\prod_{i=1}^Na_i}{\frac{b_1}{a_1}\prod_{i=1}^Na_i}
\end{equation}
with $\Sigma=\sum_{i=1}^{N}\frac{b_i}{a_i}$. The last equation simplifies to
\begin{equation}
\label{quadratic}
-B\left(\frac{b_1}{a_1}\right)^2-4\frac{b_1}{a_1}+B=0.
\end{equation}
The same relation can be now obtained for the inverse of the fraction, $\frac{a_i}{b_i}$. However, 
one can easily verify that the product of the two solutions of (\ref{quadratic}) is $-1$. QED.

For the sake of the numerical minimization of $\Delta^2(H_{Ising})$ we consider the the states that have a period of two sites:
\begin{eqnarray}
\label{isingstate}
|\theta_1,\theta_2\rangle=\left((\cos\theta_1/2|0\rangle+\sin\theta_1/2|1\rangle)(\cos\theta_2/2|0\rangle+\sin\theta_2/2|1\rangle)\right)^{\otimes N/2}.
\end{eqnarray}
Such a form of the state can be justified in a following way: when we minimize the energy of the Ising model over factorisable states, we want to allow the local states to have the same spin projection onto the direction of the field and in the same time every two nearest neighbors have should have the opposite components in the interaction direction. This can be done with states (\ref{isingstate}). $\langle\sigma_2^{[i]}\rangle$ is not important for any $i$, thus we assume that the whole vector lies in the $xz$-plane. However, calculating $\Delta^2(H_{Ising})$ we also deal with means like $\langle\sigma_3^{[i]}\sigma_3^{[i+2]}\rangle$ (since the state is factorisable, all other cross terms, like $\langle\sigma_3^{[i]}\sigma_3^{[i+2]}\sigma_3^{[i+3]}\sigma_3^{[i+4]}\rangle$ will vanish). It would seem that one needs to consider states of the periodicity of four sites. Nevertheless, numerical calculations show that increasing the period does not allow to mimimize $\Delta^2(H_{Ising})$ more than by using the period of two sites.

Let us introduce a short-hand notation $x_i=\langle\sigma_1^{[i]}\rangle=\sin\theta_i$ and $z_i=\langle\sigma_3^{[i]}\rangle=\cos\theta_i$.  Under the above assumptions, the variance of the energy per spin can be written as
\begin{eqnarray}
\label{isingvarh}
\Delta^2(H_{Ising})=&N\left(1+z_1^2+z_2^2-3z^2_1z_2^2\right)\nonumber\\
\pm&2BN(z_1z_2(x_1+x_2)\nonumber\\
+&B^2\frac{N}{2}\left(2-x_1^2-x_2^2\right),
\end{eqnarray}
The expression has been subjected to a numerical minimization over $\theta_1$ and $\theta_2$, the results of which are plotted in Figure 5.15.

\begin{figure}
\centering
\includegraphics[width=5cm]{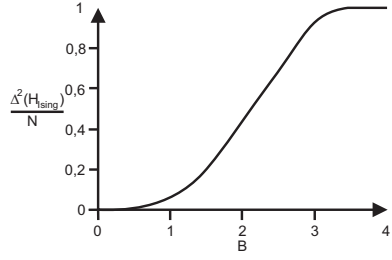}
\caption{The minimal variance per site $\bar{\Delta}^2(H_{Ising})$ of the Hamiltonian of a transverse Ising ring over 2-translation invariant product states $\left(|\phi_{1}\rangle|\phi_{2}\rangle\right)^{\otimes N/2}$ versus
the magnetic field $B$ ($J=1$).} 
\label{Isingplot}
\end{figure}

We can see that the variance vanishes for any case, except for $B=0$. When the field-interaction energy becomes significantly larger than the inter-spin interaction, the ground state becomes completely determined by the field, thus a factorisable state. Even then the variance is non-zero as we vary the magnetic field, not the interaction strength. Note that the same results apply to both FM and AFM cases of the Ising model.

These results can be confronted with the specific heat of the system found in \cite{katsura}. Therein, the heat capacity per spin was found to be
\begin{eqnarray}
\label{heatkatsura}
\frac{C_{Ising}}{N}=\frac{1}{\pi T^2}\int_0^\pi \frac{f(B,\omega)}{\cosh^2\frac{f(B,w)}{T}}d\omega,
\end{eqnarray}
where $J=k=1$ and $f(B,\omega)=\sqrt{1-2B\cos\omega+B^2}$.



Figure 5.16 presents the heat capacity per spin (\ref{heatkatsura}) for a given magnetic field $B=2$. The curve represents $a/T^2$, where $a=0.4197$ is the minimal Hamiltonian variance over factorisable for this value of the field. Below this line the thermal state is definitely entangled.
\begin{figure}
\centering
\label{isingplot3}
\includegraphics[width=5cm]{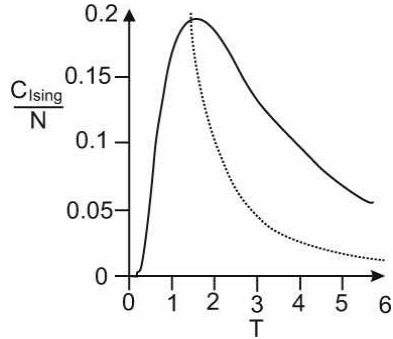}
\caption{The heat capacity per spin ($\kappa=1$) of a transverse antiferromagnetic Ising ring versus the temperature $T$ for $B=2$. The dotted curve, $0.4197/T^2$, is the heat capacity entanglement criterion.} 
\end{figure}

Thus for the Ising model, and for many other models without any rotational invariance it sufficies to show that $\Delta^2(H)$ does not vanish for product states to use the heat capacity as an entanglement witness. However, more interesting from the physical point of view models, like $xx$ or $xxx$ systems, fail to be applied to the same criterion. At least one of the eigenstates of their Hamiltonians, though not necessarily the ground state is a product state, e.g., $|1...1\rangle$. That is, the energy in the state is exactly determined, hence the variance can be set 0 within the set of product states.

In \cite{w5} we argue basing on fundamental concepts of Thermodynamics that the low-temperature heat capacity, which is an easily measurable quantity, can still reveal thermal entanglement given the certainty that the ground state is entangled. One also needs to know the structure of the spectrum, if it is gap-less, or the ground-state energy is separated from the first excited state energy by an energy gap $\Delta$. The last information to be known is the bound on the internal energy, below which the state is certainly entangled \cite{qph0406040,dowling, wu}, $E_B$, as well as the ground state energy $U_0$.

{\em Gap-less systems:} let us first demonstrate our argument for system with a continuous bottom part of the spectrum per spin. Examples of such systems are half-odd integer spin chains, taken in the thermodynamical limit. Their internal energy per spin can be written as $\bar{U}=U_0+(\kappa T)^{\gamma}(c_0+c_1\kappa T+...)$, where $\gamma>0$ and $c_0,c_1,...$ are some material-dependent constants. At low temperatures we may drop higher-order terms and approximate $\bar{U}\approx U_0+c_0(\kappa T)^{\gamma}$. Consequently, at these temperatures the heat capacity shall be given as
\begin{equation}
\label{gapless1}
\bar{C}=\frac{\partial \bar{U}}{\partial T}=\gamma c_0\kappa(\kappa T)^{\gamma-1}=\gamma\frac{\bar{U}-\bar{U}_0}{T}.
\end{equation}
We can now use the bound on the internal energy to derive a similar condition for the heat capacity. Namely,
\begin{equation}
\label{gapless2}
\bar{C}<\gamma\frac{\bar{E}_B-\bar{U}_0}{T}
\end{equation}
implies entanglement within the range of the approximation. Only if the ground state is separable, that is $\bar{U}_0=\bar{E}_B$, (\ref{gapless2}) prevents $\bar{C}\rightarrow\infty$ as $T$ approaches the absolute zero. In general, however, inequality (\ref{gapless2}) shows that the heat capacity does not diverge at low temperatures only due to entanglement. The result is thus similar to the one from \cite{w4}. In both cases, the zero-temperature finiteness of respective thermodynamical quantities can only be explained with quantum correlations of the thermal state.

Note that this criterion is applicable for the whole class of gap-less models in 1+1 dimensions\footnote{Space+Time dimensions.}. For these models the conformal field theory \cite{prl56746,prl56742} predicts $C=(\pi c\kappa T)/(3\hbar v)$, with $c$ being the central charge in the corresponding Virasoro algebra and $v$ - the spin velocity. Affleck \cite{prl56746} has shown that for a chain of spins-$l$ $c=6l/(2+2l)$.

An explicit example of a gap-less model is an infinite xxx antiferromagnetic spin-$\frac{1}{2}$ chain with the Hamiltonian (\ref{hxxxb}) with $B=0$, for which $\bar{U}_0=-0.443$ \cite{hulten} and $\bar{E}_B=-0.25$ \footnote{Now we use spin operators rather than Pauli matrices, hence the values are divided  by 4.}. Hence by (\ref{gapless2}) the condition for quantum correlations is $C<0.386/T$. On the other hand, from \cite{xiang} we estimate that the linear behavior of the heat capacity $\bar{C}\approx\frac{2}{3}T$ hold for $T\leq 0.1$. In this region we can conclude non-classicality of the thermal state from $\bar{C}$.

{\em Gapped systems:} On contrary, integer antiferromgnetic spin chains, and, by definition, all quantum systems of a finite dimensionality belong to a universal class of gapped states. Haldane has argued in \cite{pla93464} that for these systems the low-temperature heat capacity per spin reads
\begin{equation}
\label{gapped1}
\bar{C}=c'\left(\frac{\kappa T}{\Delta}\right)^\delta\exp\left(-\frac{\Delta}{\kappa T}\right),
\end{equation}
where $c',\delta$ are again material-dependent constants. The internal energy can hence be found as $\bar{U}(T)=\bar{U}_0+\int_0^T\bar{C}(T')dT'=U_{0}+c'\frac{\Delta}{\kappa}\Gamma\left(-\delta-1,\frac{\Delta}{\kappa T}\right)$. Now, according to \cite{abramowitzstegun} we approximate $\Gamma(a,x)\approx e^{-x}x^{a-1}$ for large $x$ and obtain 
\begin{equation}
\bar{U}=\bar{U}_0+c'\kappa T^{\delta+2}e^{-\frac{\Delta}{\kappa T}}/\Delta.
\end{equation}
Finally, we again use the bound for separable states and argue that
\begin{equation}
\label{gapped2}
\bar{C}<\frac{\Delta(\bar{E}_B-\bar{U}_0)}{\kappa T^2}
\end{equation}
can be only due to entanglement, of course, within the range of the approximation. Let us here comment that the second approximation, $\kappa T<<\Delta$, is usually much weaker than for the gap-less systems.

To exemplify this version of the argument, we consider an infinite spin-1 one-dimensional $xxx$ antiferromagnet with $J=1$, for which $c'=\Delta^{5/2}/\sqrt{2\pi}$, $\Delta=0.411$, $\bar{U}_0=-1.401$ \cite{prb509265,prb483844} and $\bar{E}_B=-1$ (the proof is a simple extension of (\ref{bvbound}) to spins-1). Thus all values of $\bar{C}$ below $0.165/T^2$ ensure us about thermal entanglement in the ring at $T<<0.4$.
\chapter{Summary}
\hspace*{5mm}

Entangled states \cite{schroedinger} are not only a counter-intuitive feature of Quantum Mechanics \cite{epr}, but also a key resource for many quantum information processing tasks, like the teleportation \cite{teleportation}, Quantum Cryptographic Key Distribution \cite{bennettbrassard88,ekert91,cryptography}, or Quantum Computation algorithms, e.g., \cite{grover,shor,deutschjozsa}. Some of this applications strongly rely on the discrepancy between Quantum Mechanics and Local Realism.

For this reason this dissertation has been devoted to the problem of detecting entanglement in various physical systems. In general, this problem can be approached from two different perspectives. First, we may want to confirm whether a state directly violates Local Realism. In some models a direct application of a Bell inequality might be not feasible due to a high complexity of the physical system. In such a case, we can apply an entanglement witness. Quantum correlations confirmed by such a criterion not necessarily falsify Local Realism, but can be post-processed \cite{distillation} in order to do so. While Chapter 2 aims to present a historical outline of the Bell theorem, Chapters 3 and 4 present more recent results, and in Chapter 5 we focus on discussing entanglement witnesses, particularly those, which can be deduced from thermodynamical properties of solids.

We have begun with presenting the apparent paradox noticed by Einstein, Podolsky, and Rosen \cite{epr}, who had considered two distant particles in a quantumly correlated state. Basing on Relativity, the Authors assumed that no useful information can propagate between two remote observers with a superluminal speed. They also had a belief of Realism, that is that results of all possible experiments are predefined. However, the position of one particle from the EPR pair could be determined by the measuring the position of the other, regardless of the distance between the two particles. Similarly, the momentum of one of the particles can be known by measuring the momentum of the other without any disturbance of the first one. Thus both the position and the momentum are seemingly elements of the reality. However, the quantum-mechanical Heisenberg uncertainty principle forbids to know precisely the values of two non-commuting observables, like the position and the momentum. This brings the alternative of two possibilities. Either Quantum Mechanics is unable to provide the most complete description of the reality, or these two quantities cannot be simultaneously elements of the same physical reality. 

To answer this question Bohr \cite{bohr} has referred to his own concept of complementarity. According to him, Quantum Mechanics is a complete theory. Nevertheless, according to him, the act of measurement disturbs the system in a uncontrollable way, hence results of future measurements are irreversibly disturbed. The problem also lies in the distinction between a quantum system and a measuring apparatus. Simply, as Bohr states, a measurement of the position corresponds to a different physical situation that a measurement of the momentum.

The possibility of completing Quantum Mechanics with additional parameters, which would determine results of measurement, yet remain unknown to observers, was considered in 1952 by Bohm \cite{Bohm1,Bohm2}. He first proposed his own interpretation of Quantum Mechanics, based on an assumption, that while the position of a particle is stochastically described by a square of moduli of the wave-function, its momentum is given by a gradient of the phase function. Then stationary states, which have their wave-functions real, do  not express any motion of the particle. In such a case the particle can be seen as an entity from an ensemble of identical systems. Elements might have a precisely defined position and the wave-function would rather now describe the density within the ensemble. Thus both the position and the momentum of the particle could be simultaneously measured with arbitrarily high precisions. In more general cases Bohm suggests that the Schr\"odinger equation could have an inhomogeneous term, which would allow to introduce additional parameters and build a new measurement theory. 

However, Bell's seminal paper \cite{bell} has shown that if such hidden parameters exist, they cannot be of a local character. Bell showed that Quantum Mechanics, or, in particular, entanglement, is in a sharp disagreement to Local Realism. The discrepancy was demonstrated by inequality (\ref{belloriginal}), the violation of which was later tested in various experiments \cite{Aspect1982,weihs,rowe}. None of the experiments was fully convincing, however. In many cases, the violation of a Bell inequality has been found related to quantum-over-classical advantages in various quantum information processing tasks, e.g.,  cryptographic key distribution \cite{cryptography} or communication complexity problems \cite{complexity}.

Greenberger, Horne, and Zeilinger showed that the violation of Local Realism can be demonstrated, even in a stronger version than for two qubits, for larger systems. Initially formulated without inequalities, their argument was rephrased by Mermin \cite{mermin} with help of a series of statistical expressions with local realistic bounds. Gradual increase of the arbitrarity of measurements performed by obsevers was presented subsequently by Ardehali  \cite{ardehali}, and Belinskii and Klyshko  \cite{belinskiiklyshko}. The further generalizations were made by Werner and Wolf \cite{wernerwolf}, Weinfurter and \.Zukowski \cite{weinfurterzukowski}, \.Zukowski and Brukner \cite{zukowskibrukner}, Wu and Zong \cite{wuzong}, and Laskowski, Paterek, \.Zukowski, and Brukner \cite{patlaszukbruk}. 
On the other hand, there were many attempts to find Bell inequalities by finding hyperplanes containing faces of a certain convex hull, e.g. \cite{Froissart,pitowskysvozil,sliwa}. A convenient way of analyzing such a hull for correlations between many qubits was presented recently by \.Zukowki \cite{qph0611086} and developed and applied directly to $N=3$ by Wie\'sniak, Badzi\k{a}g, and \.Zukowski  in \cite{w8}. We focused on the simplest non-trivial case three observables per site. The found inequalities are special cases of (\ref{plzb3qa}). We also show that a Bell inequality, in which {\em all} three observers choose between three observables. 

In Section 3.12 we consider effects of taking the rotational invariance of the correlation function as an addition constraint on theories with LHV \cite{w3}. As it appears, for GHZ states the inequalities which utilize this assumption are exponentially more robust against the white noise admixture than MAKB, WWW\.ZB, or WZLP\.ZB inequalities. Expressions derived in \cite{w3} utilize a continuum of possible local observables, nevertheless the amount of violation can by computed from the  condition (\ref{rotinvcond}) with the knowledge of only $2^N$ mean values.

Another trend in the studies on the Bell theorem is a discussion on realizations of a Bell test with a single photon, rather than a set of entangled particles, as a carrier of non-classicality. This idea was initially proposed by Tan, Walls, and Collett \cite{tanwallscollett}. One of our results \cite{w7} is showing that the realization of Tan-Walls-Collett-type proposal described by Bj\"ork, Jonsson, and S\'anchez-Soto \cite{bjorkjonssonsanchezsoto} can be suprizingly robust against photon loss, which in two-photon experiments is equivalent to the detection efficiency. The required detection efficiency in standard Bell experimental set-ups was found by Garg and Mermin \cite{gargmermin} to be $82.8\%$ for a maximally entangled state, whereas Eberhard \cite{eberhard} has shown that in passing from the singlet state to the limit of a factorisable state the efficiency can be lowered to $66.7\%$. We have obtained that it is necessary to preserve at least $17.2\%$ of photons from the source in the BJSS scheme. Interestingly, this proposal is the case of a great advantage of the CH inequality (\ref{chinequality}) over CHSH (\ref{chshinequality}). Another interesting feature of the schehme is that the optimal realization of the BJSS experiments requires a strong coherent beam, unlike the Tan-Walls-Collett scheme, which works out best in the limit of a very weak coherent part of the field. The considered decoherence model turned out to be analogous in its role to the depolarizing channel acting on a pair of entangled photons. One should mention, however, that with the current state-of-the-art technology we are not able perform the experiment, as it requires detectors, which distingush between different photon numbers. Nevertheless, there are no reasons to assume that the scheme will never be experimentally feasible.

Chapter 5 focuses on the problem of detecting thermal entanglement is solids. Conducting a Bell-type experiment on a set of about $10^{23}$ spins does not seem possible. Thus one detects entanglement in bulk bodies with witnesses, operators or state functions, which can take some values only for entangled state. Entanglement detected by witnesses can falsify Local Realism, but can also fall into the category of bound entanglement, which cannot be even distilled to a useful form \cite{boundentanglement,distillation}. We begin with explaining the concept of an entanglement witness \cite{peresppt,horodeckippt,witness} and showing that for specific systems low-temperature values of the internal energy cannot be explained only with separable states \cite{pla301,tothposter,qph0406040,tothinternalenergy}.

Nevertheless, we stress that it is a technical problem to learn the value of the internal energy at a given temperature. The first solution suggested by Wie\,sniak, Vedral, and Brukner \cite{w4} is to measure the magnetic susceptibility, whenever the Hamiltonian is rotationally invariant. This is a well-known routine in solid-state physics. We have proven that for these models the susceptibility converges to 0, or, more generally, a finite value at $T\rightarrow 0$ only because of entanglement of the thermal state. Notably, in contrast to the internal energy, the bound on which for separable states must be computed in each case individually, the magnetic response of the material is the most universal thermodynamical entanglement witness known up to date. As the coupling constants do not enter the expression, the quantity, thus its entanglement witnessing properties are independent of an exact form of the Hamiltonian. Our criterion can be as well applied to a simple spin-$\frac{1}{2}$ $xxx$ chain as to dimerized, frustrated, or higher-dimensional lattices. Even lattices with spins of various magnitudes can be tested for entanglement with our method. The other feature of the magnetic susceptibility is its advantage over the internal energy in a similar role. 
On the other hand, however, it is not possible to detect genuine multipartite entanglement with $\chi$, as it was shown possible for $U$ by G\"uhne, T\'oth, and Briegel \cite{guehnetothbriegel}. It is sufficient to recall that $|\Psi\rangle=\left(\frac{1}{\sqrt{2}}(|01\rangle-|10\rangle)\right)^{\otimes N/2}$ is an $N$-partite state of the total angular momentum 0, which is only bipartite-entangled. 

In \cite{w4} we have also presented a non-trivial complementarity relation between the magnetization and its variance (which in the absence of the magnetic field can be deduced from the susceptibility), which expresses macroscopic quantum information sharing. The left hand side of (\ref{wvbcomplementarity}) has been divided into two terms and only one of them can be maximized at a time. The first is dependent on pairwise correlations between qubits, while other is based on their local and individual properties. Thus the quantum information can be stored either in the magnetization, or the pairwise correlations, or in a form irrelevant to (\ref{wvbcomplementarity})

Finally, we show in \cite{w5} that whenever there is a certainty that the ground state is entangled, also the heat capacity can serve as an entanglement witness. Without purely quantum correlations $C$ would diverge to infinity at low temperatures. However, the low-temperature behavior of the heat capacity is under requirements of the Third Law of Thermodynamics \cite{nernst}. We have also based our argument on the {\em universality} of thermodynamical quantities. In this context the universality means that the thermodynamical functions have similar forms for various physical systems. Thermal entanglement can be revealed by the heat capacity for both general universality classes, i.e., systems with gap-less and gapped spectra. We, however require the knowledge of the ground-state energy and the bound for separable states. For this reason our considerations cannot be treated as a derivation of new argument, but rather a version of arguments based on the internal energy. 

In conclusions, we hope that our research has brought an impact in detecting entanglement in various physical systems. We have presented new Bell inequalities, methods of their derivation and conditions for their violation \cite{w3,w7,w8}. The other problem addressed by Wie\'sniak, Vedral, and Brukner \cite{w4,w5} was thermal entanglement in bulk systems. In one of the papers we have derived thus far the most general macroscopic entanglement witness, the magnetic susceptibility. The other article associates quantum correlation of the ground state with the Third Law of Thermodynamics. We believe that all these results have enlarged our knowledge about the role of entanglement in Nature. We hope that in future they will also find more specific applications in processing quantum information. 
\chapter{Acknowledgements}
I would like to thank my supervisor, prof. Marek \.{Z}ukowski, as well as my family and my love Magda for giving me their precious support and patience. Without them it would surely be impossible to finalize this and earlier stages of my education. I gratefully acknowledge members of all groups, in which I had worked, especially Wies\l aw Laskowski, Tomasz Paterek, \v{C}aslav Brukner, Vlatko Vedral, Johannes Kofler, Aires Ferreira, and Dagomir Kaszlikowski, for their collaboration. Then I thank all the nice people, who give a meaning to working for the society. I shall also thank the good spirit of the Quantum Information Technology community, who is uses his great power to help weaker.

\end{document}